\documentclass[9pt,journal,twocolumn]{IEEEtran}

\usepackage{framed,xcolor}
\usepackage{morefloats}
\usepackage{txfonts}
\usepackage{url}
\usepackage{graphicx}
\usepackage{threeparttable}
\usepackage{bm}
\usepackage{arydshln}
\usepackage[T1]{fontenc}

\definecolor{shadecolor}{gray}{0.90}

\DeclareMathAlphabet{\mathrm}{OT1}{antt}{li}{it}
\DeclareMathAlphabet{\mathrm}{OT1}{pzc}{m}{it}

\begin{document}
\title{Multigrid-based inversion  for volumetric  radar  imaging with asteroid interior reconstruction as a potential application}
\author{Mika Takala, Defne Us, Sampsa Pursiainen 
\thanks{M.\ Takala,  D.\ Us and S.\ Pursiainen are with Laboratory of Mathematics, Tampere University of Technology, PO Box 553, 33101 Tampere, Finland.}
\thanks{M.\ Takala is with Laboratory of Pervasive Computing, Tampere University of Technology.}
\thanks{D.\ Us is with Laboratory of Signal Processing, Tampere University of Technology.}
\thanks{Copyright  \textcircled{c} 2018 IEEE. Personal use of this material is permitted. However, permission to use this material for any other purposes must be obtained from the IEEE by sending a request to pubs-permissions@ieee.org.}
\thanks{Digital Object Identifier 10.1109/TCI.2018.2811908}}
\maketitle
\begin{abstract}
This study concentrates on advancing mathematical and computational methodology for radar tomography imaging in which the unknown volumetric velocity distribution of a wave within a bounded domain is to be reconstructed. Our goal is to enable effective simulation and inversion of a large amount of full-wave data within a realistic 2D or 3D geometry. For propagating and inverting the wave, we present a rigorous multigrid-based forward approach which utilizes the finite-difference time-domain method and a nested finite element grid structure. We also introduce and validate a multigrid-based inversion algorithm which allows regularization of the unknown distribution through a coarse-to-fine inversion scheme. Using this approach, sparse signals can be effectively inverted, as the coarse fluctuations are reconstructed before the finer ones. Furthermore, the number of nonzero entries in the system matrix can be compressed and, thus, the inversion procedure can be speeded up. As the test scenario, we investigate satellite-based asteroid interior reconstruction. We use both full-wave and projected wave data and estimate the accuracy of the inversion under different error sources: noise and positioning inaccuracies. The results suggest that the present inversion technique allows recovering the interior with a single satellite recording backscattering data. Robust results can be achieved, when the peak-to-peak signal-to-noise ratio is above 10 dB. Furthermore, the robustness for the deep interior part can be enhanced if two satellites can be utilized in the measurements. 
\end{abstract}

\IEEEpeerreviewmaketitle

\section{Introduction}

Modeling and inverting a full wave in order to estimate an unknown parameter distribution inside a target object \cite{kaipio2004, tarantola2005} is a computationally challenging imaging approach which has a wide range of applications, e.g., in biomedical microwave or ultrasonic tomography  \cite{grzegorczyk2012,meaney2010,fear2002, ruiter2012, opielinski2013, ranger2012}, non-destructive material testing \cite{yoo2003,chai2010,chai2011,acciani2008} and astro/geophysical radar imaging \cite{pursiainen2016, su2016, herique2016, kofman2015, kofman2007}. Waveform (radar) imaging is rapidly becoming a standard  approach in several fields of science and technology for the constantly increasing computational resources available for an ordinary user. For example, ultrasonic/microwave detection of breast lesions has recently been approaching the reliability of the classical computed X-ray tomography (CT) and magnetic resonance imaging (MRI)  \cite{opielinski2013, opielinski2015}. 

In this study, we evaluate and compare different inversion schemes for a hierarchical (nested)  multigrid  mesh \cite{braess2007} resulting from the finite element  (FE)  discretization of the spatial domain. This multigrid mesh structure is also an essential part of the forward simulation process in which geometrically complex interior and boundary structures need to be modeled. The internal relative permittivity of the target object determining the wave speed is reconstructed using a total variation regularized inversion technique \cite{clarkson1933,chambolle2004,scherzer2008, stefan2008, kaipio2004}.  In forward simulation,  a combination of the finite element method (FEM) \cite{braess2007} and finite-difference time-domain (FDTD) method \cite{schneider2016} is utilized. The multigrid mesh enables effective forward and inverse computations in connection with nested finite element mesh structures. In particular, we introduce and validate a coarse-to-fine  reconstruction algorithm that utilizes the multigrid approach to invert a  linearized wave equation. In this approach, the coarse-level estimate can be utilized to regularize sparsity-based reconstructions and also to filter out the uninteresting parts of the permittivity distribution to reduce the computational cost  for a  large  computational domain or  three-dimensional geometry. 

As a potential future application of the presented methodology, we consider reconstructing the interior structure of a small asteroid based on measurements performed by small orbiting spacecraft. The first attempt to reconstruct the interior structure of an small solar system body (SSSB) was made as a part of the Rosetta mission of the European Space Agency (ESA) in 2014. This Comet Nucleus Sounding Experiment by Radio-wave Transmission \cite{kofman2007, kofman2015} (CONSERT) aimed at finding out the permittivity of the comet P67/Churyumov-Gerasimenko based on a radio transmission between the mothership Rosetta and its lander Philae.  At the moment,  future missions are being planned as several organizations aim to explore SSSBs with probes. There is also a growing interest to use small spacecraft as a part of such a mission. In 2018, the Hayabusa-2 probe \cite{kawaguchi2008, tsuda2013} by Japan Aerospace Exploration Agency (JAXA) will deploy Mascot-1 \cite{lange2015}  (German Aerospace Center, DLR) which will land on the asteroid 162173 Ryugu (1999 JU3). The preliminary plans of the ESA's unrealized Asteroid Impact Mission (AIM) \cite{michel2016} to asteroid 65803 Didymos (1996 GT)  included at least three small spacecraft: the Mascot-2 asteroid lander by DLR as well as two or more CubeSats, which would be used to record radio-frequency data of the deep interior \cite{herique2016}.   

\begin{figure}
\begin{center}
\begin{minipage}{6.5cm} \begin{center} \includegraphics[width=5.0cm]{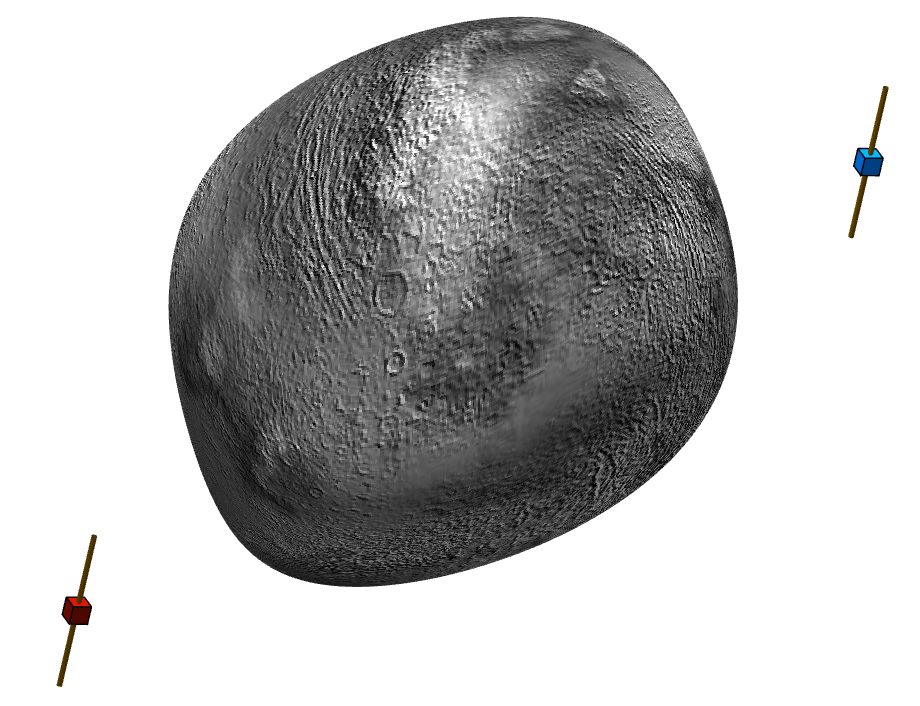} \end{center} \end{minipage}  
\end{center}
\caption{ Two small spacecraft exploring the interior of an asteroid. Each one is equipped with a  single-rod dipole antenna similar to those utilized in low-frequency georadar surveys. \label{satellites}}
\end{figure}

Motivated by these plans,  we simulate and compare: (1) dense vs.\ sparse measurements,  (2)  sparse monostatic and bistatic measurement configuration, (3) full wave vs.\ projected (compressed) data, and (4) single- and dual-resolution inversion approach.  In the numerical experiments, a cover layer and and internal voids of a 135 m diameter two-dimensional target are  to be reconstructed utilizing a 10 MHz radio-frequency signal \cite{binzel2005} with several different levels of total noise. The accuracy of the inversion results is analyzed via relative overlap and value error (ROE and RVE) measures.  The results suggest that the present  full-wave approach allows robust reconstruction of both the cover and the voids with a single satellite recording backscattering data (Figure \ref{satellites}), if the peak-to-peak signal-to-noise ratio is above 10 dB. Reconstructing the deep part of the interior can be enhanced, if,  instead of one, two or more satellites can be  used to gather the measurements. That is, if also the through-going wave can be recorded in addition to the backscattering part. The  present tomography approach seems promising for various volumetric waveform imaging contexts. For the generality of the results, we present an alternative scaling of the parameters for 50 MHz radio and 10 GHz microwave frequencies, and give examples of related tomography applications. 

This paper is organized as follows. Section 2 describes the mathematical forward and inverse approaches of this study.  Section 3 and 4 describe the numerical experiments and the results obtained, respectively, and Section 5 includes the discussion. 

\section{Materials and methods}

\subsection{Wave equation}
\label{section:forward_model} 

In this study, we model the scalar electric potential distribution $u$ in the spatio-temporal set $[0, T] \times \Omega$.  This distribution is assumed to satisfy the following hyperbolic wave equation
\begin{equation} 
\label{pde1}
\varepsilon_r \frac{\partial^2 u}{\partial t^2} + \sigma \frac{\partial u}{\partial t}  - \Delta_{\vec x} u = \frac{\partial f}{\partial t} \quad \hbox{for all} \quad (t,{\vec x}) \in  [0, T] \times \Omega 
\end{equation}
with a given real-valued relative electric permittivity $\varepsilon_r$, real conductivity distribution $\sigma$, and the initial conditions $u|_{t = 0} = u_0$ and $(\partial u/ \partial t) |_{t  = 0} = u_1$. The right hand side is a signal source of the form $\partial f(t, \vec{x} ) / \partial t= \delta_{\vec{p}}(\vec{x}) \partial \tilde{f}(t) / \partial t $ transmitted at the point $\vec{p}$.  Here,  $\tilde{f}(t)$ denotes  the time-dependent part of $f$ and $\delta_{\vec{p}}(\vec{x})$ a Dirac's delta function with respect to  $\vec{p}$. The governing equation (\ref{pde1}) can be formulated as a first-order system of the form 
 \begin{equation}\label{first-order_system}
\varepsilon_r \frac{\partial u}{\partial t} + \sigma {u} - \nabla \cdot {\vec g}  =   f \quad \hbox{and} \quad 
\frac{\partial {\vec g}}{\partial t}  -   \nabla u  =  0, \quad \hbox{in} \quad \Omega \times [0,T], 
\end{equation}
where ${\vec g} = \int_0^t \nabla u(\tau, \vec{x}) \, d \tau$, ${\vec g} |_{t = 0}  = \nabla u_0$ and $u |_{t = 0} = u_1$. Integrating (\ref{first-order_system}) multiplied by $v  : [0,T] \to H^1(\Omega)$ and ${\vec w} : [0,T] \to L_2(\Omega)$ and applying the rule of partial integration, one can obtain the weak form 
{\setlength\arraycolsep{1 pt} \begin{eqnarray} \label{weak_form_1}
\frac{\partial }{\partial t} \! \int_\Omega \! {\vec g} \cdot {\vec w} \, \hbox{d} \Omega \! - \! \int_\Omega \! {\vec w} \cdot \nabla u \, \hbox{d}  \Omega & = & 0, \\ 
\frac{\partial}{\partial t} \!  \int_\Omega \! \varepsilon_r  \, u \, v \, \hbox{d}  \Omega \! + \! \int_\Omega \!  \sigma  \, u \, v \, \hbox{d}  \Omega \! + \! \int_\Omega \!  {\vec g} \cdot \nabla v \, \hbox{d}  \Omega & = & \left\{ \begin{array}{ll} \tilde{f}(t), & \hbox{if } \vec{x} \! = \! \vec{p}, \\ 0, &  \hbox{else}.  \end{array} \right. \label{weak_form_2}
\end{eqnarray}} 
Here, it is assumed that the domain and the parameters are regular enough, so that the weak form has a unique solution $u : [0,T] \to H^1(\Omega)$ \cite{evans1998}. 

\subsection{Multigrid forward approach}
\label{mutigrid_approach}

In order to solve (\ref{pde1}) numerically, we cover the $\mathrm{d}$-dimensional ($\mathrm{d} = 2 $ or $3$) spatial domain $\Omega$  with a  $\mathrm{d}$-simplex mesh  $\mathcal{T}$ consisting of $m$ elements $ \mathrm{T}_1, \mathrm{T}_2, \ldots, \mathrm{T}_m $.  Each element $\mathrm{T}_i$ is associated with an indicator function $\chi_i  \in L_2(\Omega)$. The set of $n$ mesh nodes $\vec{r}_1, \vec{r}_2, \ldots, \vec{r}_n$ is identified with piecewise linear (nodal) basis functions $\varphi_1,\varphi_2, \ldots, \varphi_n \in H^1(\Omega)$ \cite{braess2007}.  The potential and gradient fields are approximated as the finite sums $u = \sum_{j = 1}^n p_j  \, \varphi_j$ and ${\vec g } = \sum_{k = 1}^\mathrm{d} g^{(k)} {\vec e}_k$ with $g^{(k)} = \sum_{i = 1}^m q^{(k)}_i \, \chi_i$ which are associated with the coordinate vectors ${\bf p} = (p_1,p_2,\ldots, p_n)$, ${\bf q}^{(k)} = (q^{(k)}_1, q^{(k)}_2, \ldots, q^{(k)}_m)$, respectively. The numerical solution of the wave equation yielding ${\bf p}$ and ${\bf q}^{(k)}$ has been briefly reviewed in Section \ref{section:forward_simulation} (Appendix).

The relative permittivity distribution is reconstructed utilizing an approximation   $\varepsilon_r = \sum_{j = 1}^M {c}_j {\chi}'_j$ which is defined for the coarse and nested $\mathrm{d}$-simplex mesh ${\mathcal{T}'}$  with  $M$ elements $\mathrm{T}'_1, \mathrm{T}'_2, \ldots, \mathrm{T}'_M $  and $N$ nodes $\vec{r}_1, \vec{r}_2, \ldots, \vec{r}_N$ shared by the dense mesh  $\mathcal{T}$ ($N < n$). In the reconstruction procedure, the relation between the coordinate vector of the permittivity ${\bf x} = (c_1, c_2, \ldots, c_M)$  and discretized potential field ${\bf p}$  is approximated via the following linearized  forward (data prediction) model
\begin{equation}
{\bf p} [{\bf x}] =  {\bf p}[{{\bf x}_0}] + {\bf J}[{{\bf x}_0}] ({\bf x}  - {{\bf x}_0} ).
\end{equation}
Here, ${{\bf x}_0}$ is an initial {\em a priori} guess for the permittivity and ${\bf J}[{{\bf x}_0}]$ is a Jacobian matrix, which  consists of partial derivatives of the form ${\partial {\bf p}_\ell}/{\partial c_j}$ evaluated at ${\bf x}_0$. The partial derivative ${\partial {\bf p}_\ell}/{\partial c_j}$ can be formulated as the sum \begin{equation} \label{approx_1} \frac{\partial {\bf p}_\ell}{\partial c_j}  = \sum_{\vec{r}_i \in \mathrm{T}'_j, \,  i \leq n}  {\bf d}_\ell^{(i,j)} \end{equation}  in which  $ {\bf d}_\ell^{(i,j)}$ can be obtained by solving the auxiliary system of the form (\ref{linearized1}) and  (\ref{linearized2})  as shown in Appendix \ref{section:forward_simulation}. 

The number of the terms in the sum ${\partial {\bf p}_\ell}/{\partial c_j}  = \sum_{\vec{r}_i \in \mathrm{T}'_j, \,  i \leq n}  {\bf d}_\ell^{(i,j)}$ depends on the density of the finite element mesh $\mathcal{T}$, that is, the number of nodes $\vec{r}_i \in \mathcal{T}$ belonging to $ \mathrm{T}'_j$. In order to lower this number, and thereby also reduce the computational work in forming the model, we redefine (\ref{approx_1}) with respect to the coarse mesh $\mathcal{T}'$ as follows:  \begin{equation} \label{approx_2} \frac{\partial {\bf p}_\ell}{\partial c_j}  \approx \sum_{\vec{r}_i \in \mathrm{T}'_j, \,  i \leq N}  {\bf d'}_\ell^{(i,j)} =  \sum_{k = 1}^{\mathrm{d}+1}  {\bf d'}_\ell^{(i_k,j)}, \end{equation} where the number of terms is $\mathrm{d} + 1$,  that is, the number of nodes in $\mathrm{T}'_j$. This estimate is, in this study, regarded as the multigrid approach to forward modeling, as it is based on the nested multigrid mesh structure (Appendix  \ref{section:forward_simulation}). 

\subsection{Reconstruction procedure}
\label{inversion}

The linearized forward model can be written as the linear system ${\bf y}  - {{\bf y}_0}  =  {\bf L} \, ( {\bf x} - {{\bf x}_0})  + {\bf n}$, where ${\bf y}$ is the actual data vector, ${{\bf y}_0}$ and ${\bf L}$ denote the  simulated data ${\bf p}[{{\bf x}_0}]$  and Jacobian matrix ${\bf J}[{{\bf x}_0}]$ for a constant {\em a priori} guess ${{\bf x}_0}$, and ${\bf n}$ is a noise vector containing both modeling and measurement errors. In this study, a regularized solution of  ${\bf x}$ is obtained via the iteration \begin{equation} \label{tv_iteration} {\bf x }_{\ell+1}  = {{\bf x}_0} + ({\bf L}^T {\bf L} + \alpha {\bf D} {\bf \Gamma}_{\ell} {\bf D} )^{-1} {\bf L}^T ( {\bf y} - {{\bf y}_0}) \end{equation} in which ${\bf \Gamma}_{\ell}$ is a weighting matrix satisfying ${\bf \Gamma}_0 = {\bf I}$  and ${\bf \Gamma}_{\ell} = \hbox{diag} ( |{\bf D} [{\bf x_{\ell}}-{\bf x_{0}}]|)^{-1} $ for $\ell \geq 1$, and ${\bf D}$ denotes a  regularization matrix  of the form
\begin{equation}
\label{d_mat}
D_{i,j} =  \beta \delta_{i,j}  + \frac{(2\delta_{i,j} - 1) \int_{\mathrm{T}'_i \cap \mathrm{T}'_j} \, \hbox{d} s }{ \max_{i,j}  \int_{\mathrm{T}'_i \cap \mathrm{T}'_j} \, \hbox{d} s  } , \quad \! \! \! \delta_{i,j} =
\left\{ \begin{array}{ll}  1, & \hbox{if } j = i, \\ 0, & \hbox{otherwise}.  \end{array} \right. 
\end{equation} 
The first regularization term limits the magnitude of ${\bf x}$ and the second one penalizes the jumps over the edges of $\mathcal{T}'$. The iteration (\ref{tv_iteration}) can be shown \cite{pursiainen2016} to minimize $F({\bf x}) =  \| {\bf L} ({\bf x}-{{\bf x}_0})  - ({\bf y} -  {{\bf y}_0} ) \|^2_2 + 2 \sqrt{\alpha} \|{\bf D}({\bf x} - {{\bf x}_0}) \|_1$ in which the second term equals to the total variation of ${\bf x}$, if $\beta = 0$ \cite{scherzer2008, stefan2008, kaipio2004,chambolle2004}.  Here, the total variation regularization strategy produces distributions with large connected subsets close to constant, which helps to avoid noise in the  reconstructions. 

\subsection{Coarse-to-fine approach}
\label{coarse_to_fine}

\begin{figure*}
\begin{center} 
\begin{minipage}{3.9cm}
\begin{center}
\includegraphics[height=1.7cm]{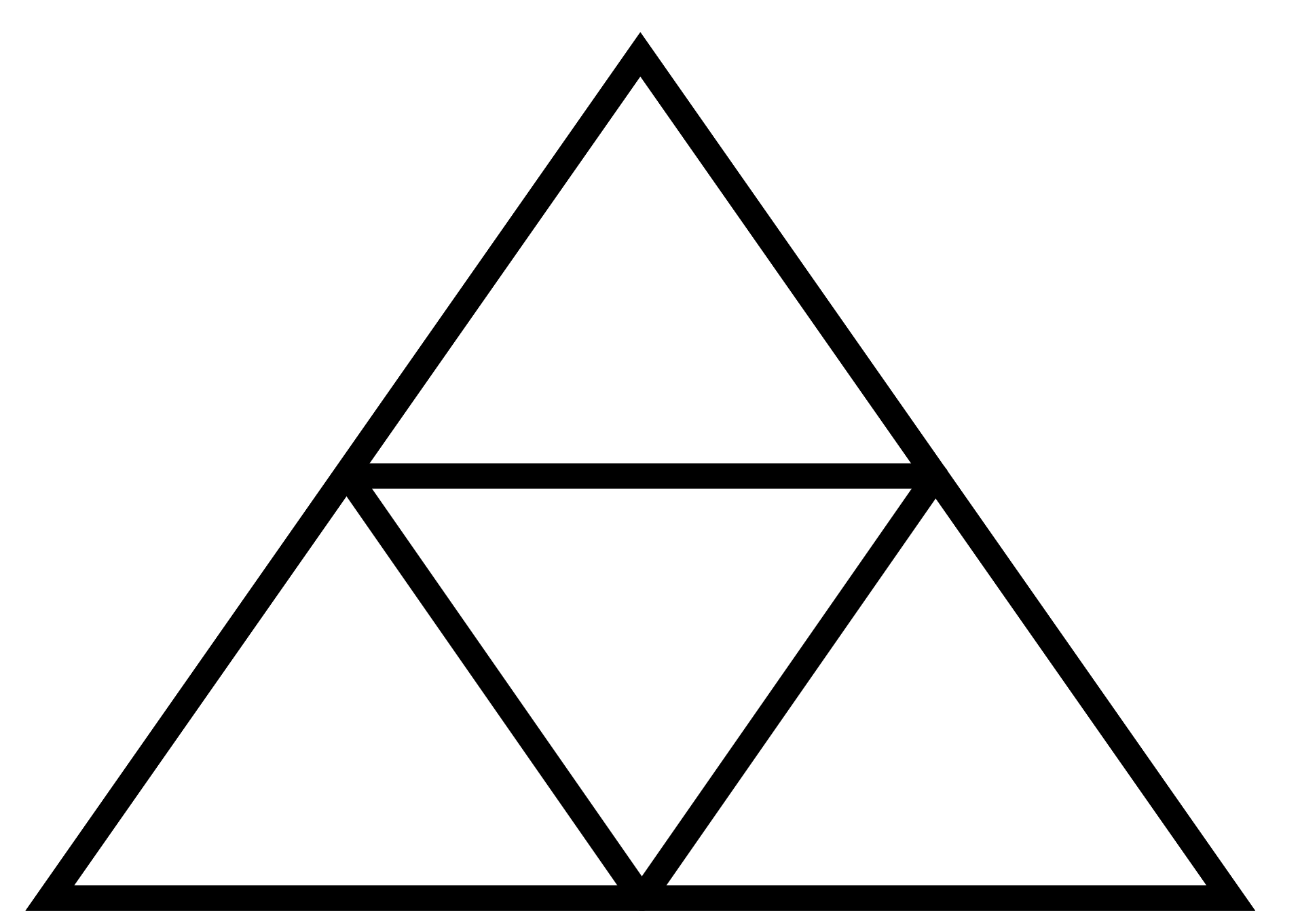} \\ 
$\mathcal{S}_{\mathcal{T}'}$
\end{center}
\end{minipage}
\begin{minipage}{3.9cm}
\begin{center}
\includegraphics[height=1.7cm]{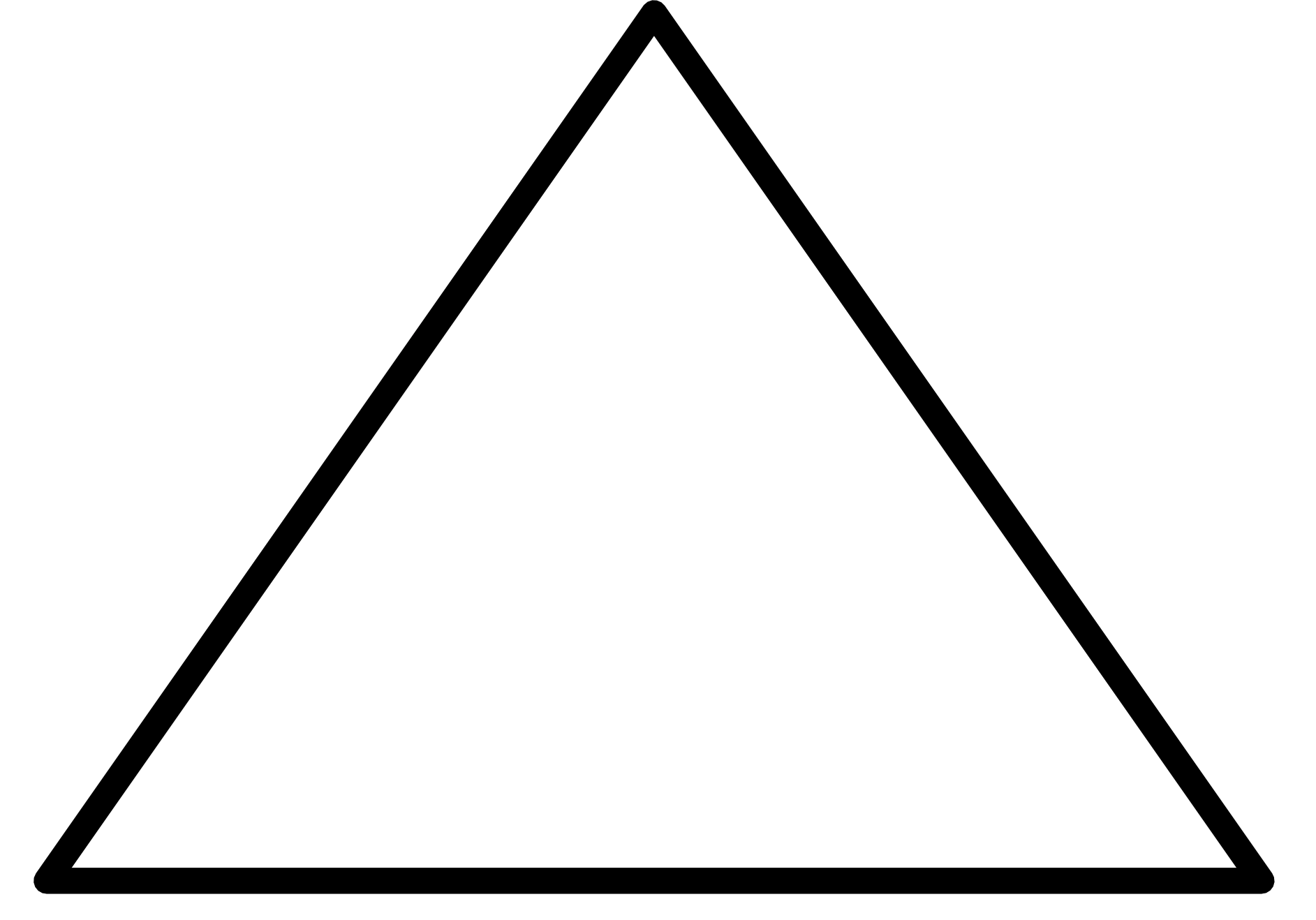} \\ 
$\mathcal{S}_{\mathcal{T}''}$
\end{center}
\end{minipage}
\begin{minipage}{3.9cm}
\begin{center}
\includegraphics[height=1.7cm]{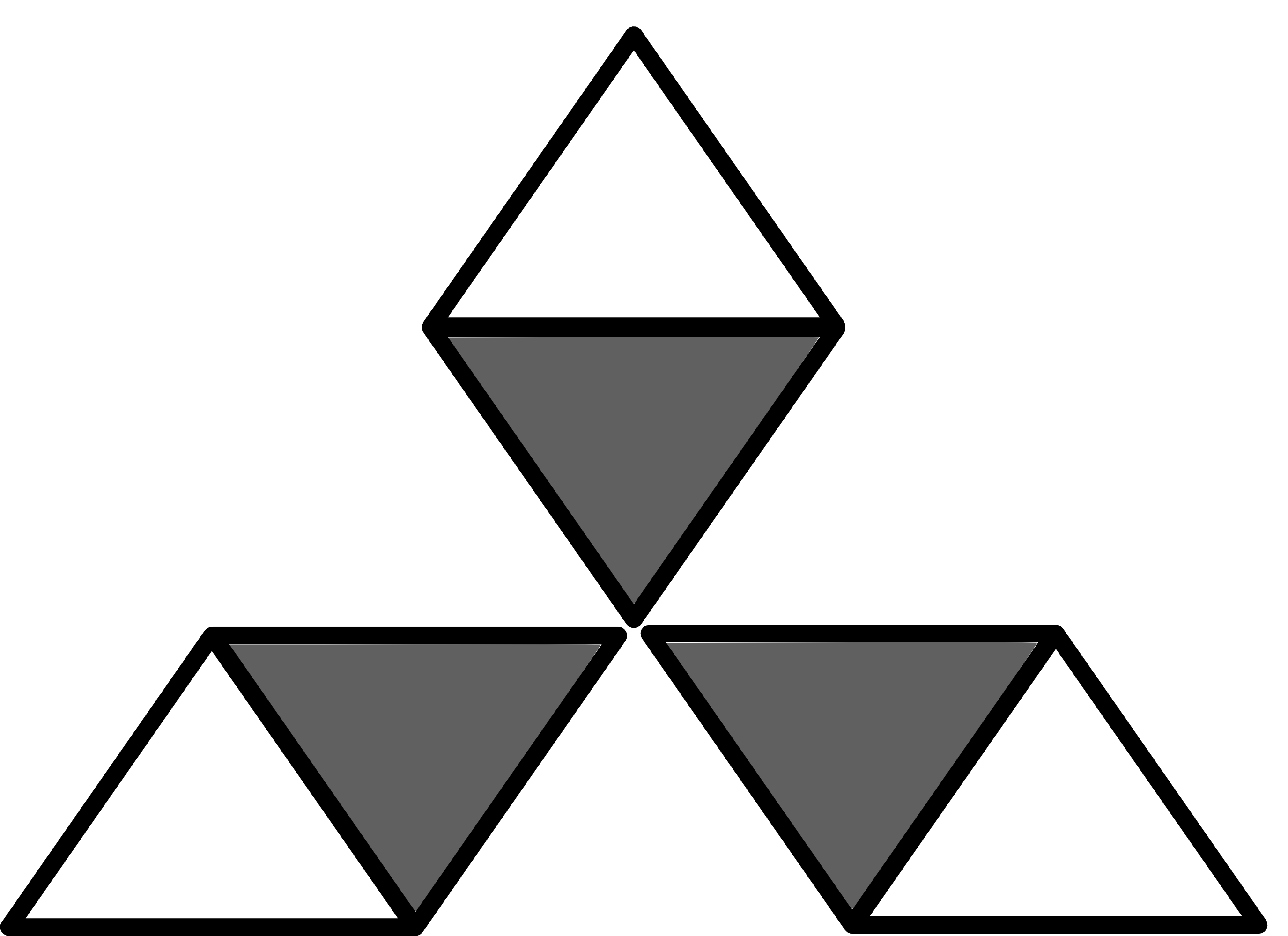} \\ 
$\mathcal{S}_{\mathcal{T}' \setminus \mathcal{T}''}$
\end{center}
\end{minipage}
\end{center}
\caption{Each triangle of the coarse mesh $\mathcal{T}''$ contains four  nested triangles of the fine one $\mathcal{T}'$. The triangles of  $\mathcal{T}'$ and  $\mathcal{T}''$ span, respectively, the spaces  $\mathcal{S}_\mathcal{T}'$ (left)  and $\mathcal{S}_\mathcal{T}''$ (center), of  trianglewise constant basis functions. The space  $\mathcal{S}_{\mathcal{T}' \setminus \mathcal{T}''}$ (right) contains those fine fluctuations of $\mathcal{S}_\mathcal{T}'$ that do not belong to $\mathcal{S}_\mathcal{T}''$. Each basis function of $\mathcal{S}_{\mathcal{T}' \setminus \mathcal{T}''}$ obtains the values 1 and -1 in an adjacent pair of triangles (white and black, respectively) and is zero elsewhere. Consequently, each linear combination of the fine resolution basis functions of  $\mathcal{S}_\mathcal{T}'$  can be presented as a linear combination of the coarse basis functions of $\mathcal{S}_\mathcal{T}''$ those of $\mathcal{S}_{\mathcal{T}' \setminus \mathcal{T}''}$. \label{mesh_hierarchy}}
\end{figure*}

Using two triangular nested inversion meshes  $\mathcal{T}'$ and $\mathcal{T}''$ , one can obtain a coarse-to-fine version of the inversion algorithm (\ref{tv_iteration}) in which the coarse  details are reconstructed before the finer ones.   Here, the  split between the two resolution levels is given by the decomposition  \begin{equation} \label{decomposition} \mathcal{S}_\mathcal{T'} =  \mathcal{S}_\mathcal{T''} \oplus \mathcal{S}_{\mathcal{T}'  \setminus \mathcal{T}''} \end{equation} in which   $\mathcal{S}_\mathcal{T'}$ and  $\mathcal{S}_\mathcal{T''}$ (Figure \ref{mesh_hierarchy}) denote the spaces of all piecewise (trianglewise) constant distributions of  $\mathcal{T}'$ and $\mathcal{T}''$ and $\mathcal{S}_{\mathcal{T}'  \setminus \mathcal{T}''}$ is the space of those fine fluctuations of $\mathcal{S}_\mathcal{T''}$ that do not  belong to $\mathcal{S}_\mathcal{T''}$. The basis  for  $\mathcal{S}_{\mathcal{T}'  \setminus \mathcal{T}''}$   can be chosen in various ways.  In this study, each basis function  obtains the constant values 1 and -1 in an adjacent pair of triangles and is zero elsewhere (Figure \ref{mesh_hierarchy}). Consequently, the support is minimal. For a given permittivity distribution and set of basis functions, both coarse and fine fluctuations are uniquely determined by (\ref{decomposition}). 

Denoting by $\mathcal{R}_c$ and $\mathcal{R}_f$ two linear  coordinate transforms  from  $\mathcal{S}_\mathcal{T''}$ and  $\mathcal{S}_{\mathcal{T}'  \setminus \mathcal{T}''}$  to  $\mathcal{S}_\mathcal{T'}$, respectively, one can define a coarse and fine level matrix ${\bf L}_c  = {\bf L} {\mathcal R}_c$ and ${\bf L}_f  = {\bf L} {\mathcal R}_f$ which determine the forward model for  the coarse and fine details. That is, if ${\bf x} = \mathcal{R}_c {\bf }{\bf x }^{(c)} + \mathcal{R}_f {\bf x }^{(f)}$, where ${\bf x }^{(c)}$ and ${\bf x }^{(f)}$ are  coordinate vectors for functions in $\mathcal{S}_\mathcal{T''}$ and  $\mathcal{S}_{\mathcal{T}'  \setminus \mathcal{T}''}$, respectively, then the forward model can be decomposed as  ${\bf L} {\bf x}= {\bf L}_c {\bf x}^{(c)} + {\bf L}_f {\bf x}^{(f)}$.  

With these definitions, the proposed coarse-to-fine inversion routine can be written as follows: 
\begin{description}
\item[(1)] Set the initial guess ${\bf x}_0 = (0,0,\ldots,0)$.   
\item[(2)] Find a coarse resolution estimate belonging to $\mathcal{S}_\mathcal{T''}$ as given by 
\begin{equation}
\label{tv_iteration_0}
{\bf x }^{(c)}_{\ell+1}  = {\bf x }_{0}^{(c)} + ({\bf L}_c^T {\bf L}_c + \alpha {\bf D}_c {\bf \Gamma}_{\ell} {\bf D}_c )^{-1} {\bf L}_c^T  ({\bf y} - {\bf y}_0) 
\end{equation} 
with ${\bf \Gamma}_{\ell} = \hbox{diag} ( |{\bf D}_c [{\bf x}^{(c)}-{\bf x}^{(c)}_{0}]|)^{-1}$.
\item[(3)] Find a fine resolution correction that belongs to $\mathcal{S}_{\mathcal{T}'  \setminus \mathcal{T}''}$ as given by
\begin{equation}
\label{tv_iteration_2}
{\bf x }^{(f)}_{\ell+1}  = {\bf x }^{(f)}_{0}  +  ({\bf L}_f^T {\bf L}_f + \alpha {\bf D}_f {\bf \Gamma}_{\ell} {\bf D}_f )^{-1} {\bf L}_f^T  ( {\bf y} - {\bf y}_0 -  {\bf L}_c [{\bf x}_{\ell+1}^{(c)} - {\bf x}^{(c)}_0 ])
\end{equation} 
with $ {\bf \Gamma}_{\ell} = \hbox{diag} ( |{\bf D}_{f} [{\bf x}^{(f)}_{\ell} - {\bf x}^{(f)}_{0}]|)^{-1}$.
\item[(4)] Set ${\bf x}_{\ell+1} = {\bf x}^{(c)}_{\ell+1} + {\bf x}^{(f)}_{\ell+1}$ and $\ell \to \ell + 1$. 
\item[(5)] If $\ell$ is smaller than the desired number of iterations, then repeat the steps  {\bf (2)}--{\bf (5)}.
\end{description}
If  more than two nested meshes are used, then  the correction step {\bf (2)} can go through multiple resolution levels $f_1, f_2, \ldots, f_{n_f}$ via the recursive process: 
\begin{eqnarray}
\label{tv_iteration_3}
{\bf x }^{(f_i)}_{\ell+1}  & = &  ({\bf L}_{f_i}^T {\bf L}_{f_i} + \alpha {\bf D}_{f_i} {\bf \Gamma}_{\ell} {\bf D}_{f_i} )^{-1} {\bf L}_{f_i}^T  ( {\bf y} -{\bf y}_0 - {\bf L}_c [{\bf x}_{\ell+1}^{(c)} - {\bf x}_{0}^{(c)}] \nonumber \\  &  & - \sum_{k = 1}^{i-1} {\bf L}_{f_i} [{\bf x}_{\ell+1}^{(f_i)}- {\bf x}_{0}^{(f_i)}])
\end{eqnarray} 
with $ {\bf \Gamma}_{\ell} = \hbox{diag} ( |{\bf D}_{f_i} [{\bf x}^{(f_i)}_{\ell} - {\bf x}^{(f_i)}_{0}]|)^{-1}$ for $i = 1, 2, \ldots, n_f$.

\subsection{Compression of coefficients}

An important aspect in the above coarse-to-fine algorithm is that the size of ${\bf L}_f$ is multiple times that of ${\bf L}_c$. For the general structure of image information, low- and high-frequency fluctuations are often concentrated on the same image areas. Hence, the coarse scale structures in the image can be used as a basis to filter and compress also the finer details. In particular, the parts  in which an image is close to zero can often be recognized based on its  coarse structures. In this study, ${\bf L}_f$ is compressed on each step of the coarse-to-fine routine:  The matrix $\mathcal{R}_f$ and thereby also ${\bf L}_f$ are defined with respect to those basis functions of  $\mathcal{S}_{\mathcal{T}' \setminus \mathcal{T}''}$ that correspond to the largest third of the coefficients of $|{\bf x }^{(c)}_{\ell+1} - {\bf x }^{(c)}_{0}|$. Since in 2D each element of $\mathcal{S}_\mathcal{T}''$ covers three basis functions of $\mathcal{S}_{\mathcal{T}' \setminus \mathcal{T}''}$, the compressed size of ${\bf L}_f$ is identical to that of ${\bf L}_c$.  Compressing the system might be necessary for large computational domains and three dimensional geometries.

\begin{figure*}
\begin{center}
\begin{minipage}{3.9cm}
\begin{center}
\includegraphics[height=2.6cm]{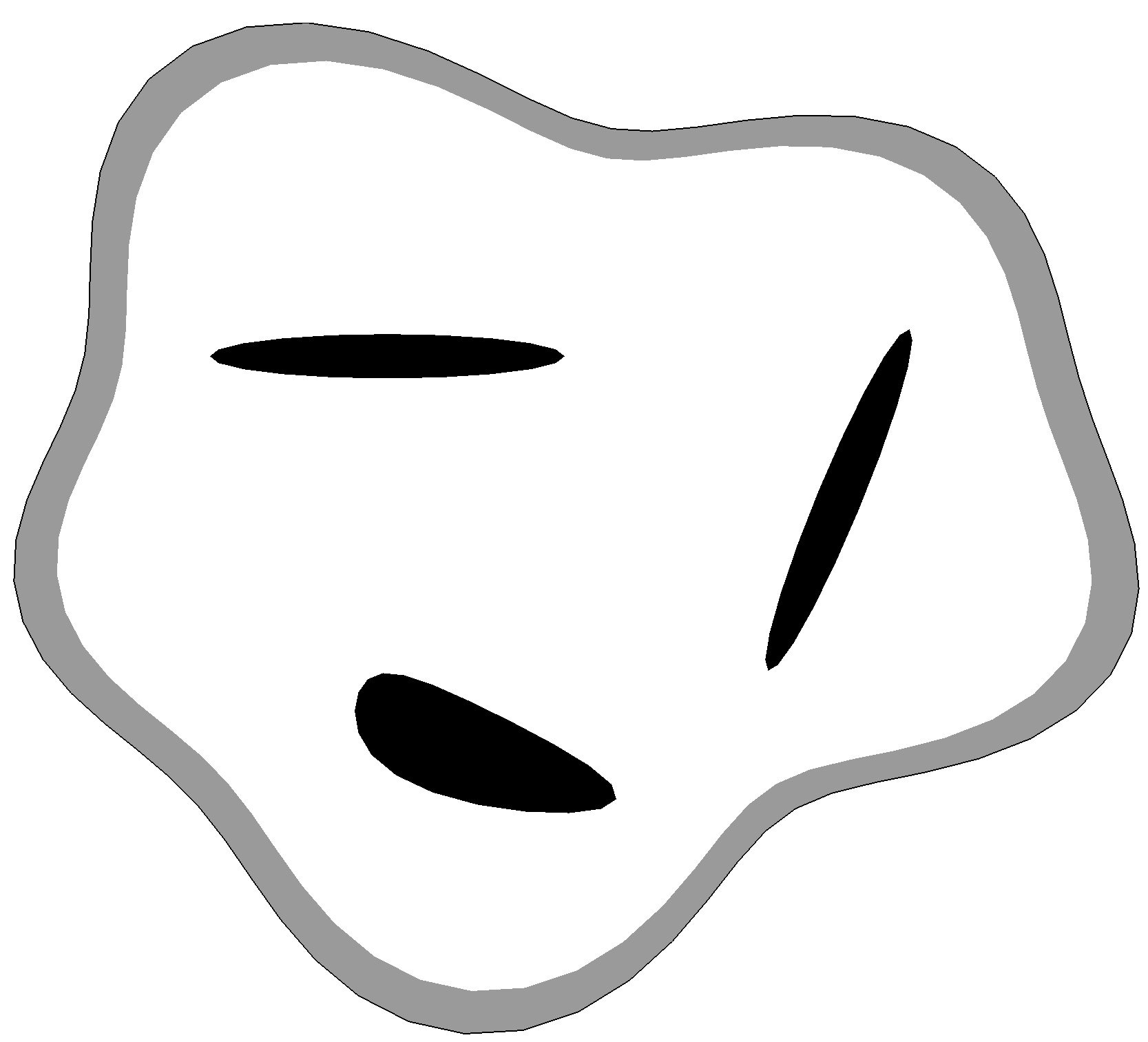} 
\end{center}
\end{minipage} 
\begin{minipage}{3.9cm}
\begin{center}
\includegraphics[height=2.6cm]{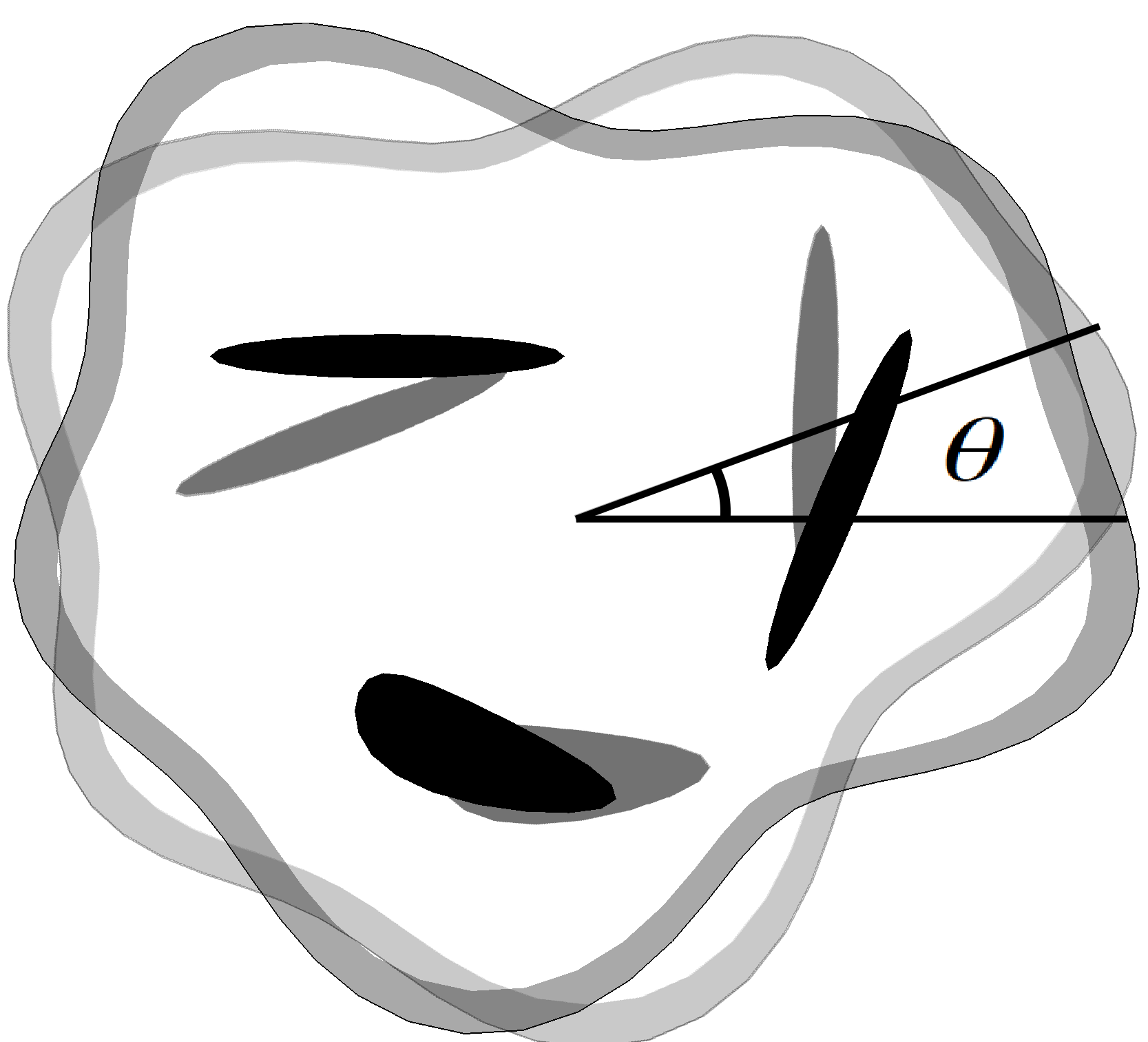} 
\end{center}
\end{minipage} 
\begin{minipage}{3.9cm}
\begin{center}
\includegraphics[height=2.6cm]{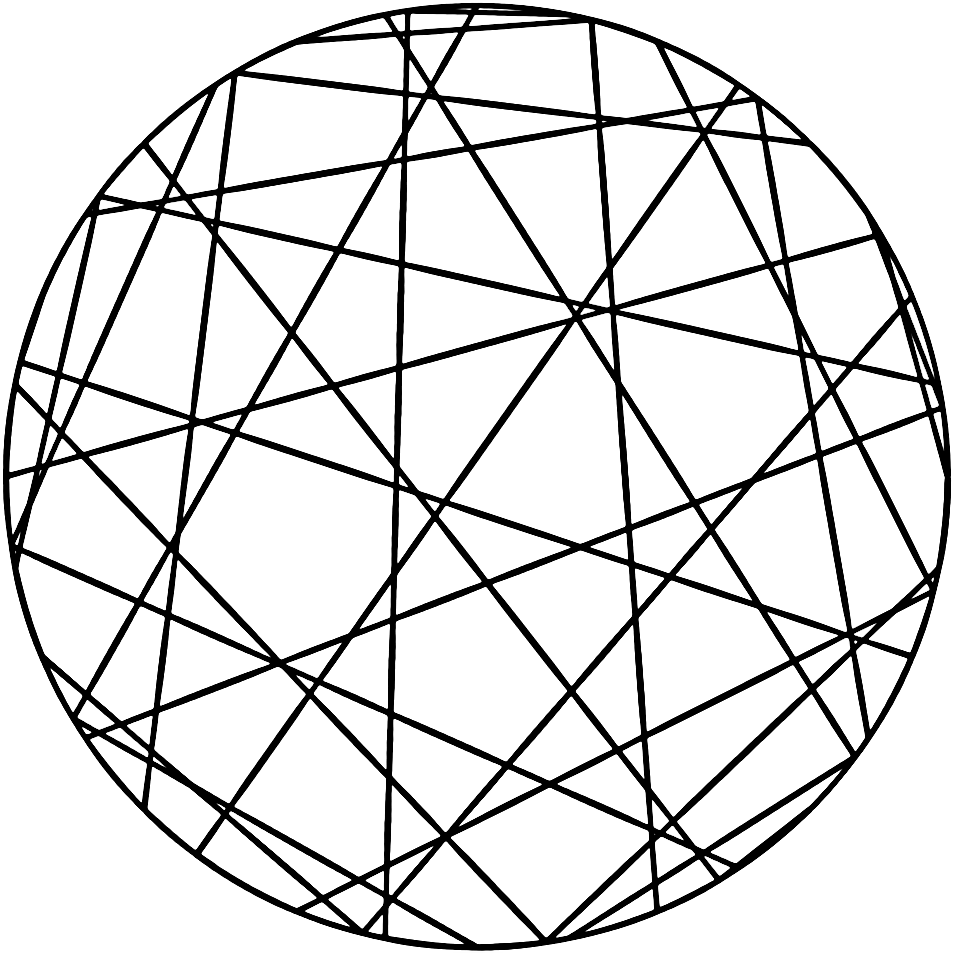} 
\end{center}
\end{minipage} 
\end{center}
\caption{{\bf Left:} The  target $\mathcal{D}$ utilized in the numerical experiments. {\bf Center:} The sensitivity to orientation errors is investigated with different values of the polar angle $\theta$. {\bf Right:}  Line segments illustrating  the orbiter-to-orbiter signal paths  in the bistatic signal configuration ({\bf C}). \label{test_domain}}
\end{figure*}

\subsection{Projected data}

Inversion of projected (compressed) data was investigated by defining the following inner products (projections)
\begin{equation}
\langle a, b \rangle_1 =  \int_0^T a(t) \, b(t) \, dt \quad  \hbox{and} \quad \langle a, b \rangle_2 =  \frac{1}{T}  \int_0^T t \, a(t) \, b(t) \, dt
\end{equation}
and the projected data vectors ${\bf y} = ({\bf y}^{\, (1)},  {\bf y}^{\,(2)})$ and ${\bf y}_{0} = ( {\bf y}_{\! 0}^{\, (1)},  {\bf y}_{\! 0}^{\, (2)})$ with $
y^{\, (j)}_i =  \langle u(t, \vec{p}_i), \, u(t, \vec{p}_i)  \rangle_j $  and $y_0^{\, (j)}\mbox{}_i =  \langle u(t, \vec{p}_i), \, u_0(t, \vec{p}_i)  \rangle_j$, 
where  $\vec{p}_i$ denotes the i-th measurement point, $u_m$ denotes the measured potential field and  $u_0$ the  simulated one corresponding to the {\em a priori} guess for the  permittivity distribution. It holds that \begin{equation} \frac{y^{\, (2)}_i}{y^{\, (1)}_i} = \frac{ \int_0^T t \, u(t, \vec{p}_i)^2 \, dt}{ T \int_0^T u(t, \vec{p}_i)^2 \, dt}, \end{equation}
that is, the relative integrated travel-time of the signal pulse. The Jacobian matrix for the projected data can be formed through 
\begin{equation}
\frac{\partial }{\partial c_\ell}  \langle u(t, \vec{p}_i), \, u_0(t, \vec{p}_i)  \rangle_j =   \langle u(t, \vec{p}_i), \, \frac{\partial }{\partial c_\ell} u_0(t, \vec{p}_i)  \rangle_j .
\end{equation}
The goal in the projected inversion approach is to find a permittivity distribution  such that the corresponding data residual is orthogonal to the original data in terms of both $\langle \cdot, \cdot \rangle_1$ and $\langle \cdot, \cdot \rangle_2$. 

\section{Numerical experiments}

\begin{table*}[t] \begin{center}
\caption{The properties of the test domain $\Omega$ with three different choices of the scaling parameter $s$.}
\label{scaling_values_domain}
\begin{tabular}{lllllllllll} 
Scaling &  {$s$}  & & Circle $\mathcal{C}$ &Target $\mathcal{D}$    & Surface layer&  Voids  \\
\hline
 {No scaling}  & 1 & Diameter & 0.32 &0.28 & 0.02 & 0.01-- 45 m \\
& & $\varepsilon_r$ & & 4 & 3 & 1 \\
& & $\sigma$ & & 20 & 15 & 5 \\
 ({\bf I}) & 500 m& Diameter &160 m &140 m& 10 m & 5 -- 45 m \\
& & $\varepsilon_r$ & & 4 & 3 & 1 \\
& & $\sigma$ & & 1.1E-4 S/m &  8.0E-5  S/m& 2.7E-5 S/m \\
 ({\bf II}) & 1250 m& Diameter & 400 m & 350 m& 25 m & 12 -- 110 m \\
& & $\varepsilon_r$ & & 4 & 3 & 1 \\
& & $\sigma$ & & 4.4E-5 S/m &  3.2E-5  S/m& 1.1E-5 S/m \\
({\bf III}) & 0.5 m  & Diameter& 16 cm& 14 cm  & 1 cm & 0.5 -- 4.5 cm \\
& & $\varepsilon_r$ & & 4 & 3 & 1 \\
& & $\sigma$ & &0.11 S/m &  0.80  S/m& 0.027  S/m \\
\hline
\end{tabular}
\end{center}
\end{table*}

\begin{table*}[t] \begin{center}
\caption{Signal properties for three different scaling parameter values. For $s = 500$ m, the model can be interpreted as a monolithic asteroid and, for $s = 1250$ m, as a rubble pile, as its  diameter is above the rubble pile spin barrier  \cite{kwiatkowski_photometric_2010} and conductivity is relatively low (due to a higher porosity).  For $s = 0.5$ m, the model has a correspondence to microwave range applications. }
\label{scaling_values_signal}
\begin{tabular}{lllllllllll} 
 &   &    &  & Pulse & Simulated   &  \\
Scaling {$s$}  & Target & $\mathcal{D}$ &  Bandwidth$\mbox{}^{1}$ &duration $T$  & wavelength in $\mathcal{D}$ & Class \\
\hline
 ({\bf I}) & 500 m &  Asteroid (monolith) & 10 MHz &  2.2  $\mu$s & 15 m    & Radio wave  \\
 ({\bf II}) & 1250 m & Asteroid (rubble pile)  &   4 MHz  & 5.5  $\mu$s & 3--7.5 m    & Radio wave  \\
({\bf III}) & 0.5  m &  Wood / & 10 GHz &  2.2 $n$s    &15 mm & Microwave \\
 &   &  Fatty tissue&     & &   &  \\
 &   &  Bone marrow&     & &   &  \\
\hline
\end{tabular}
\end{center}
$\mbox{}^{1}$ The minimum radar bandwidth required for lossless pulse transmission.
\end{table*}

\begin{table}[t] \begin{center}
\caption{The signal configurations ({\bf A})--({\bf C}): The number of measurement points, oversampling rate w.r.t.\ Nyquist criterion, the number of the orbiters, and the orbiting velocity ratio between the satellites. }
\label{signal_configurations}
\begin{tabular}{llllllllll} 
 &  Meas.\  & Oversampling  & & Velocity \\
Positioning &  points  &  rate & Orbiters  & ratio\\
\hline
 ({\bf A}) & 128  &  1.6  &  1 & -  \\
({\bf B}) & 32 &  0.4 & 1 & - \\
({\bf C}) & 32  &  0.4  &  2 & 10 \\
\hline
\end{tabular}
\end{center}
\end{table}

\begin{table}[t] \begin{center}
\caption{The mean squared error (MSE) values obtained for coarse, dual and single resolution reconstruction for the signal configuration ({\bf A}) at noise levels $\eta = -25$ and $\eta = -11$ (decibels). The first one of these is the final coarse level estimate (Figure \ref{reconstructions_levels}) of the dual resolution inversion scheme (Section \ref{coarse_to_fine}).}
\label{mse_table}
\begin{tabular}{lllll} 
& &  \multicolumn{3}{c}{MSE$\cdot 10^{2}$}\\ \cline{3-5} \\
Data &  Noise $\eta$ (dB) &  Coarse Res. & Dual Res. & Single Res. \\
\hline
Full  &  -25 & 1.41  &  1.39 &   1.38 \\
Full  &  -11 & 1.40   & 1.41   &  1.50  \\
Projected   & -25 & 1.51   & 1.47  &  1.15  \\
Projected   & -11 &  1.88  &  1.86  &  1.90  \\
\hline
\end{tabular}
\end{center}
\end{table}

\begin{table}[t] \begin{center}
\caption{Signal configuration ({\bf A}): The system matrices and vectors, the number of vector entries, matrix sizes, and  the inversion  computation times.}
\label{matrix_sizes}
\begin{tabular}{llllll} 
&   &  Matrices and &  Vector & Matrix   & Comp.\ \\
Approach & Data  & vectors &   entries & size (MB)  & time  (s) \\
\hline
Single res.\  & Full & ${\bf L}, {\bf x}$ &  3988 & 494 &  21.1 \\
Single res.\  & Proj.\  &  ${\bf L}, {\bf x}$ &3988  & 8.17 &16.2  \\
Dual res.\  & Full &  ${\bf L}_c, {\bf L}_f, {\bf x}^{(c)},{\bf x}^{(f)}$  & 997 &  123 &   12.9    \\
Dual res.\  & Proj.\  & ${\bf L}_c, {\bf L}_f, {\bf x}^{(c)},{\bf x}^{(f)}$ & 997 &  2.04 &10.2 \\
\hline
\end{tabular}
\end{center}
\end{table}

\begin{figure}\begin{scriptsize}
\begin{center}
\begin{minipage}{3.9cm}
\begin{center}
\includegraphics[width=2.7cm]{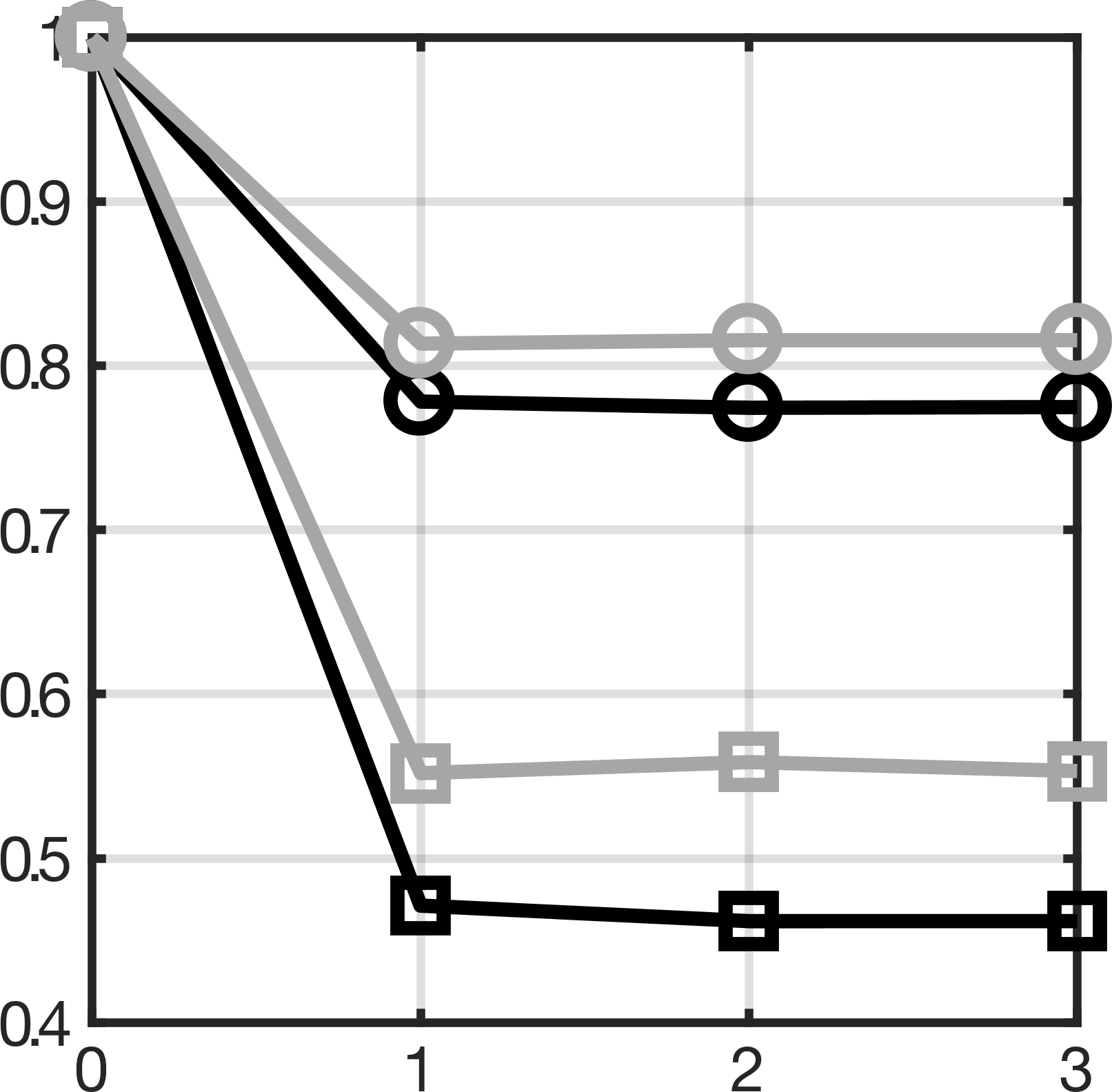} \\
Full wave 
\end{center}
\end{minipage}
\begin{minipage}{3.9cm}
\begin{center}
\includegraphics[width=2.7cm]{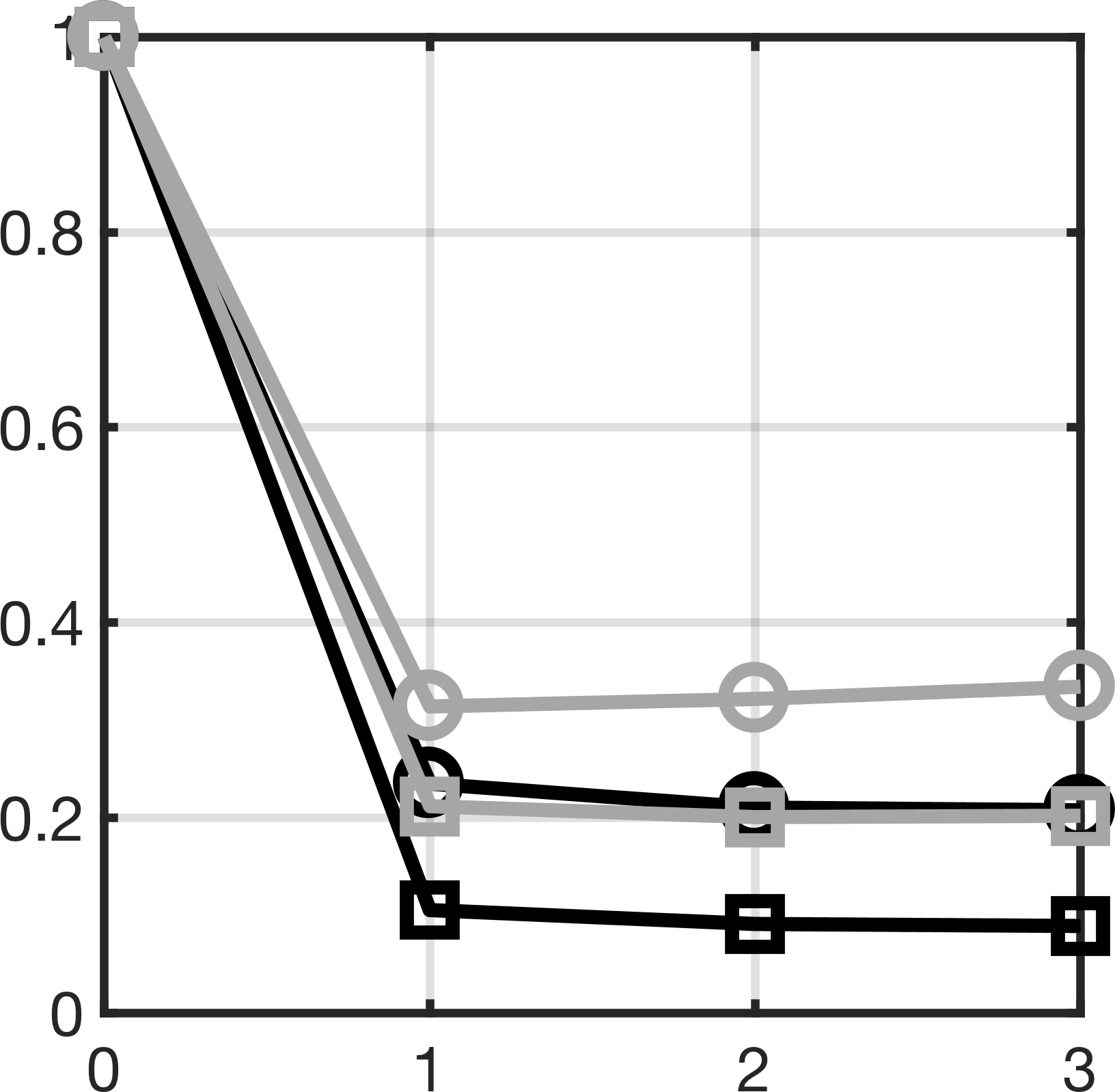} \\
Projected data
\end{center}
\end{minipage}
\end{center}
\caption{  The 2-norm of the relative residual for the single and dual resolution iteration (black and gray curves, respectively), i.e., $\| {\bf L} {\bf x} - {\bf y} \|_2/\| {\bf y }\|_2$ and $\| {\bf L}_{c} {\bf x}^{(c)} + {\bf L}_{f} {\bf x}^{(f)} - {\bf y} \|_2/\| {\bf y }\|_2$. The x-axis shows the number of the iteration steps. The curves for the full and projected data have been marked with squares and circles, respectively. The signal configuration is ({\bf A}).  \label{residual_norm_fig}}
\end{scriptsize}
\end{figure}

\begin{figure*}\begin{scriptsize}
\begin{center} 
\begin{minipage}{6.4cm}
\begin{framed} \begin{center} Full wave, $\eta = -25$ dB  \vphantom{Dual resolution}   \end{center} 
\begin{minipage}{2.8cm} \begin{center} Single resolution \\  \vskip0.1cm \includegraphics[height=2.2cm]{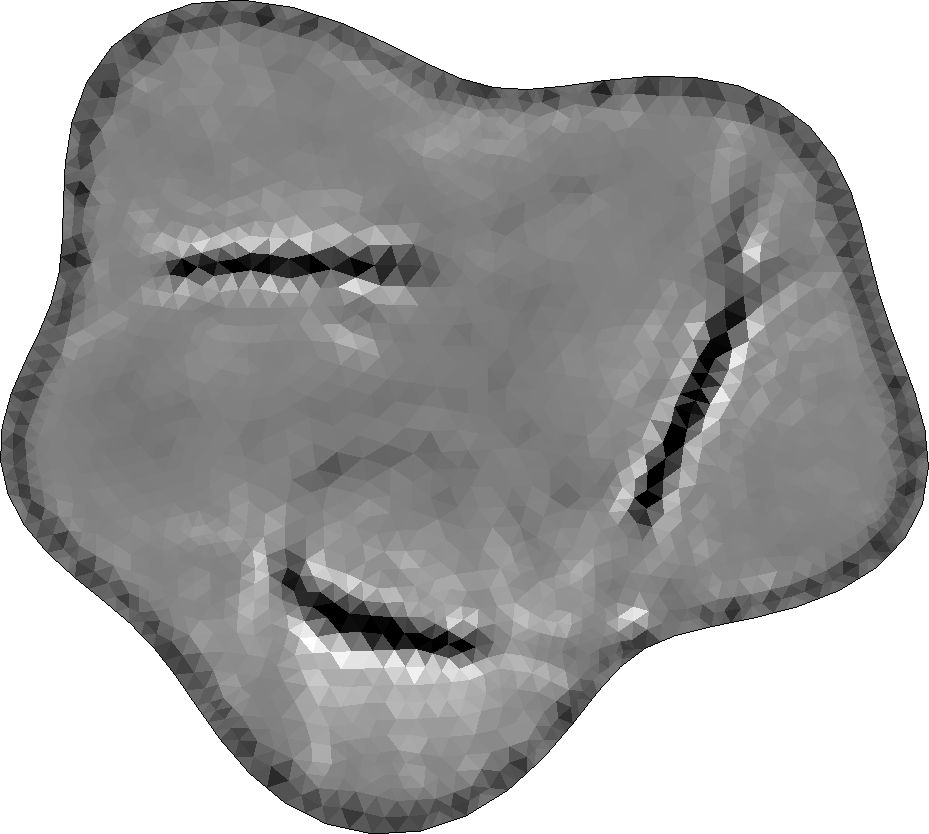}  \\ \vskip0.1cm \includegraphics[height=0.27cm]{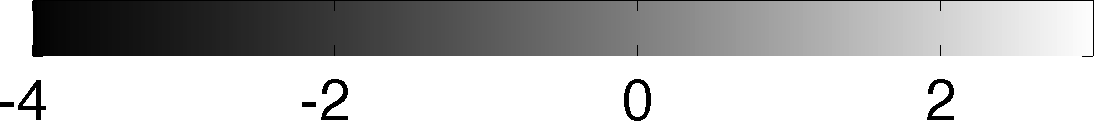} \\  Reconstruction \\ \vskip0.1cm \includegraphics[height=2.2cm]{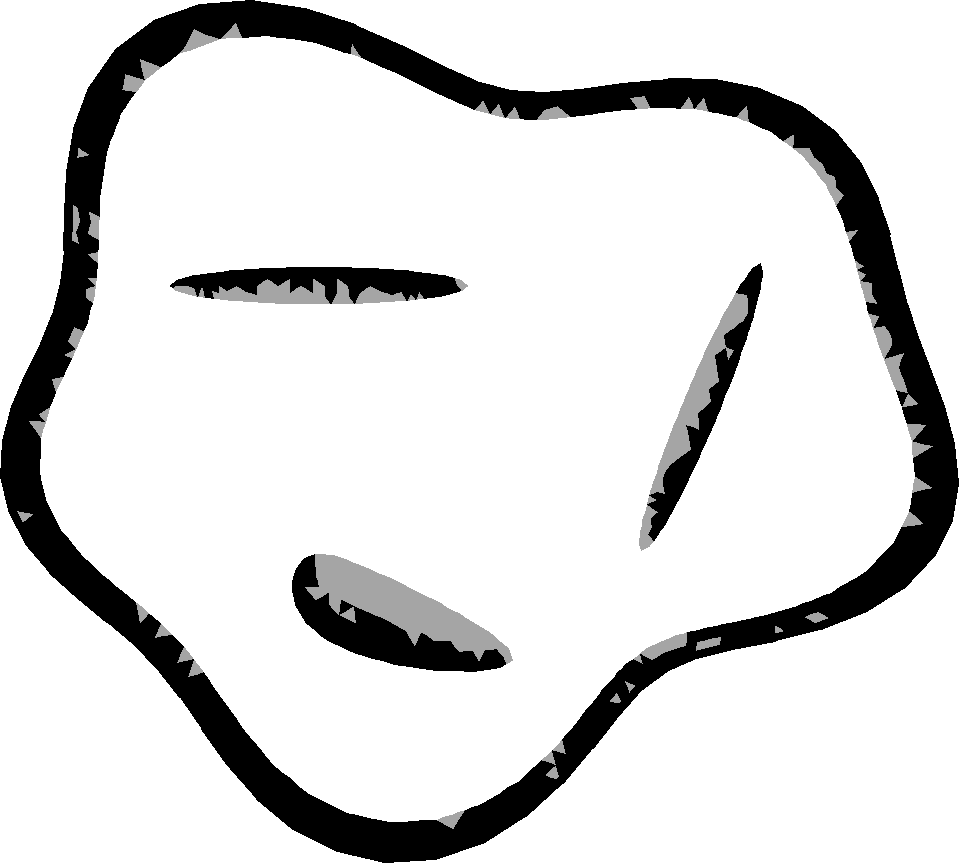} \\ Total overlap  $\mathrm{A}$\end{center}  \end{minipage}  
\begin{minipage}{2.8cm} \begin{center}  Dual resolution  \\  \vskip0.1cm\includegraphics[height=2.2cm]{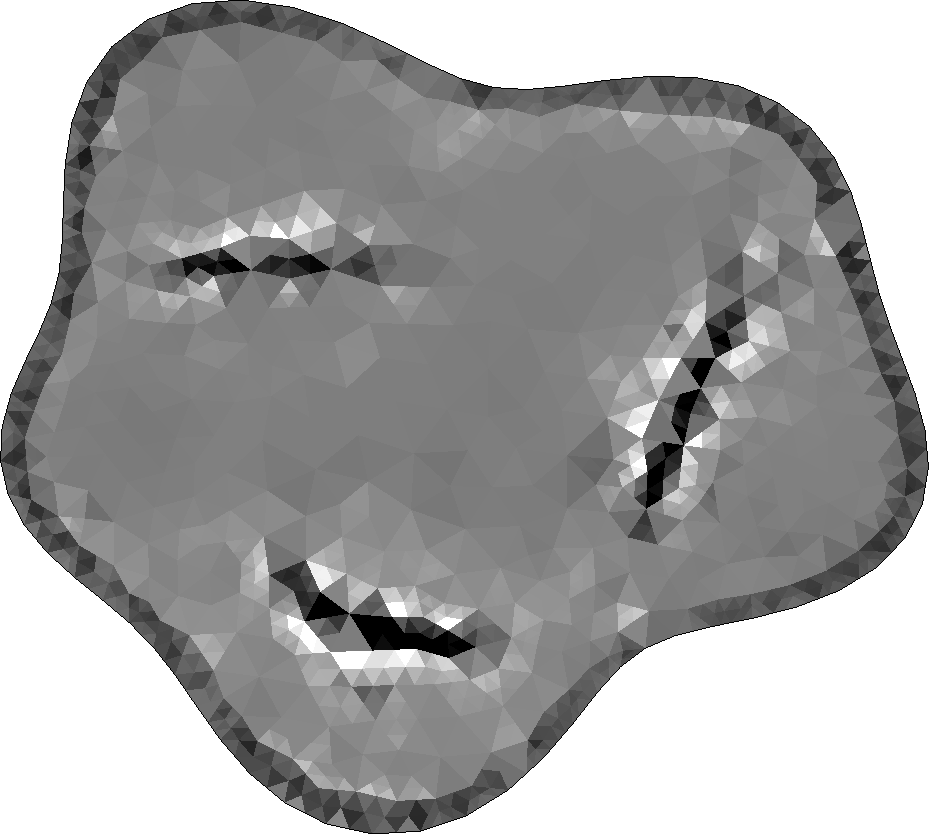}  \\ \vskip0.1cm \includegraphics[height=0.27cm]{./bar.png} \\  Reconstruction \\ \vskip0.1cm\includegraphics[height=2.2cm]{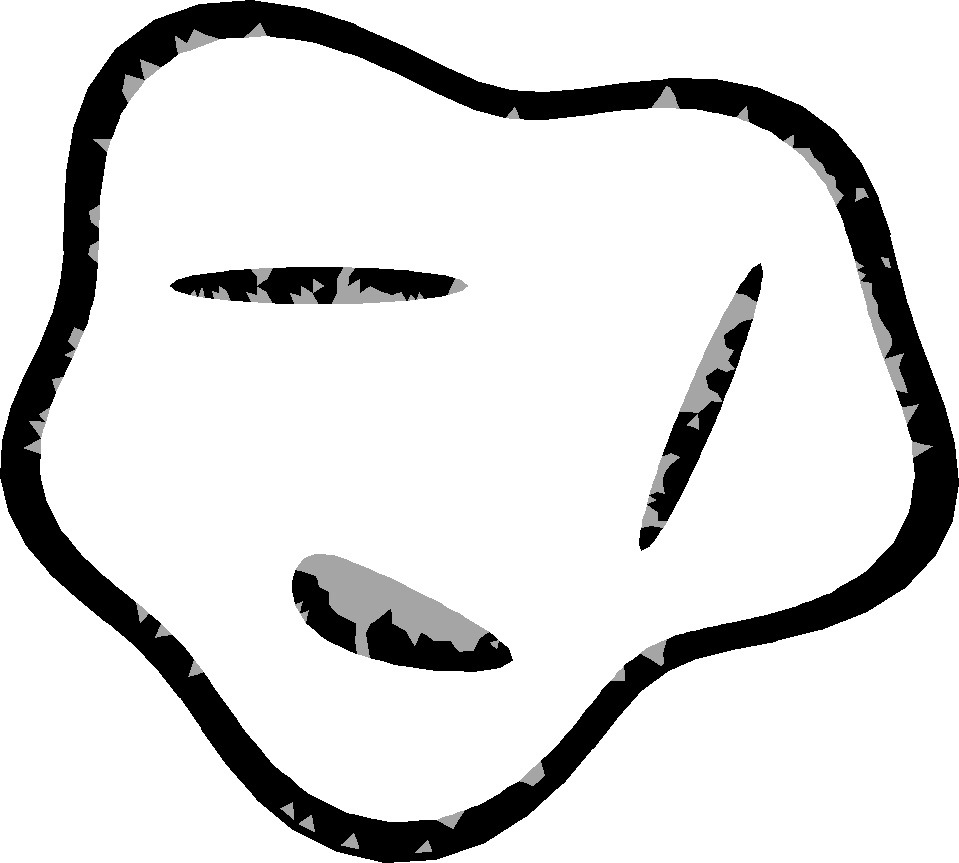} \\ Total overlap $\mathrm{A}$\end{center} \end{minipage} 
\end{framed} 
\end{minipage}
\hskip0.1cm
\begin{minipage}{6.4cm}
\begin{framed} \begin{center} Projected data, $\eta = -25$ dB  \vphantom{Single resolution}    \end{center} 
\begin{minipage}{2.8cm} \begin{center} 
Single resolution \\ \vskip0.1cm \includegraphics[height=2.2cm]{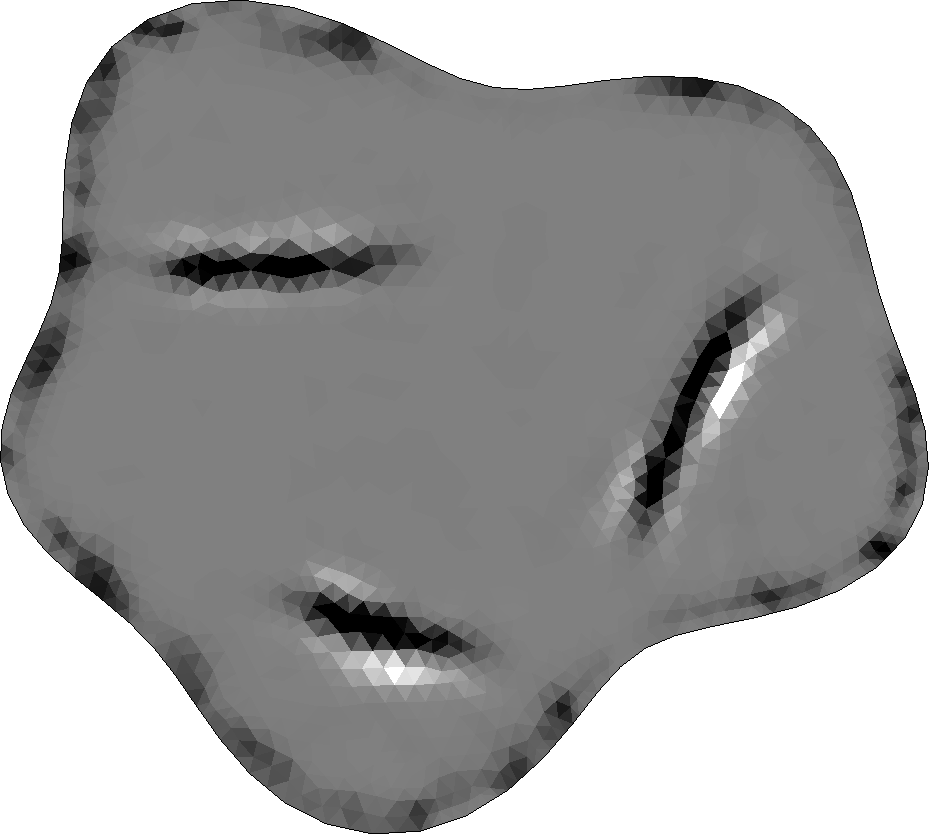} \\   \vskip0.1cm \includegraphics[height=0.27cm]{./bar.png} \\ Reconstruction \\ \vskip0.1cm \includegraphics[height=2.2cm]{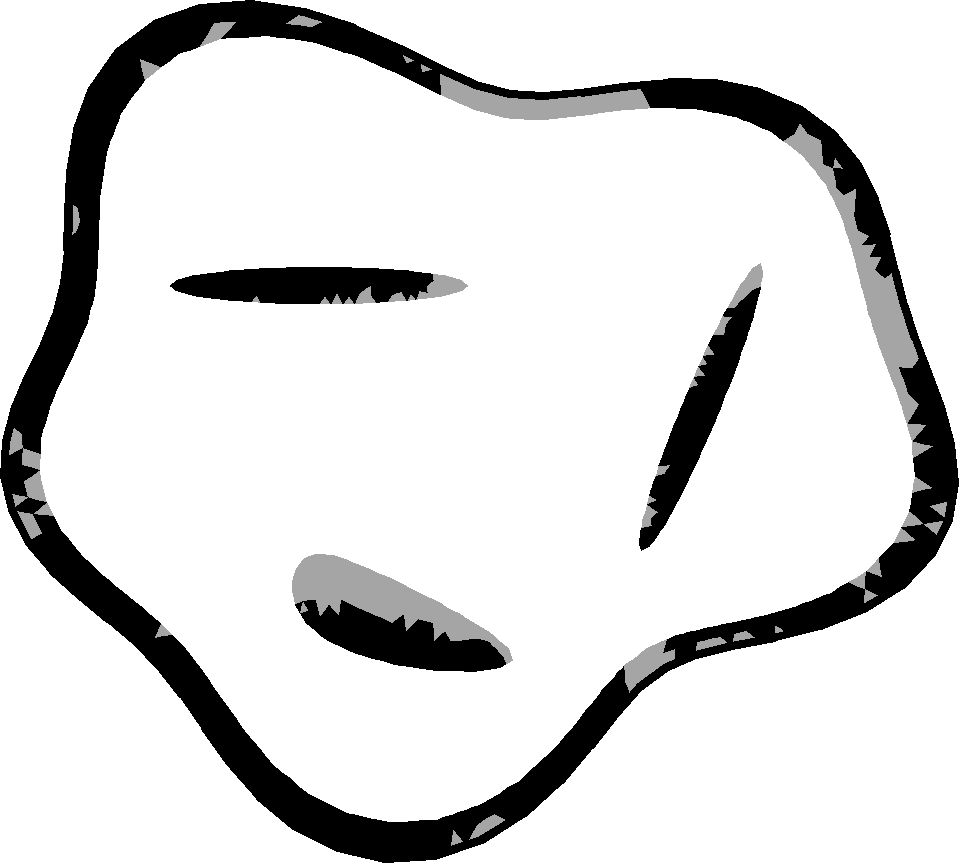} \\ Total overlap $\mathrm{A}$
   \end{center}  \end{minipage}  
\begin{minipage}{2.8cm} \begin{center} Dual resolution \\ \vskip0.1cm \includegraphics[height=2.2cm]{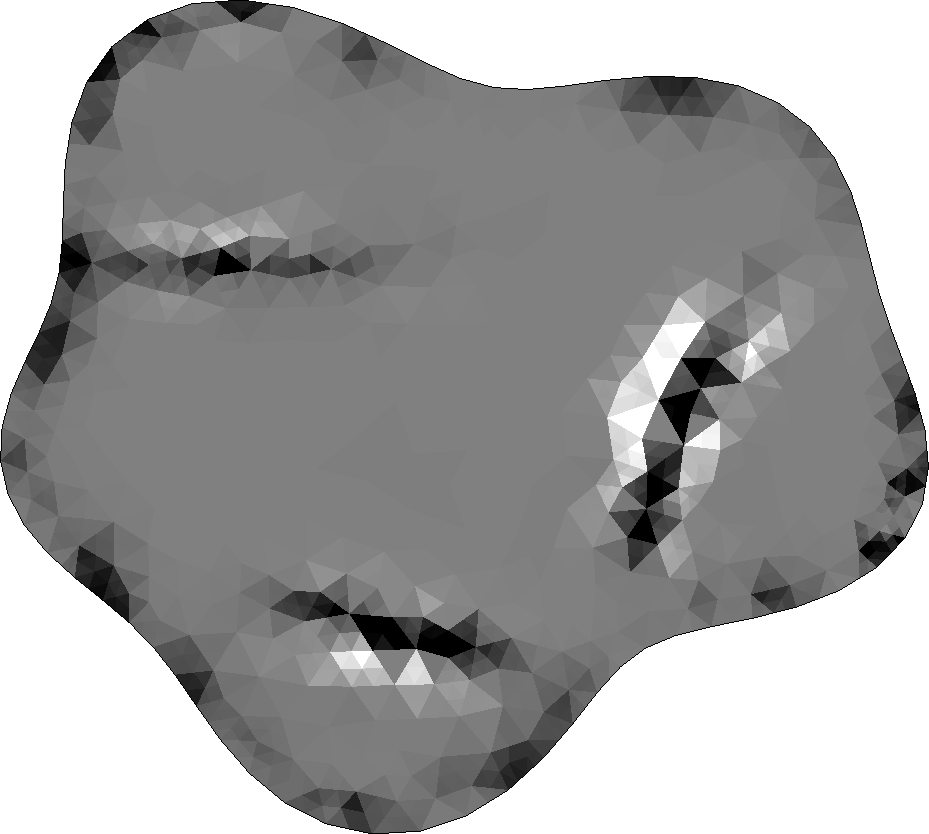} \\   \vskip0.1cm \includegraphics[height=0.27cm]{./bar.png} \\ Reconstruction \\ \vskip0.1cm \includegraphics[height=2.2cm]{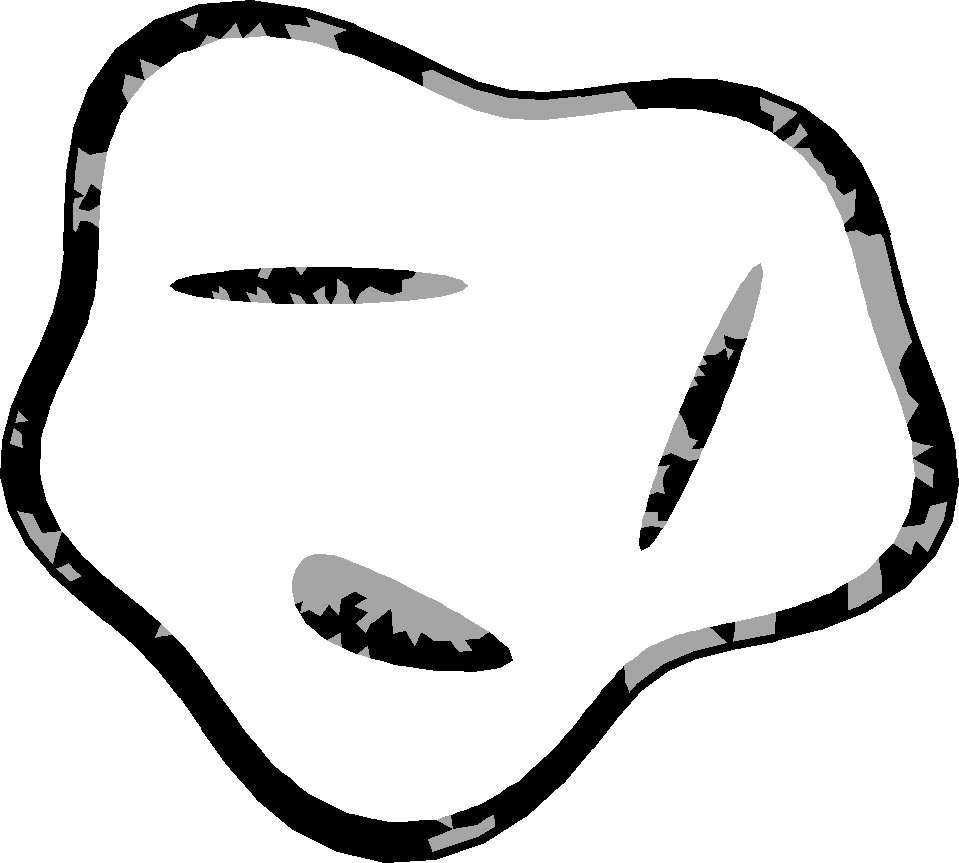} \\ Total overlap $\mathrm{A}$ \end{center} \end{minipage} 
\end{framed} 
\end{minipage} \\ 
\vskip0.1cm
\begin{minipage}{6.4cm}
\begin{framed} \begin{center}  Full wave, $\eta = -11$ dB \vphantom{Dual resolution}    \end{center} 
\begin{minipage}{2.8cm} \begin{center} Single resolution \\ \vskip0.1cm \includegraphics[height=2.2cm]{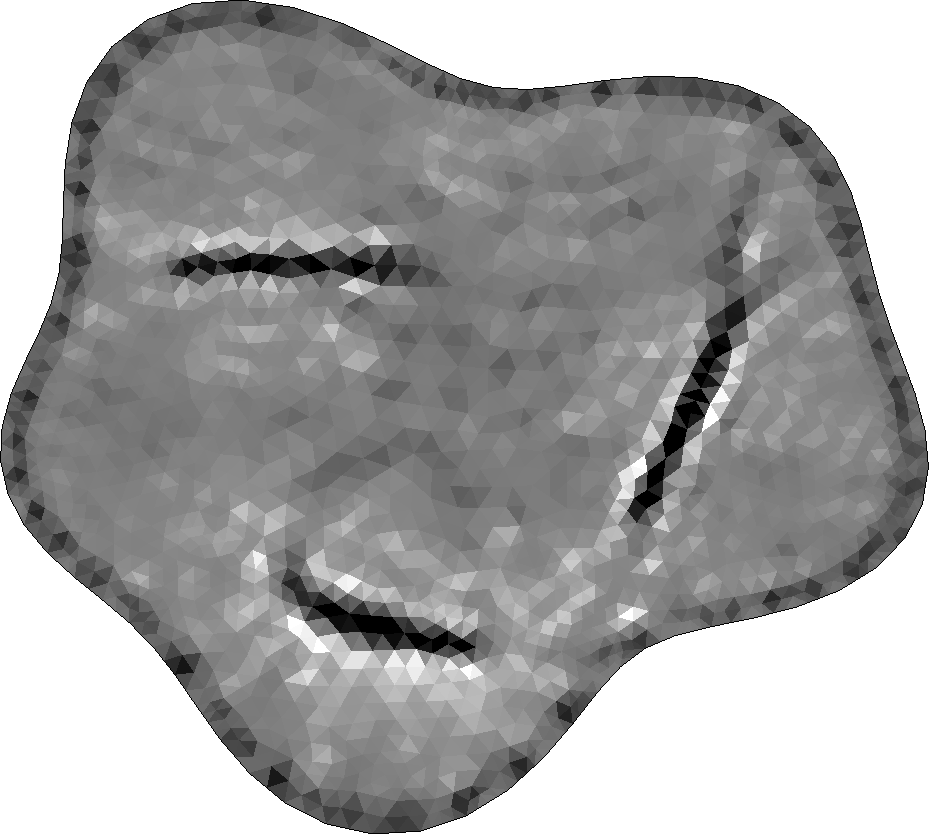}  \\ \vskip0.1cm \includegraphics[height=0.27cm]{./bar.png} \\  Reconstruction \\ \vskip0.1cm\includegraphics[height=2.2cm]{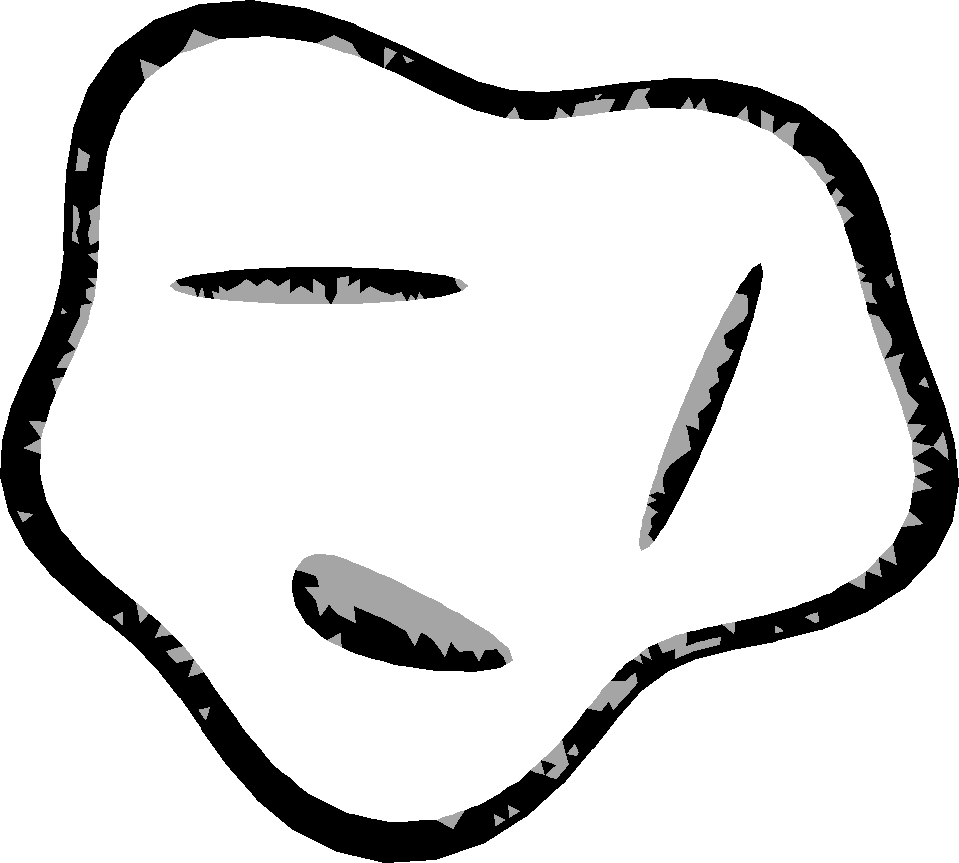} \\ Total overlap  $\mathrm{A}$ \end{center}  \end{minipage}  
\begin{minipage}{2.8cm} \begin{center} 
Dual resolution \\ \vskip0.1cm \includegraphics[height=2.2cm]{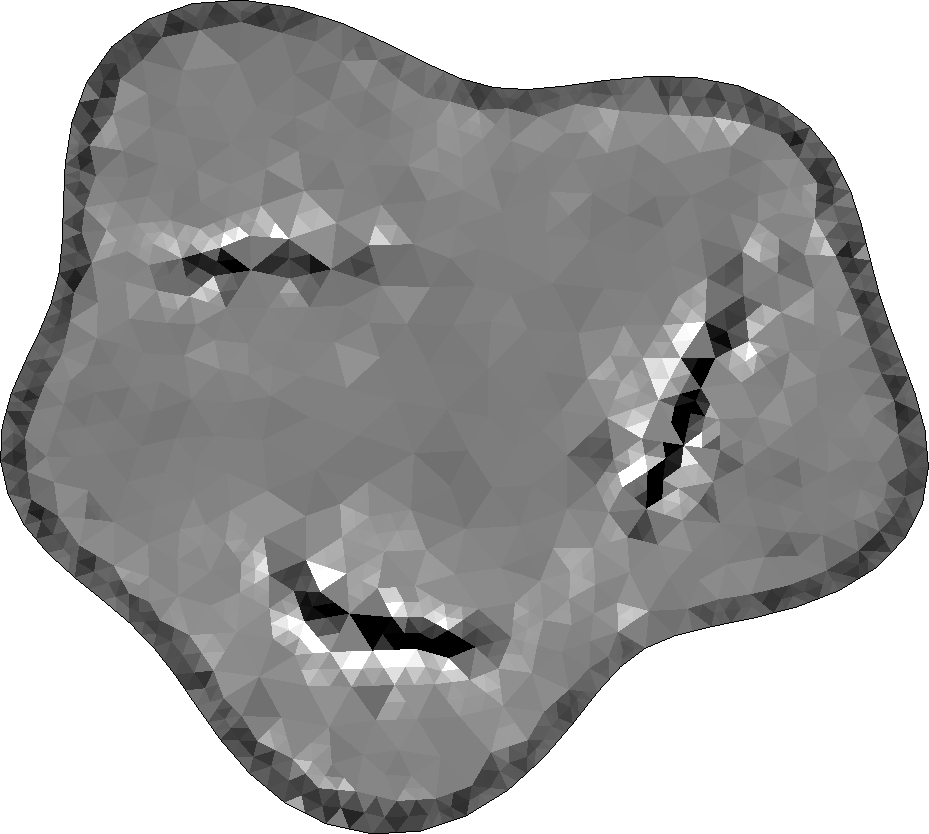}  \\ \vskip0.1cm \includegraphics[height=0.27cm]{./bar.png} \\  Reconstruction \\ \vskip0.1cm \includegraphics[height=2.2cm]{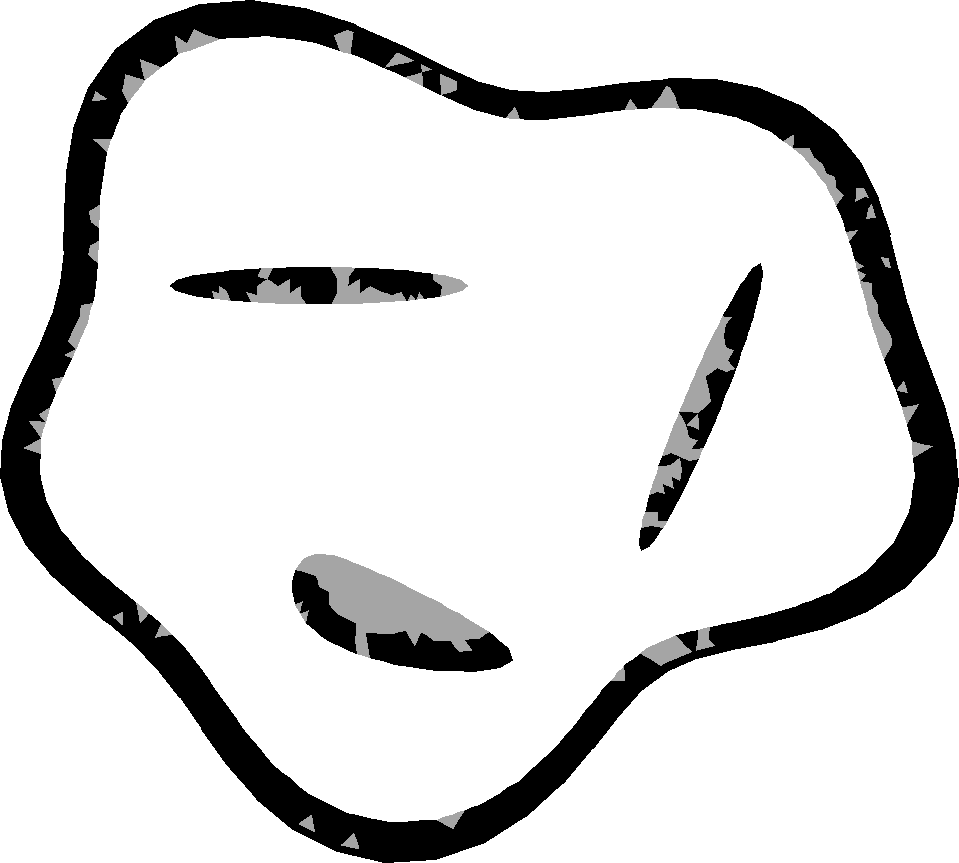} \\ Total overlap  $\mathrm{A}$
 \end{center} \end{minipage} 
\end{framed} 
\end{minipage}
\hskip0.1cm
\begin{minipage}{6.4cm}
\begin{framed} \begin{center} Projected data, $\eta = -11$ dB \vphantom{Single resolution}   \end{center} 
\begin{minipage}{2.8cm} \begin{center} 
Single resolution \\ \vskip0.1cm \includegraphics[height=2.2cm]{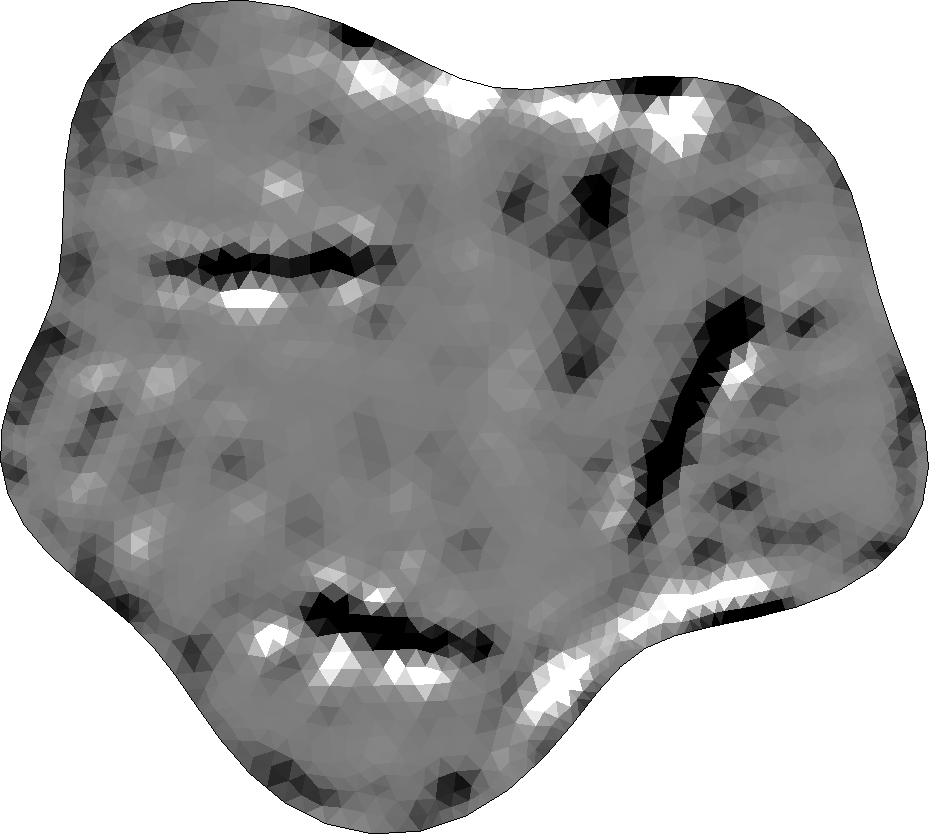} \\   \vskip0.1cm \includegraphics[height=0.27cm]{./bar.png} \\ Reconstruction \\ \vskip0.1cm \includegraphics[height=2.2cm]{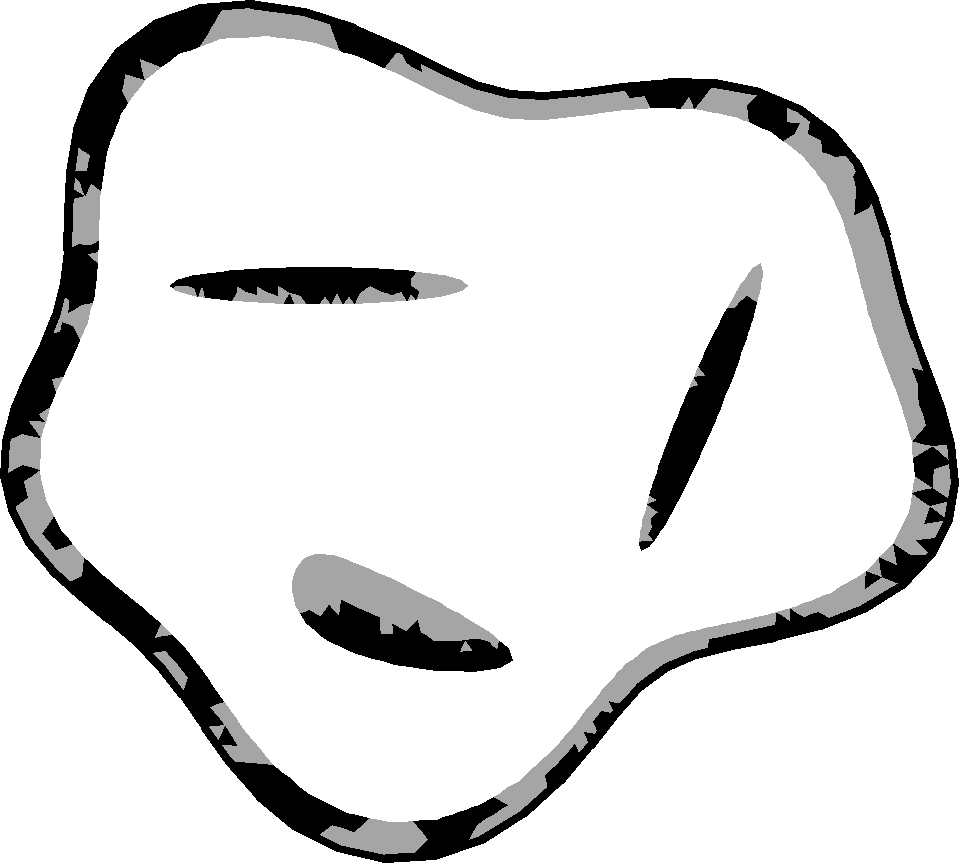} \\ Total overlap $\mathrm{A}$
 \end{center}  \end{minipage}  
\begin{minipage}{2.8cm} \begin{center} Dual resolution \\ \vskip0.1cm \includegraphics[height=2.2cm]{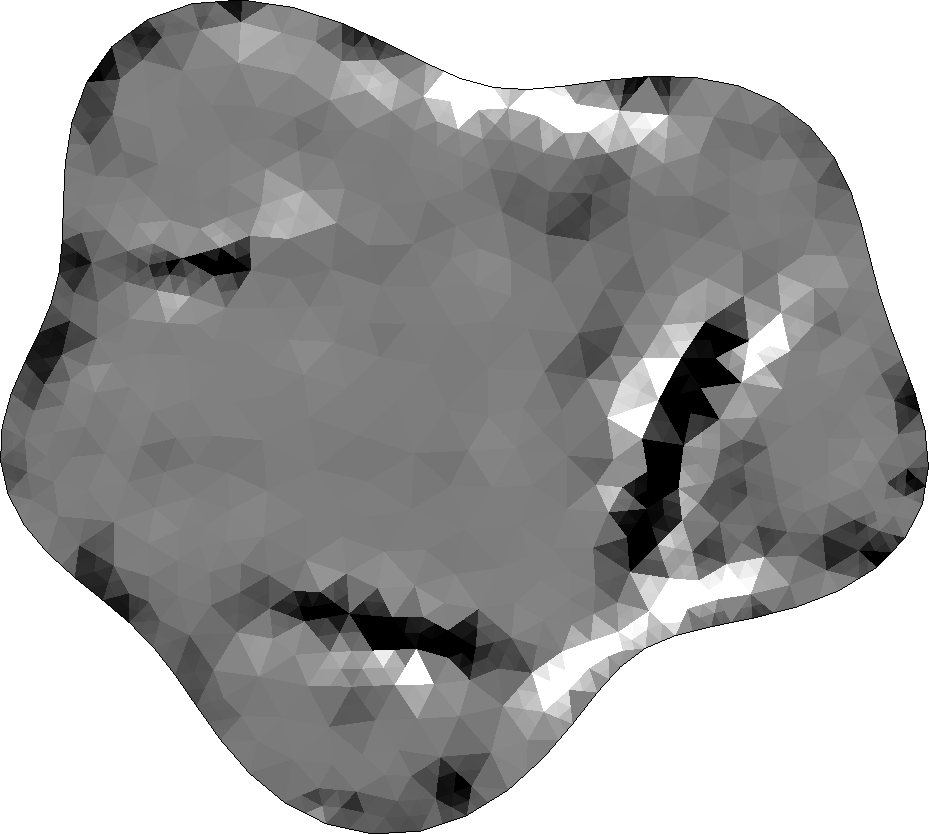} \\   \vskip0.1cm \includegraphics[height=0.27cm]{./bar.png} \\ Reconstruction \\ \vskip0.1cm \includegraphics[height=2.2cm]{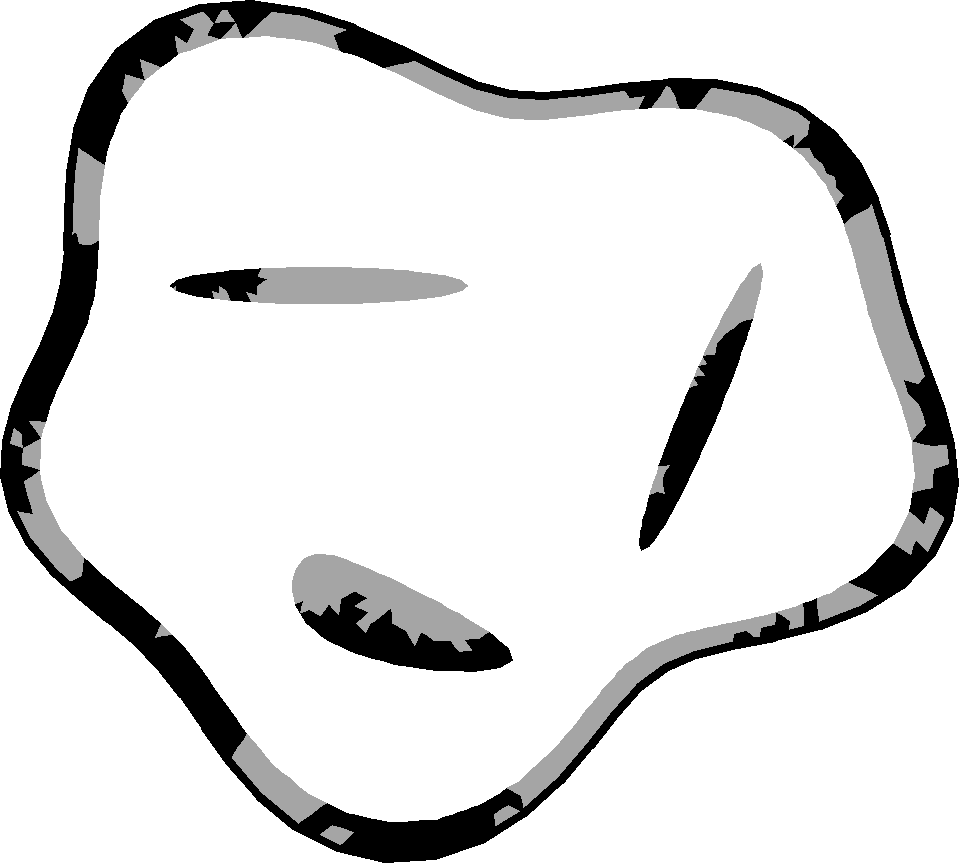} \\ Total overlap $\mathrm{A}$ \end{center} \end{minipage} 
\end{framed} 
\end{minipage}
\end{center}
\caption{ Single and dual resolution reconstructions obtained with full and projected data at noise levels $\eta = -11$ dB and $\eta = -25$ dB. The color scale shows the difference between the inverse estimate and the initial guess for the relative permittivity, i.e., the constant distribution $\epsilon_r = 4$. Below each reconstruction, the total overlap set $\mathrm{A}$ has been visualized. \label{reconstructions_comparison}}
\end{scriptsize}
\end{figure*}

\begin{figure*}\begin{scriptsize}
\begin{center} \begin{minipage}{6.4cm}
\begin{framed} \begin{center}  Full wave   \vphantom{Projected data}    \end{center} 
\begin{minipage}{2.8cm} \begin{center} Coarse \\ \vskip0.1cm \includegraphics[height=2.2cm]{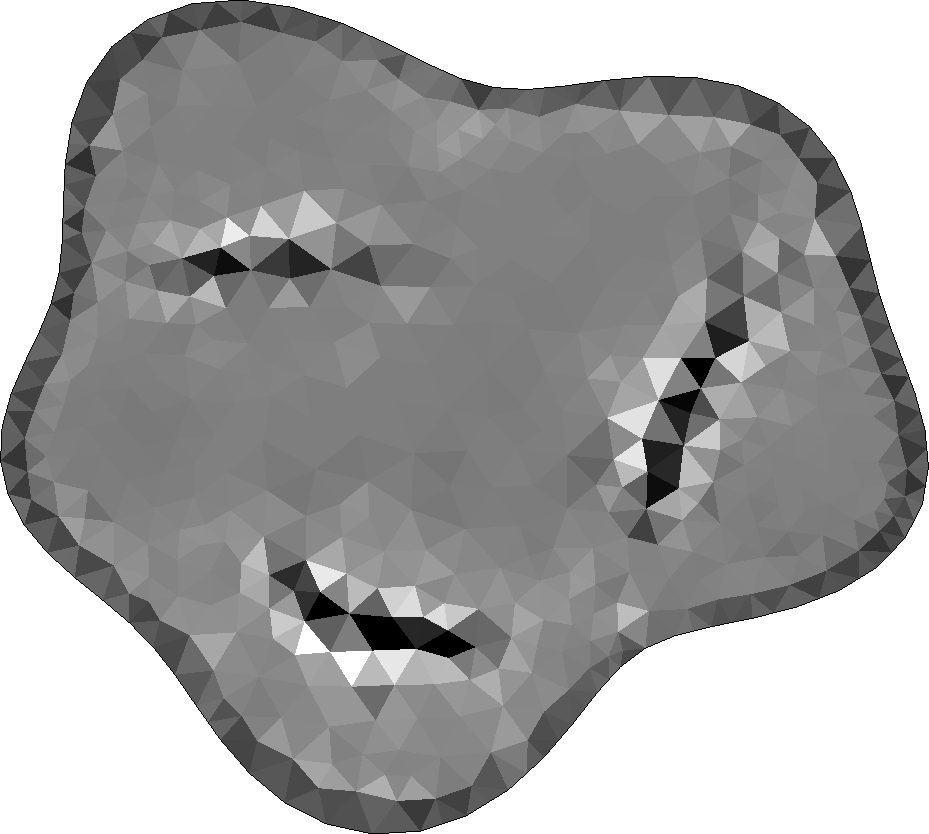}  \\  $\eta= -25$ dB \\ \vskip0.1cm\includegraphics[height=2.2cm]{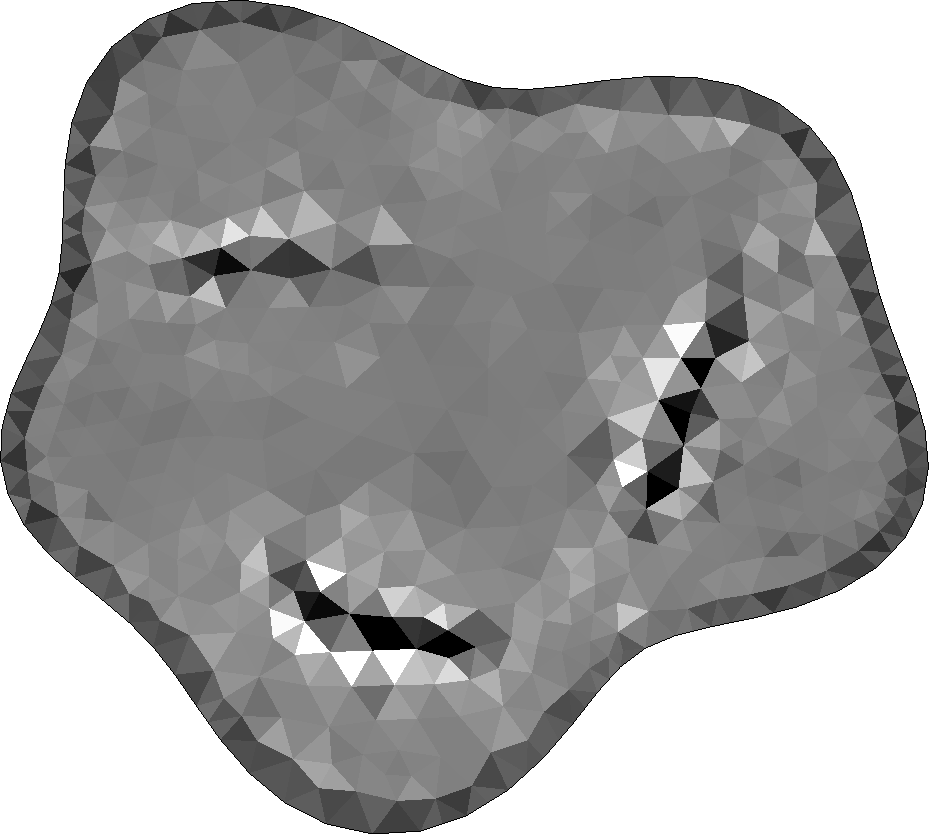} \\  $\eta= -11$ dB  \vskip0.1cm    \includegraphics[height=0.27cm]{./bar.png}  \end{center}  \end{minipage}  
\begin{minipage}{2.8cm} \begin{center} Fine  \\ \vskip0.1cm  \includegraphics[height=2.2cm]{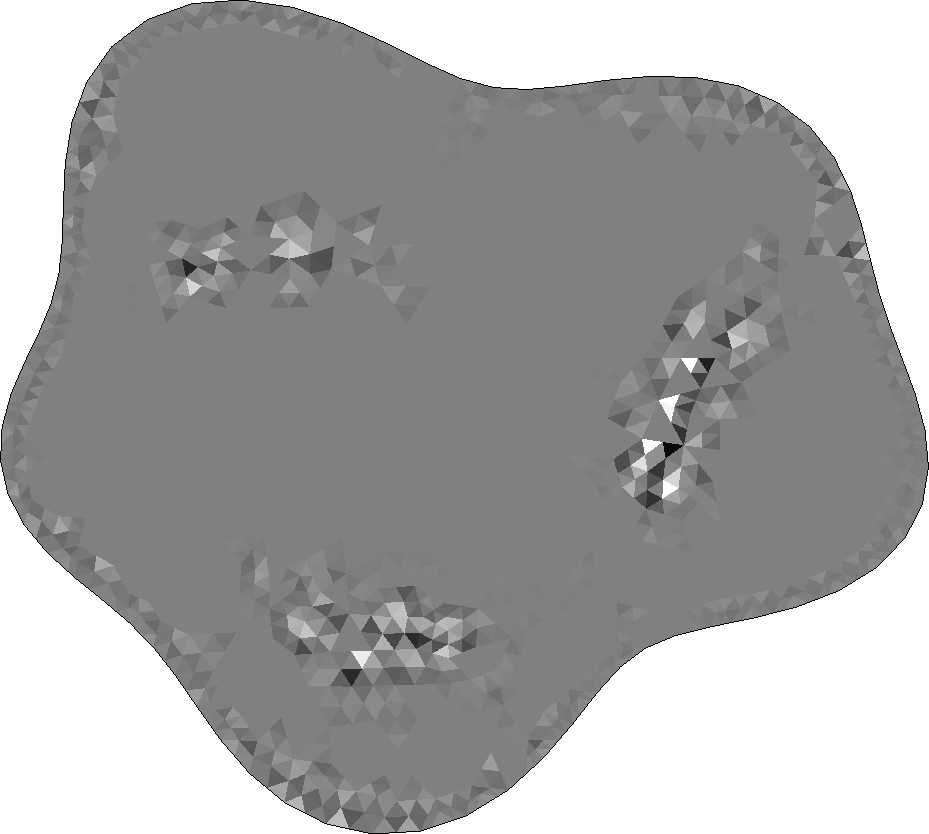} \\    $\eta= -25$ dB   \\ \vskip0.1cm\includegraphics[height=2.2cm]{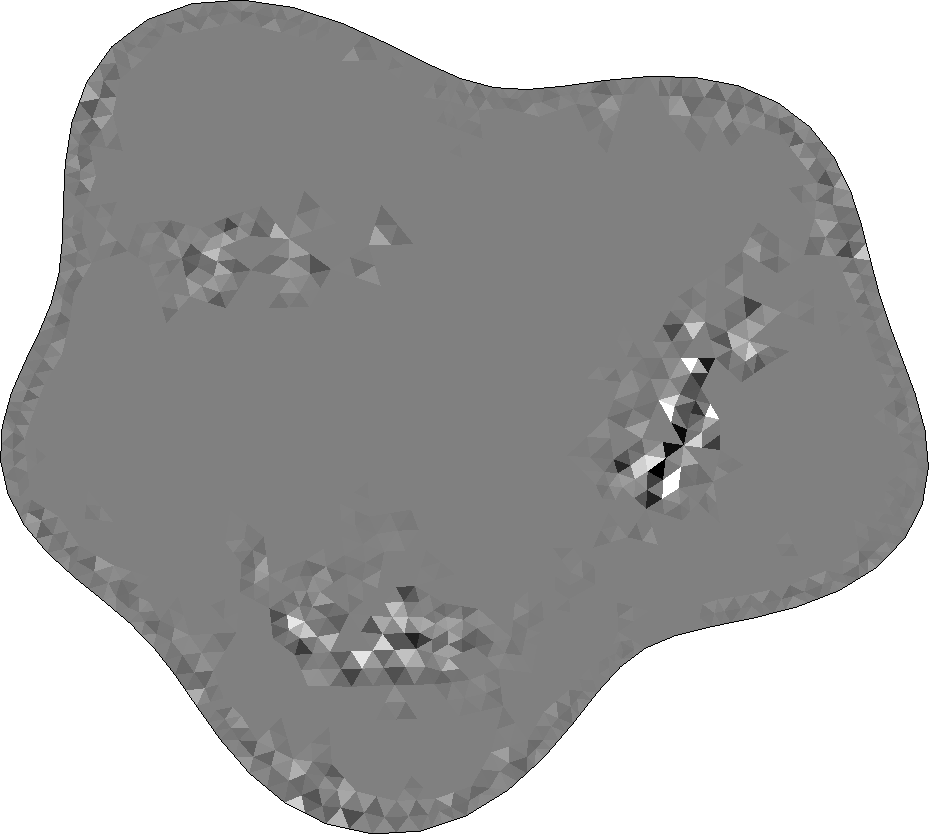} \\ $\eta=$ -11 dB \\ \vskip0.1cm    \includegraphics[height=0.27cm]{./bar.png} \end{center} \end{minipage} 
\end{framed} 
\end{minipage}
\hskip0.1cm
\begin{minipage}{6.4cm}
\begin{framed} \begin{center} Projected data \vphantom{Full wave}   \end{center} 
\begin{minipage}{2.8cm} \begin{center} Coarse \\ \vskip0.1cm  \includegraphics[height=2.2cm]{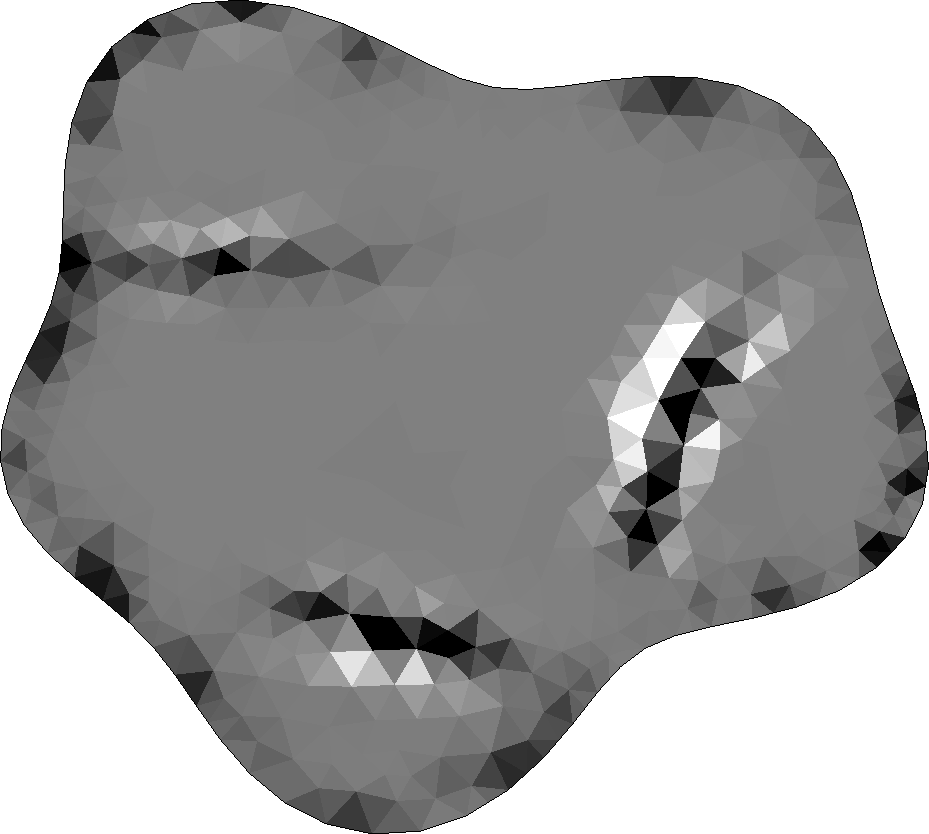}  \\ $\eta=-25$ dB \\ \vskip0.1cm\includegraphics[height=2.2cm]{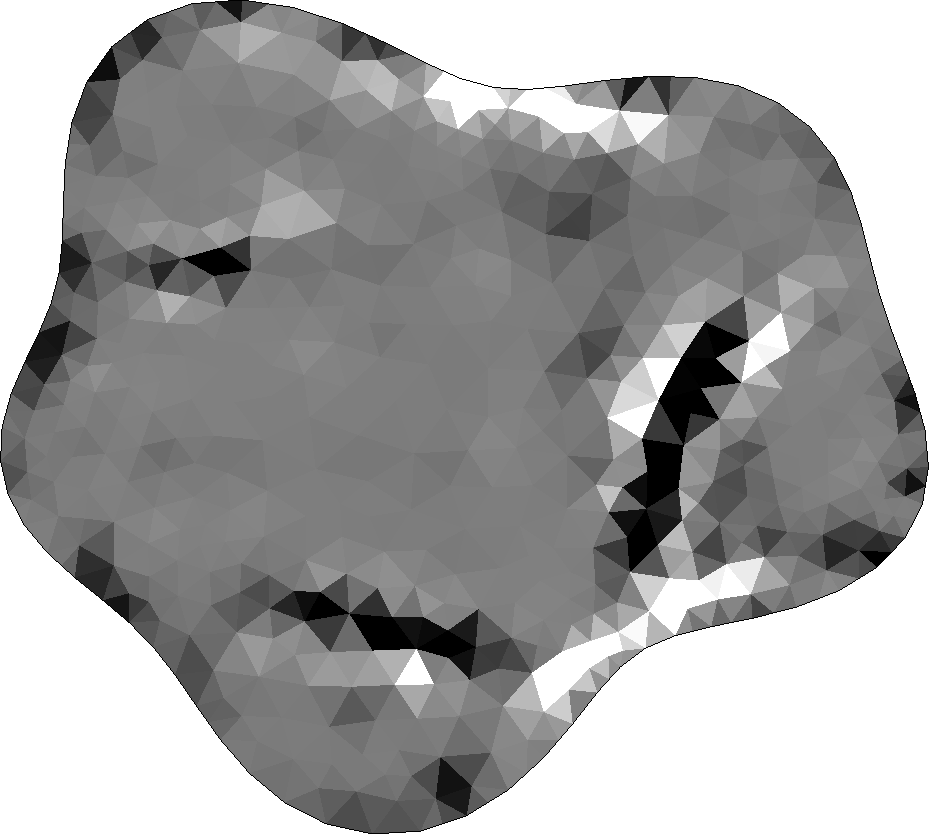} \\ $\eta=-11$ dB  \\ \vskip0.1cm    \includegraphics[height=0.27cm]{./bar.png} \end{center}  \end{minipage}  
\begin{minipage}{2.8cm} \begin{center} Fine \\ \vskip0.1cm  \includegraphics[height=2.2cm]{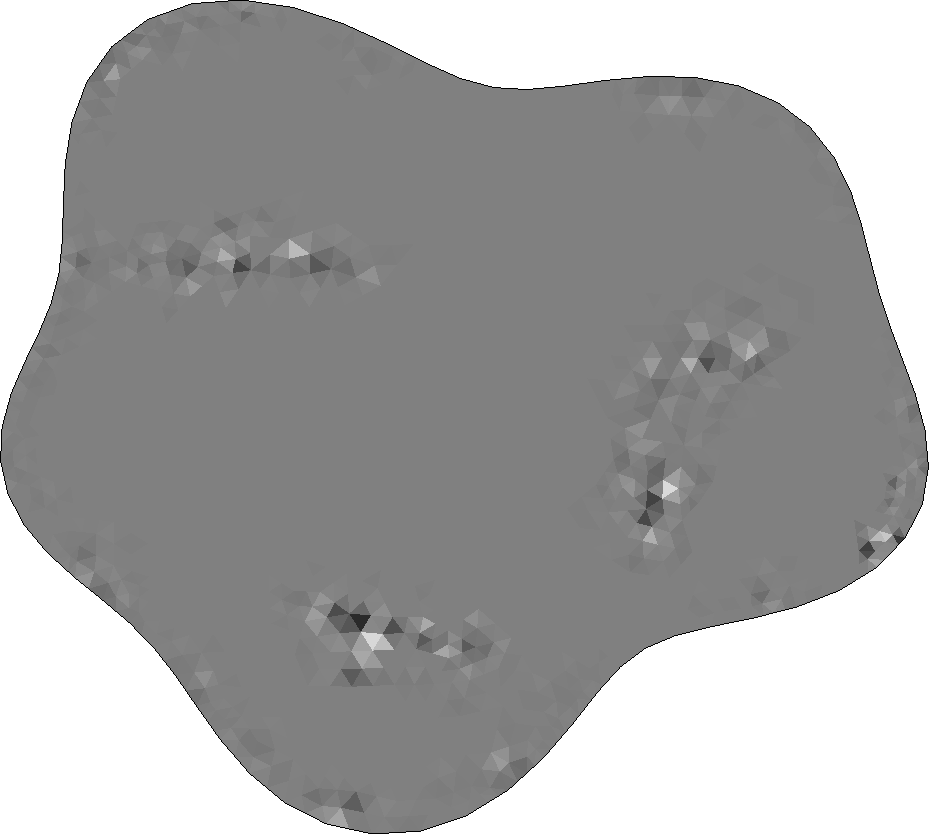}\\ $\eta=-25 dB$  \\ \vskip0.1cm \includegraphics[height=2.2cm]{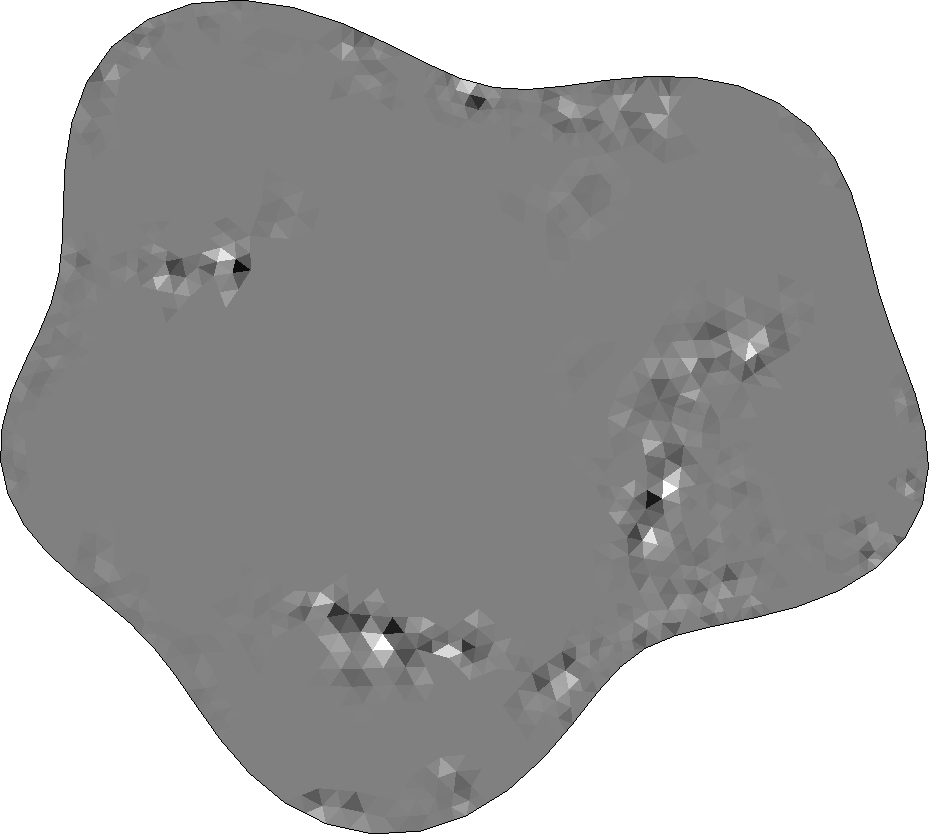} \\ $\eta=-11$ dB \\ \vskip0.1cm    \includegraphics[height=0.27cm]{./bar.png} \end{center} \end{minipage} 
\end{framed} 
\end{minipage}
\end{center}
\caption{The coarse and fine level fluctuations of the dual resolution reconstructions for full and projected data at noise levels $\eta = -25$ dB and $\eta = -11$ dB. \label{reconstructions_levels}}
\end{scriptsize}
\end{figure*}

\begin{figure*}\begin{scriptsize}
\begin{center}
\begin{minipage}{6.4cm}
\begin{center}
\begin{framed}
Full wave, Single resolution \\
\begin{minipage}{2.8cm}
\begin{center}
\includegraphics[width=2.8cm]{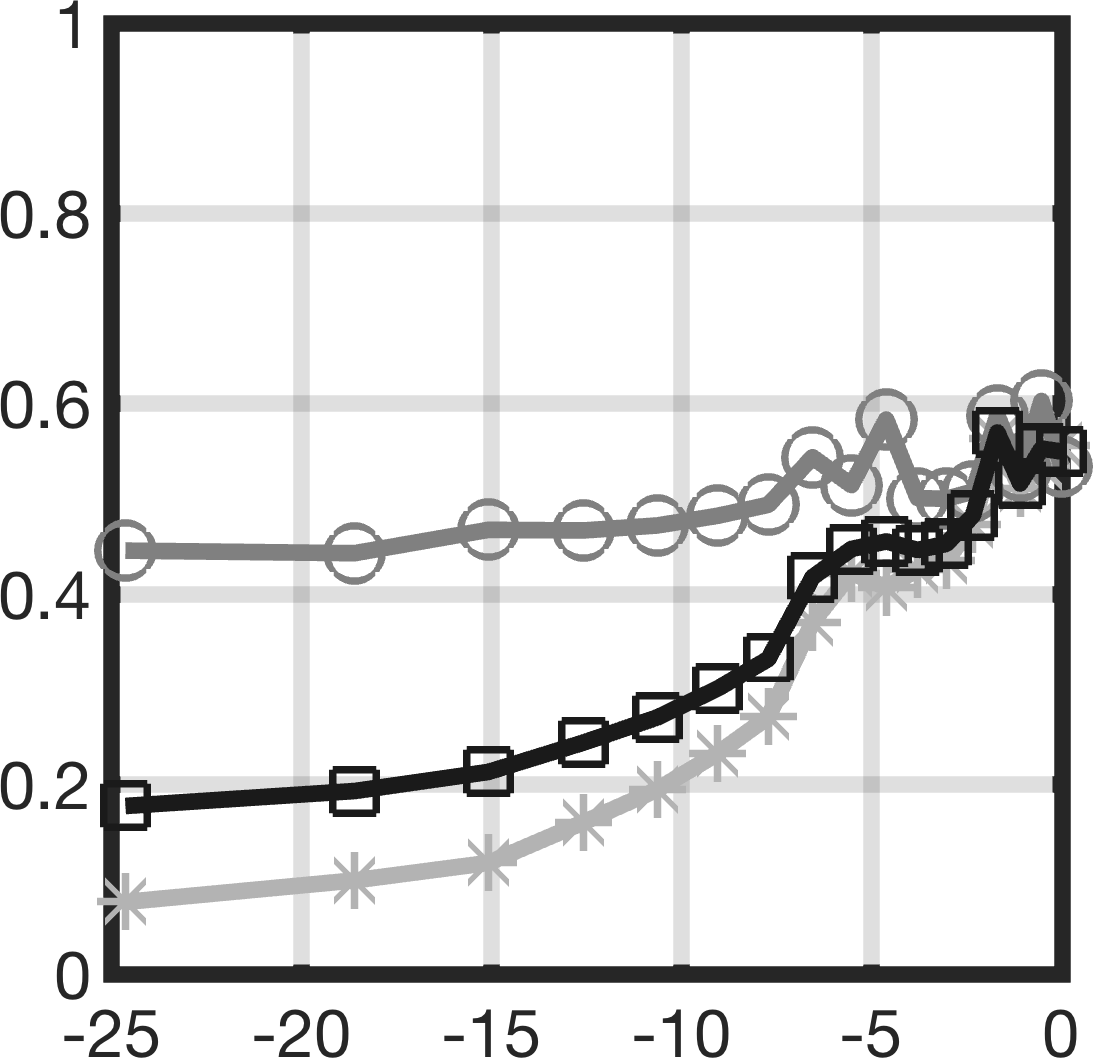}  \\
Noise sensitivity\\
x: $\eta$ (dB), y: ROE
\end{center}
\end{minipage}
\begin{minipage}{2.8cm}
\begin{center}
\includegraphics[width=2.8cm]{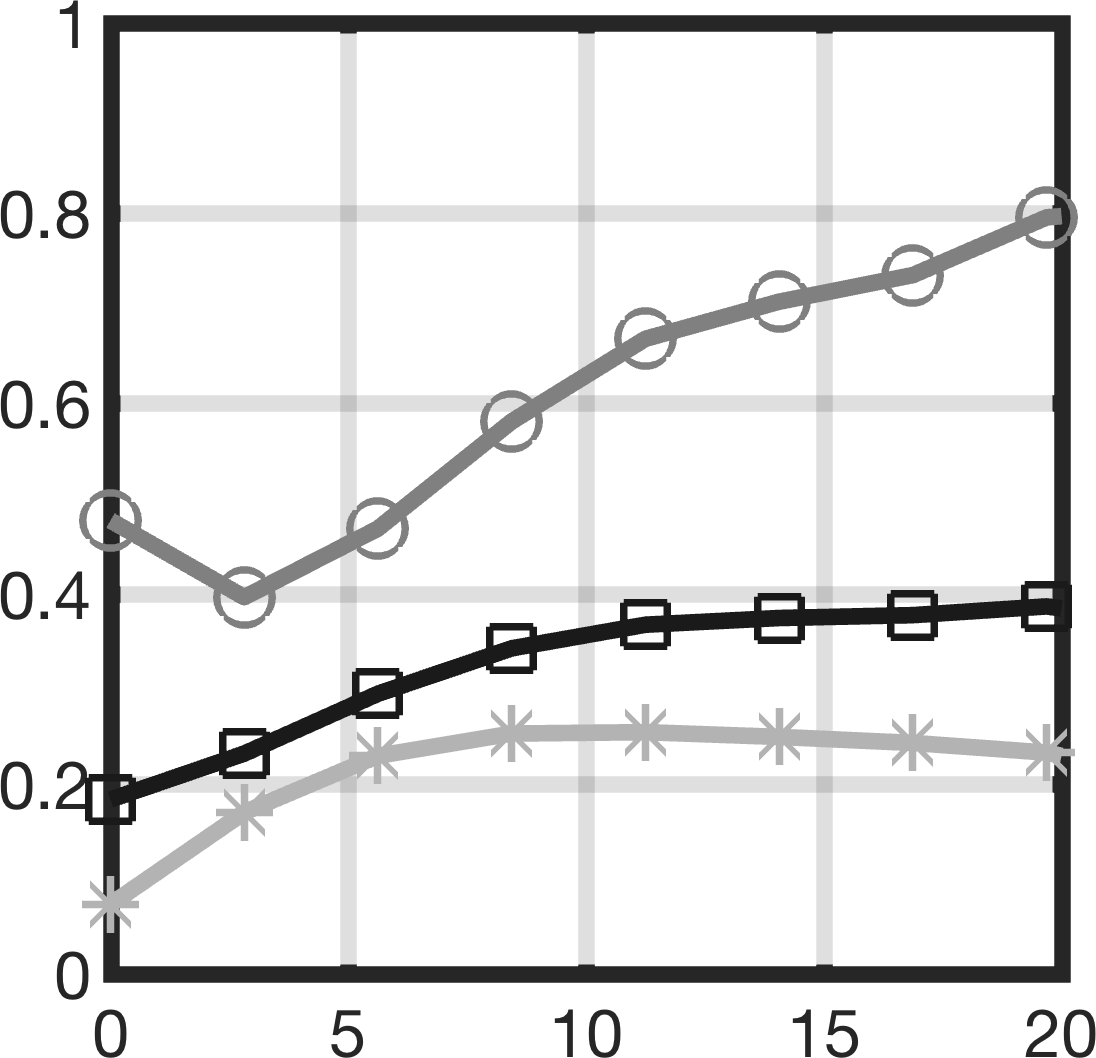} \\
Angular sensitivity \\
 x: $\theta$  (deg), y: ROE
\end{center}
\end{minipage}
\\
\begin{minipage}{2.8cm}
\begin{center}
 \includegraphics[width=2.8cm]{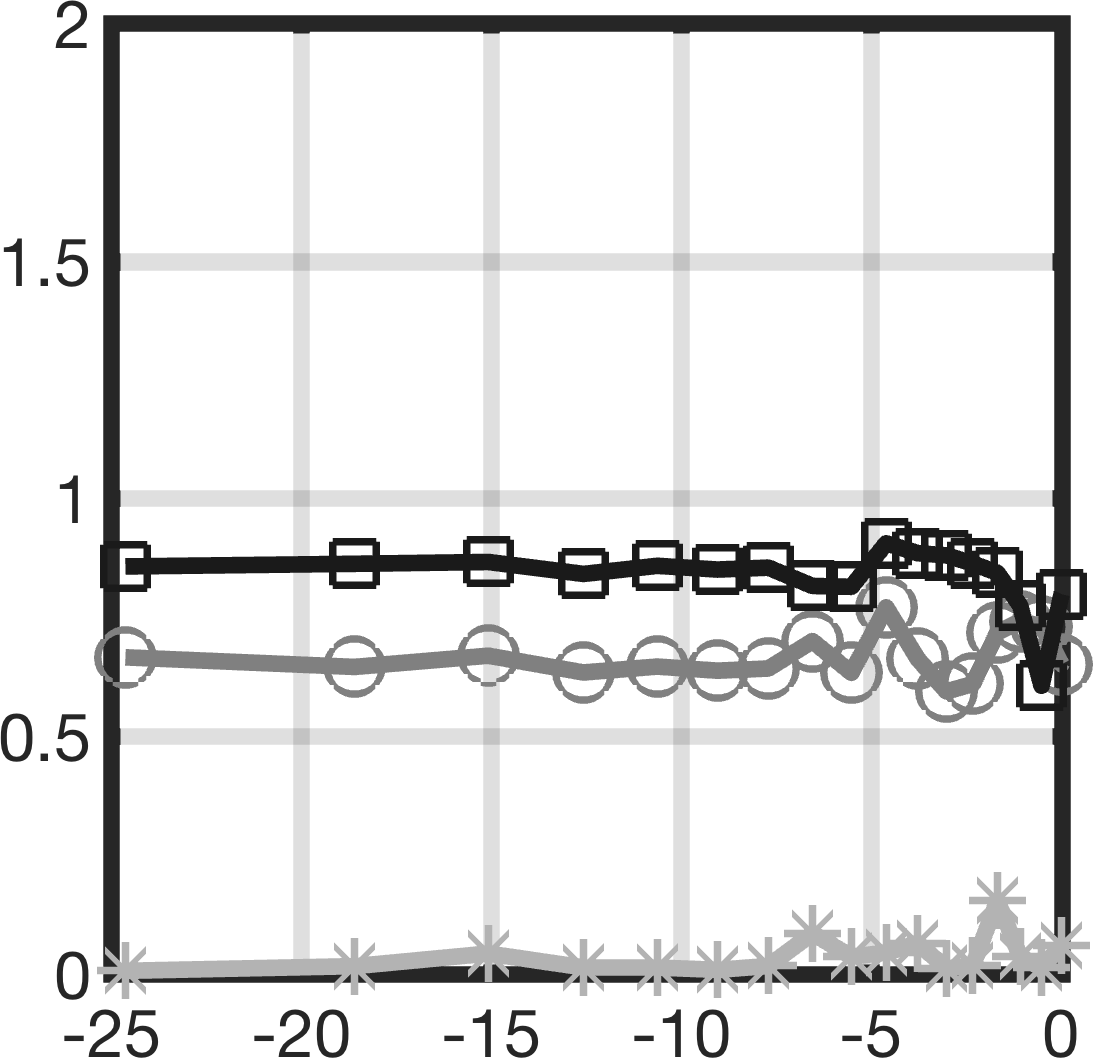}  \\
Noise sensitivity\\
x: $\eta$ (dB), y: RVE
\end{center}
\end{minipage}
\begin{minipage}{2.8cm}
\begin{center}
\includegraphics[width=2.8cm]{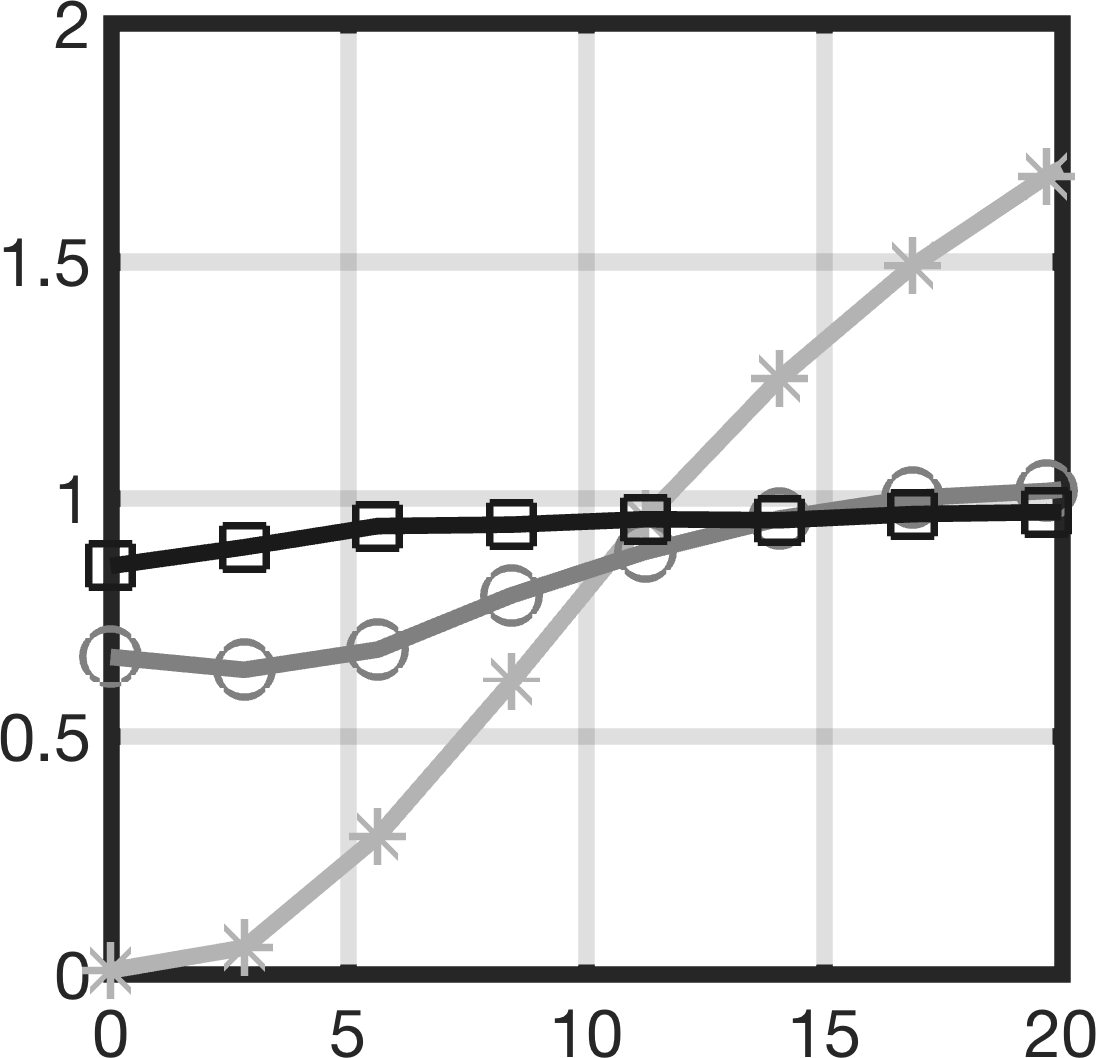} \\ 
Angular sensitivity \\
x: $\theta$  (deg), y: RVE
\end{center}
\end{minipage}
\end{framed}
\end{center}
\end{minipage}
\hskip0.1cm
\begin{minipage}{6.4cm}
\begin{center}
\begin{framed}
Full wave, Dual resolution \\
\begin{minipage}{2.8cm}
\begin{center}
\includegraphics[width=2.8cm]{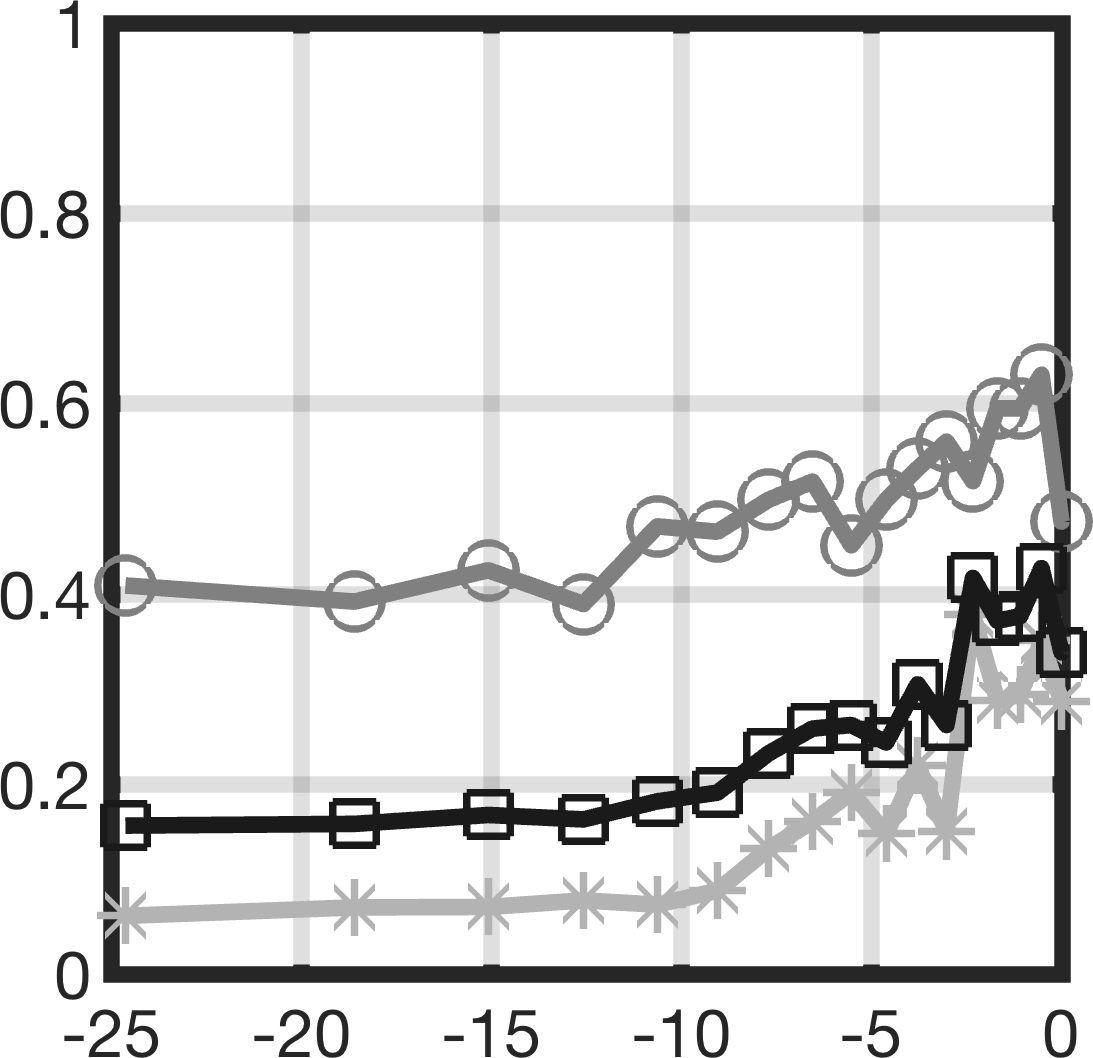}  \\
Noise sensitivity \\
x: $\eta$ (dB), y: ROE
\end{center}
\end{minipage}
\begin{minipage}{2.8cm}
\begin{center}
\includegraphics[width=2.8cm]{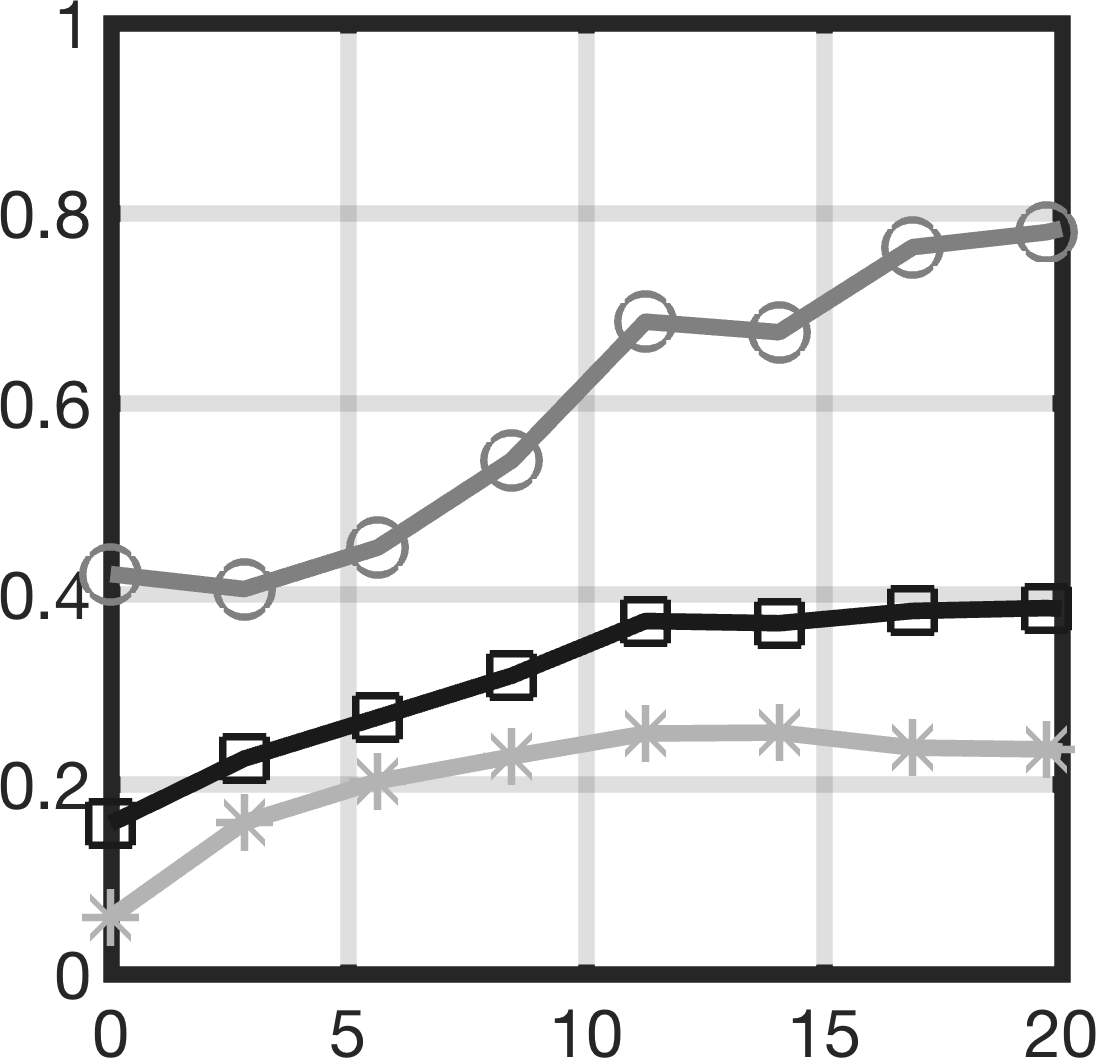} \\
Angular sensitivity \\
x: $\theta$  (deg), y: ROE
\end{center}
\end{minipage}
\\ 
\begin{minipage}{2.8cm}
\begin{center}
 \includegraphics[width=2.8cm]{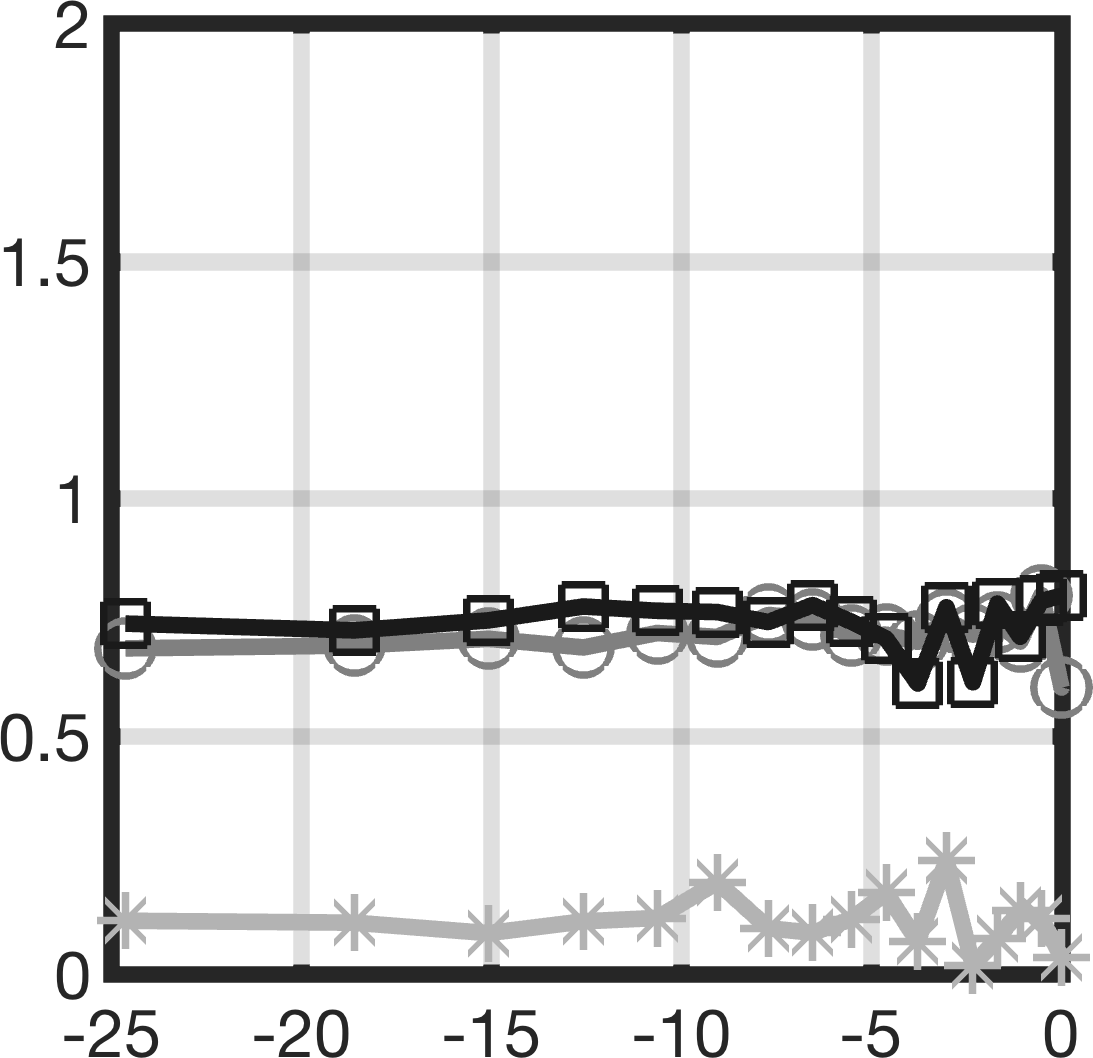}  \\
Noise sensitivity \\
x: $\eta$ (dB), y: RVE
\end{center}
\end{minipage}
\begin{minipage}{2.8cm}
\begin{center}
\includegraphics[width=2.8cm]{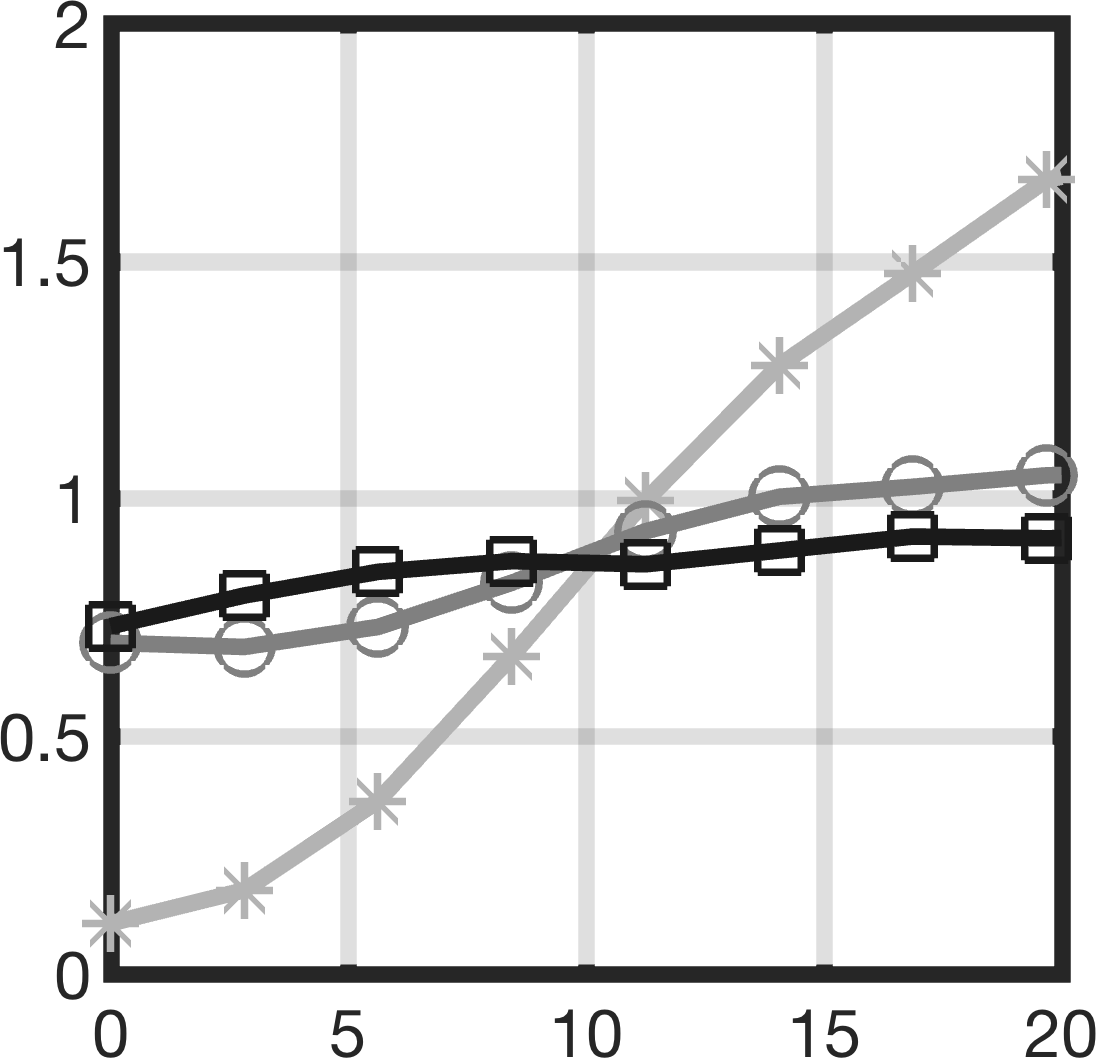} \\ 
Angular sensitivity \\
x: $\theta$  (deg), y: RVE
\end{center}
\end{minipage}
\end{framed}
\end{center}
\end{minipage} \\
\vskip0.1cm
\begin{minipage}{6.4cm}
\begin{center}
\begin{framed}
Projected data, Single resolution \\
\begin{minipage}{2.8cm}
\begin{center}
\includegraphics[width=2.8cm]{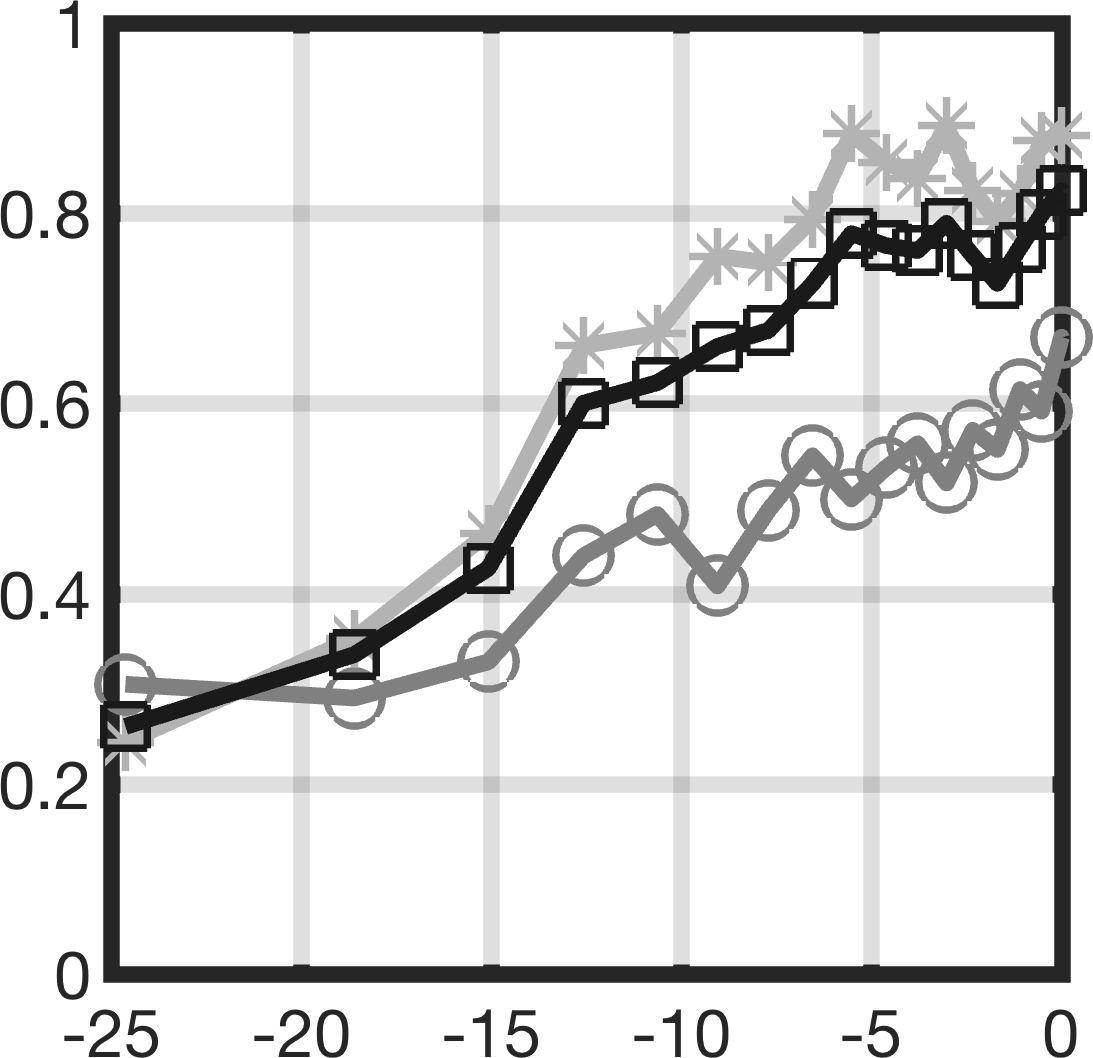}  \\
Noise sensitivity \\
x: $\eta$ (dB), y: ROE 
\end{center}
\end{minipage}
\begin{minipage}{2.8cm}
\begin{center}
\includegraphics[width=2.8cm]{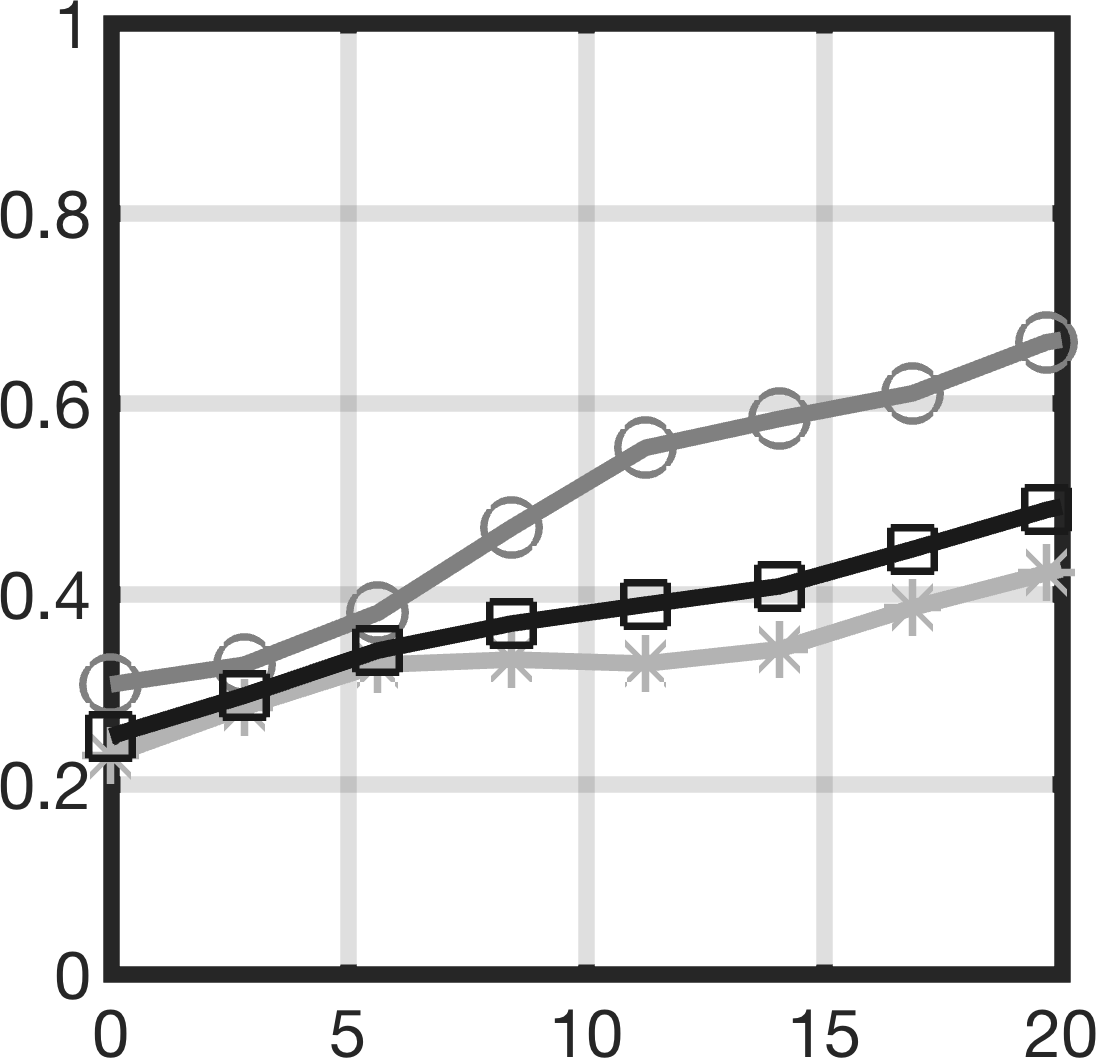} \\ 
Angular sensitivity \\
x: $\theta$  (deg), y: ROE
\end{center}
\end{minipage}
\\
\begin{minipage}{2.8cm}
\begin{center}
 \includegraphics[width=2.8cm]{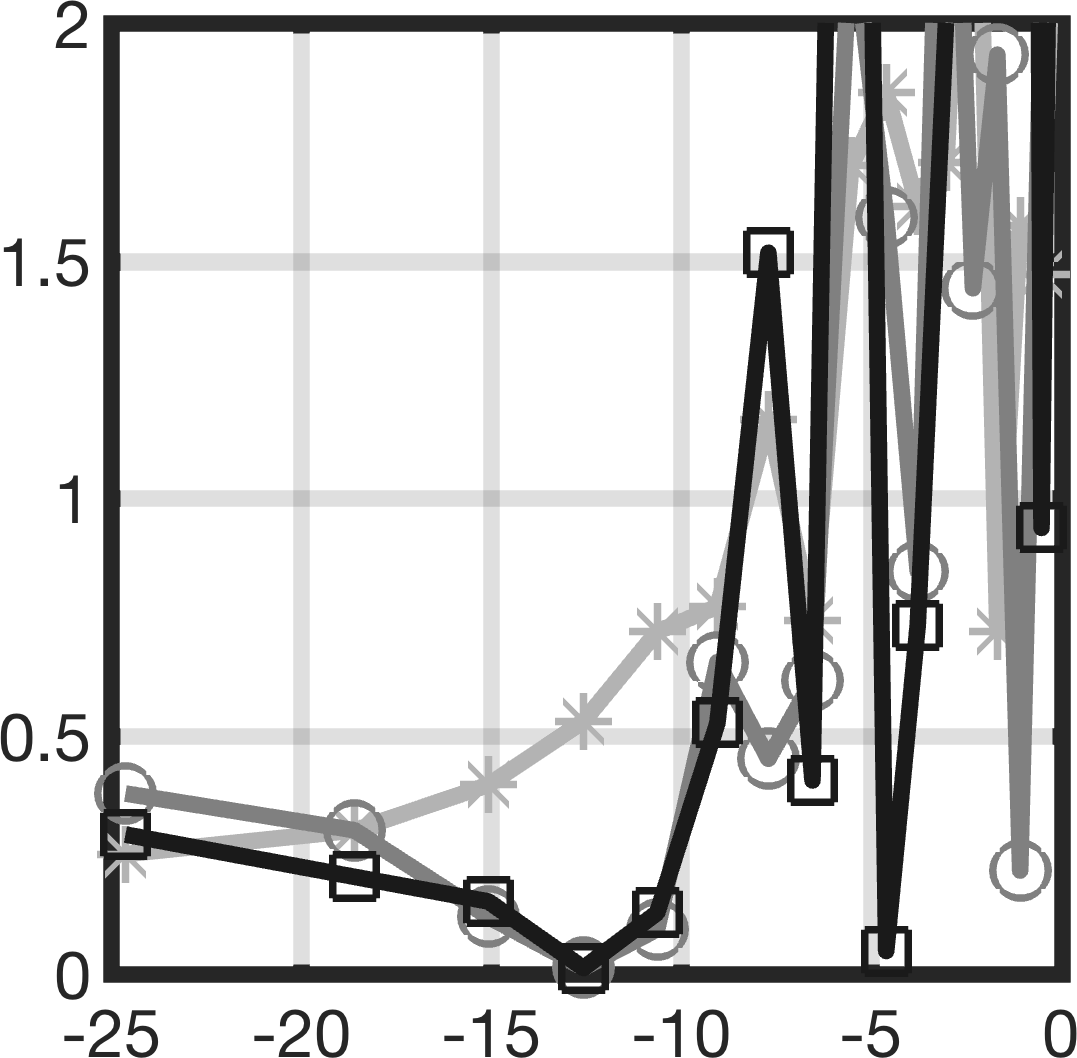}  \\
Noise sensitivity \\
x: $\eta$ (dB), y: RVE
\end{center}
\end{minipage}
\begin{minipage}{2.8cm}
\begin{center}
\includegraphics[width=2.8cm]{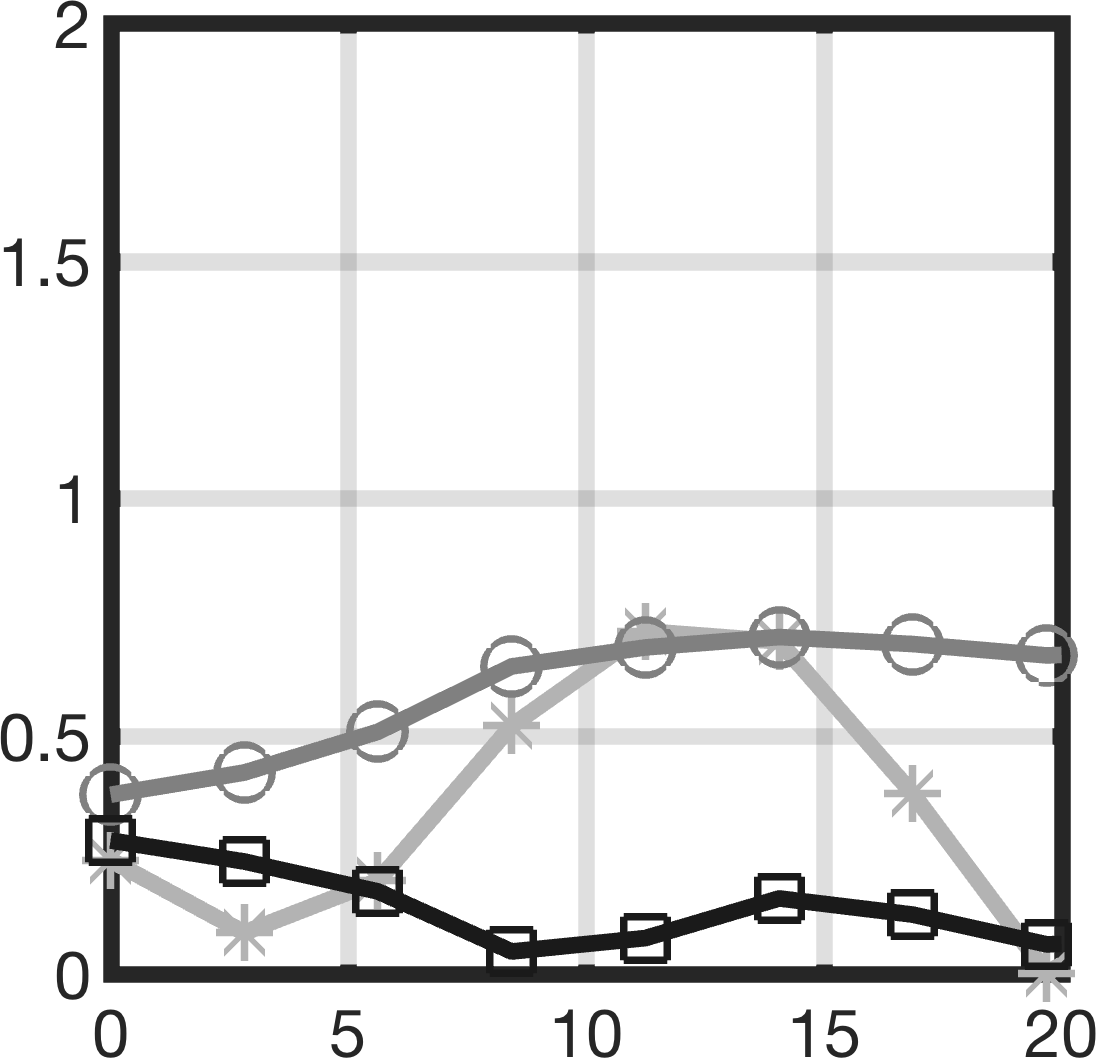} \\ 
Angular sensitivity \\
x: $\theta$  (deg), y: RVE
\end{center}
\end{minipage}
\end{framed}
\end{center}
\end{minipage}
\hskip0.1cm
\begin{minipage}{6.4cm}
\begin{center}
\begin{framed}
Projected data, Dual resolution \\
\begin{minipage}{2.8cm}
\begin{center}
\includegraphics[width=2.8cm]{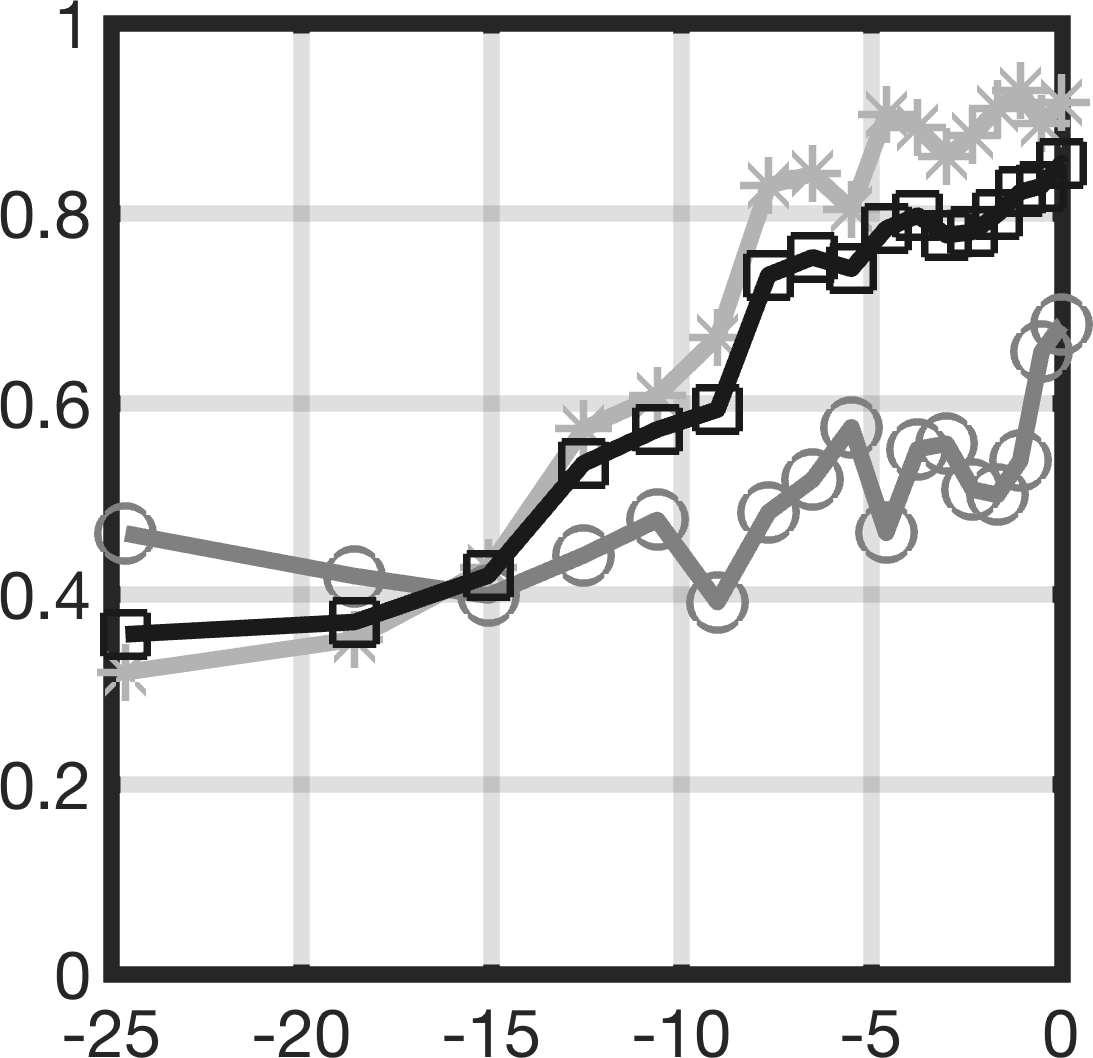}  \\ 
Noise sensitivity \\
x: $\eta$ (dB), y: ROE
\end{center}
\end{minipage}
\begin{minipage}{2.8cm}
\begin{center}
\includegraphics[width=2.8cm]{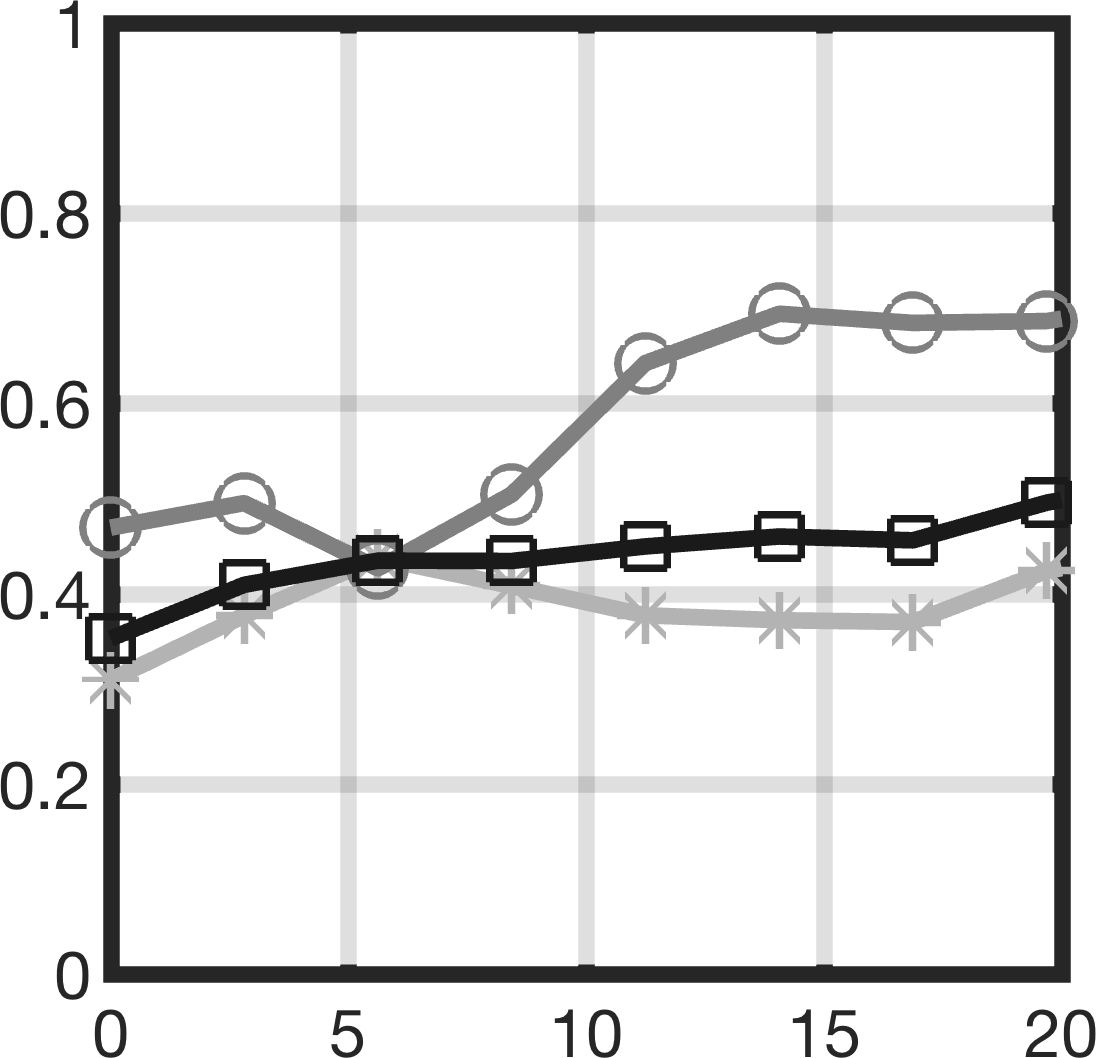} \\
Angular sensitivity \\
x: $\theta$  (deg), y: ROE
\end{center}
\end{minipage}
\\
\begin{minipage}{2.8cm}
\begin{center}
 \includegraphics[width=2.8cm]{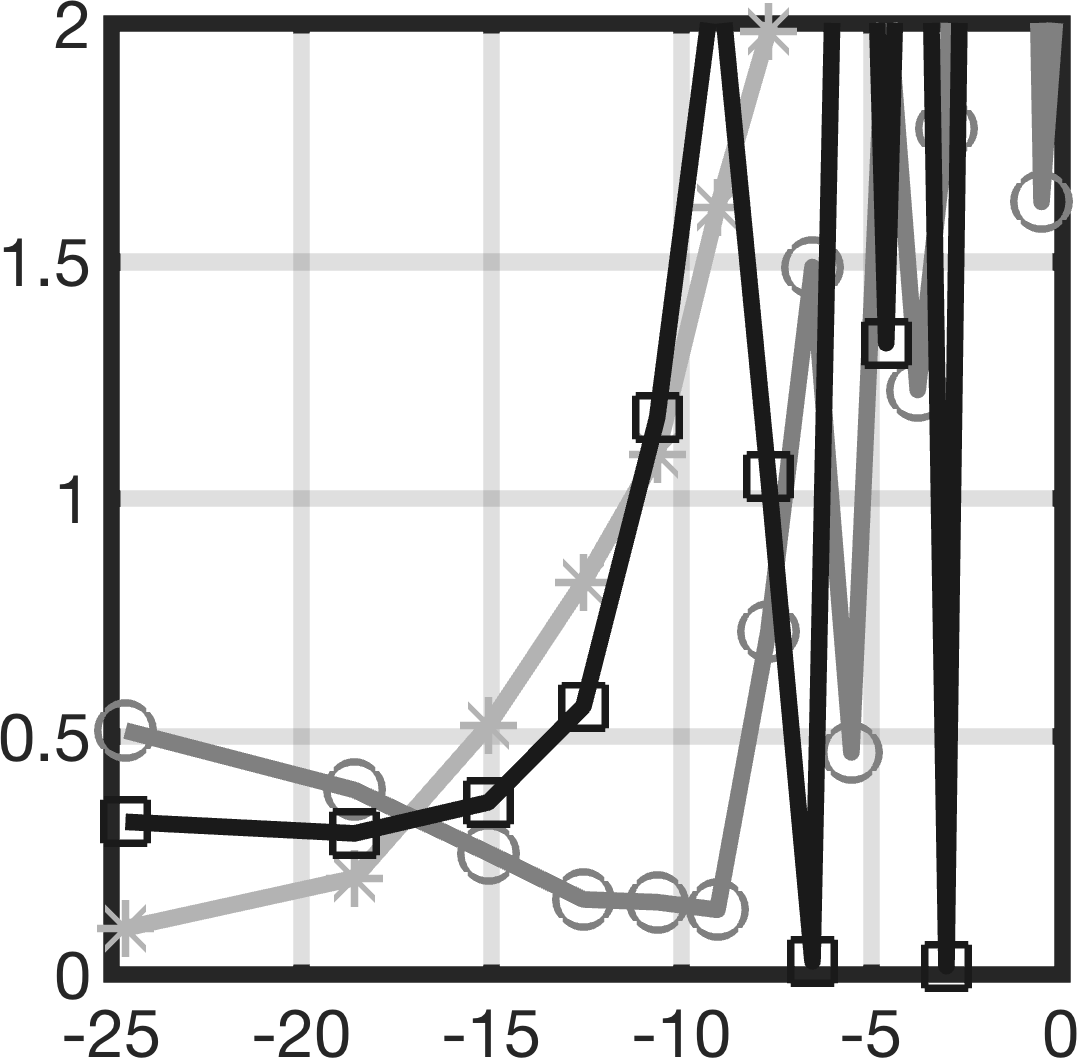}  \\
Noise sensitivity \\
x: $\eta$ (dB), y: RVE
\end{center}
\end{minipage}
\begin{minipage}{2.8cm}
\begin{center}
\includegraphics[width=2.8cm]{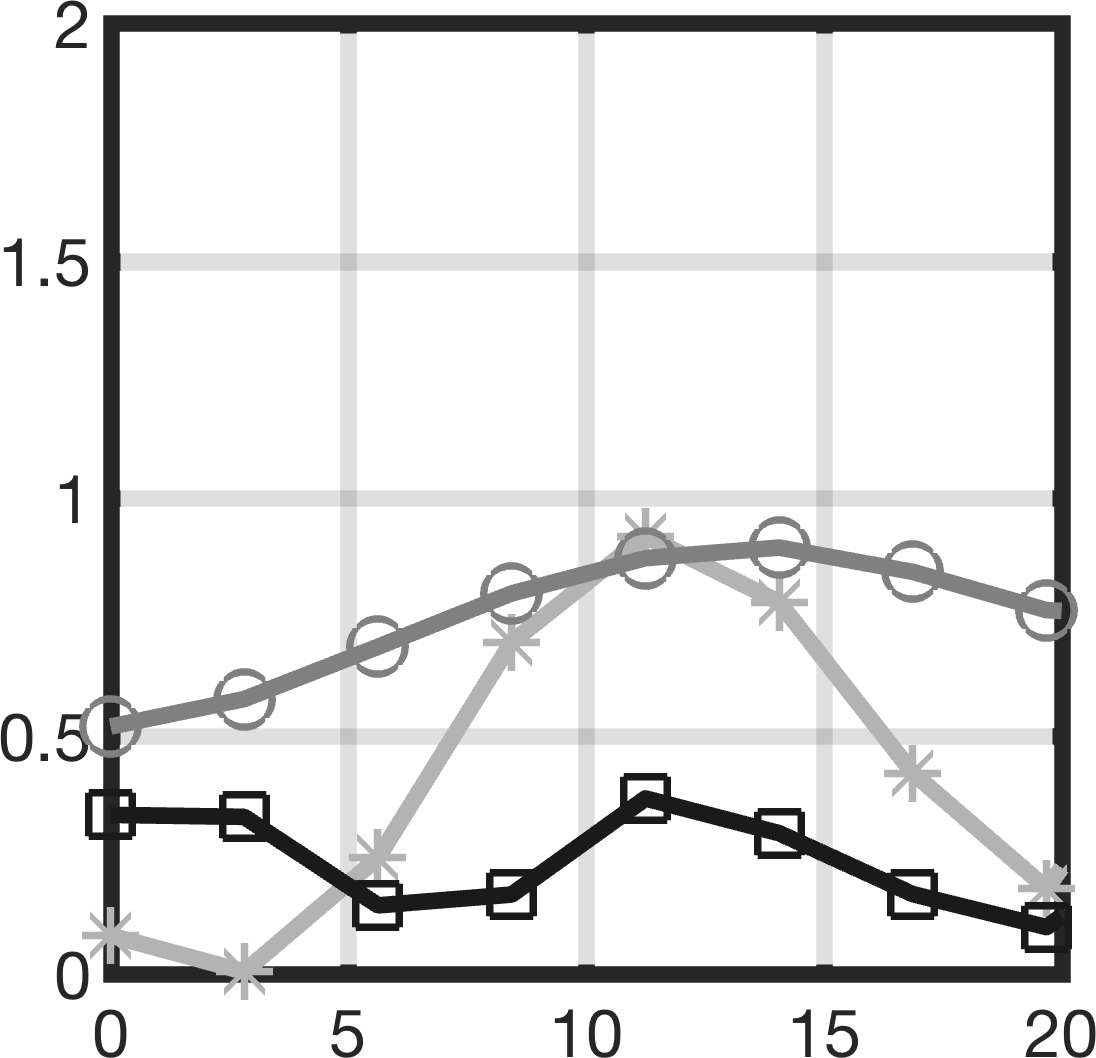} \\ 
Angular sensitivity \\
x: $\theta$  (deg), y: RVE
\end{center}
\end{minipage}
\end{framed}
\end{center}
\end{minipage}
\end{center}
\caption{  ROE and RVE  results for the signal configuration ({\bf A}). {\bf Surface layer}: light grey line, star marker,  {\bf Voids}: darker grey line, circle marker, {\bf Total value}: black line, square marker. \label{a_results}}
\end{scriptsize}
\end{figure*}

\begin{figure*}\begin{scriptsize}
\begin{center}
\begin{minipage}{6.4cm}
\begin{center}
\begin{framed}
Full wave, Single resolution \\
\begin{minipage}{2.8cm}
\begin{center}
\includegraphics[width=2.8cm]{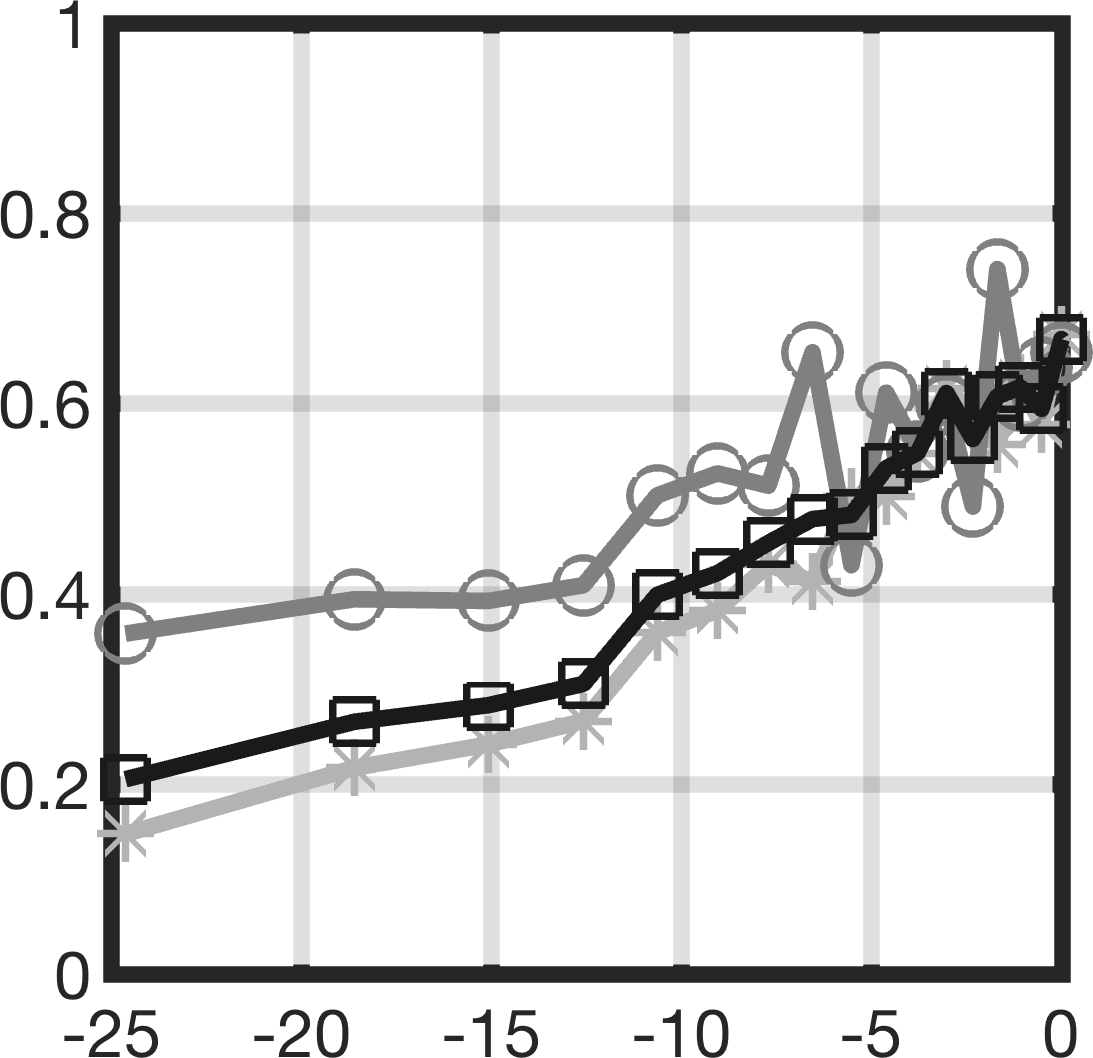}  \\
x: $\eta$ (dB), y: ROE
\end{center}
\end{minipage}
\begin{minipage}{2.8cm}
\begin{center}
\includegraphics[width=2.8cm]{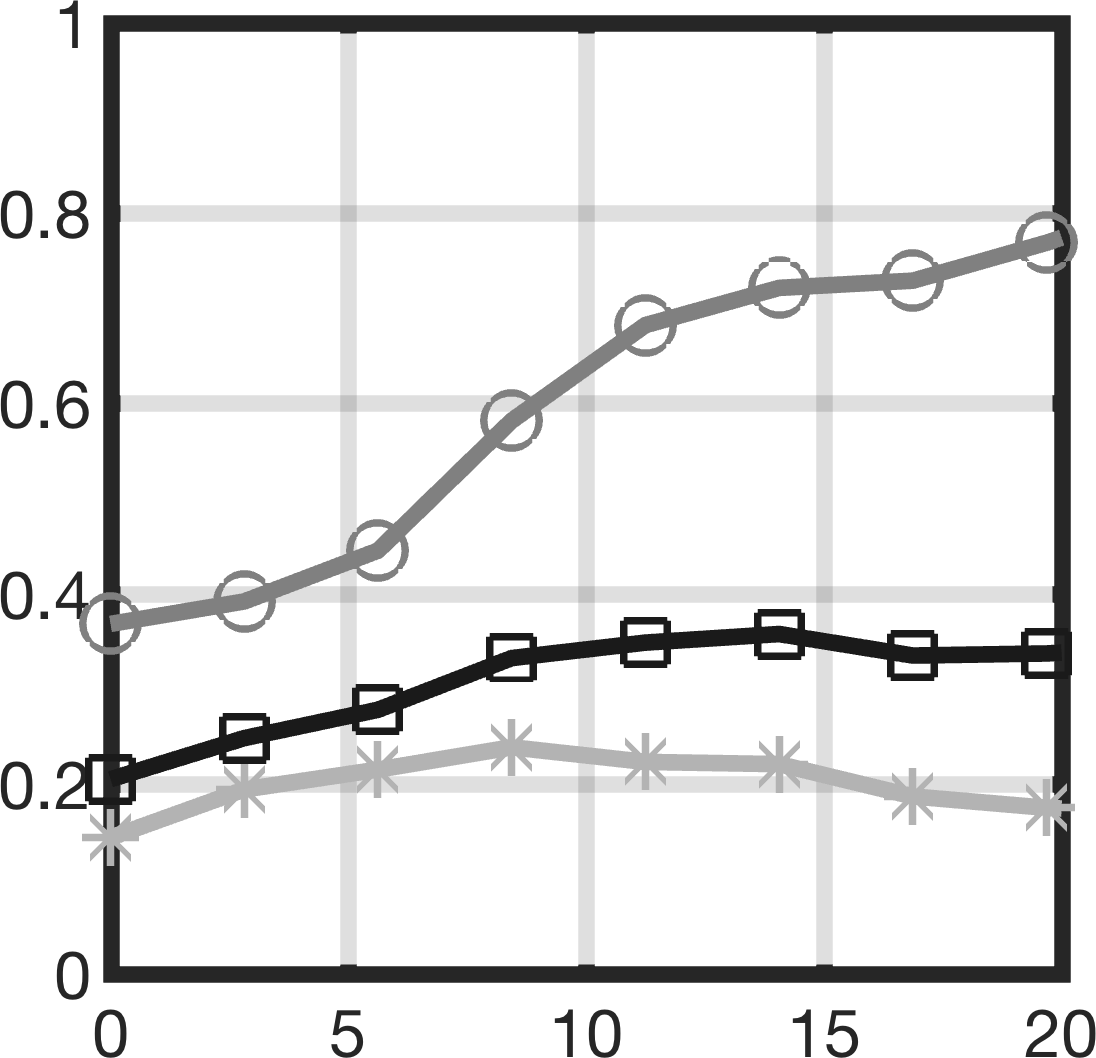} \\
x: ROE, y: $\theta$  (deg)
\end{center}
\end{minipage}
\\
\begin{minipage}{2.8cm}
\begin{center}
 \includegraphics[width=2.8cm]{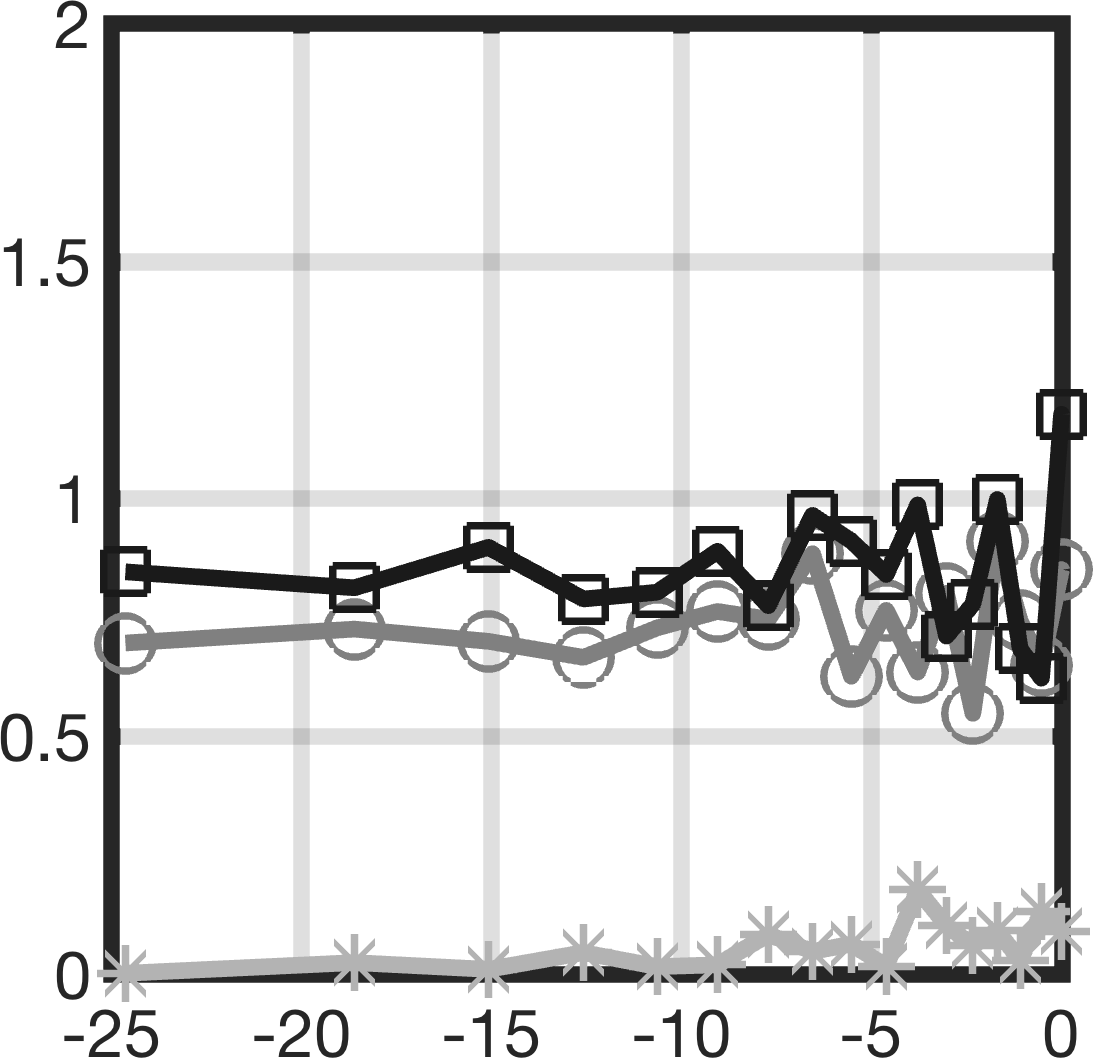}  \\
x: $\eta$ (dB), y: RVE
\end{center}
\end{minipage}
\begin{minipage}{2.8cm}
\begin{center}
\includegraphics[width=2.8cm]{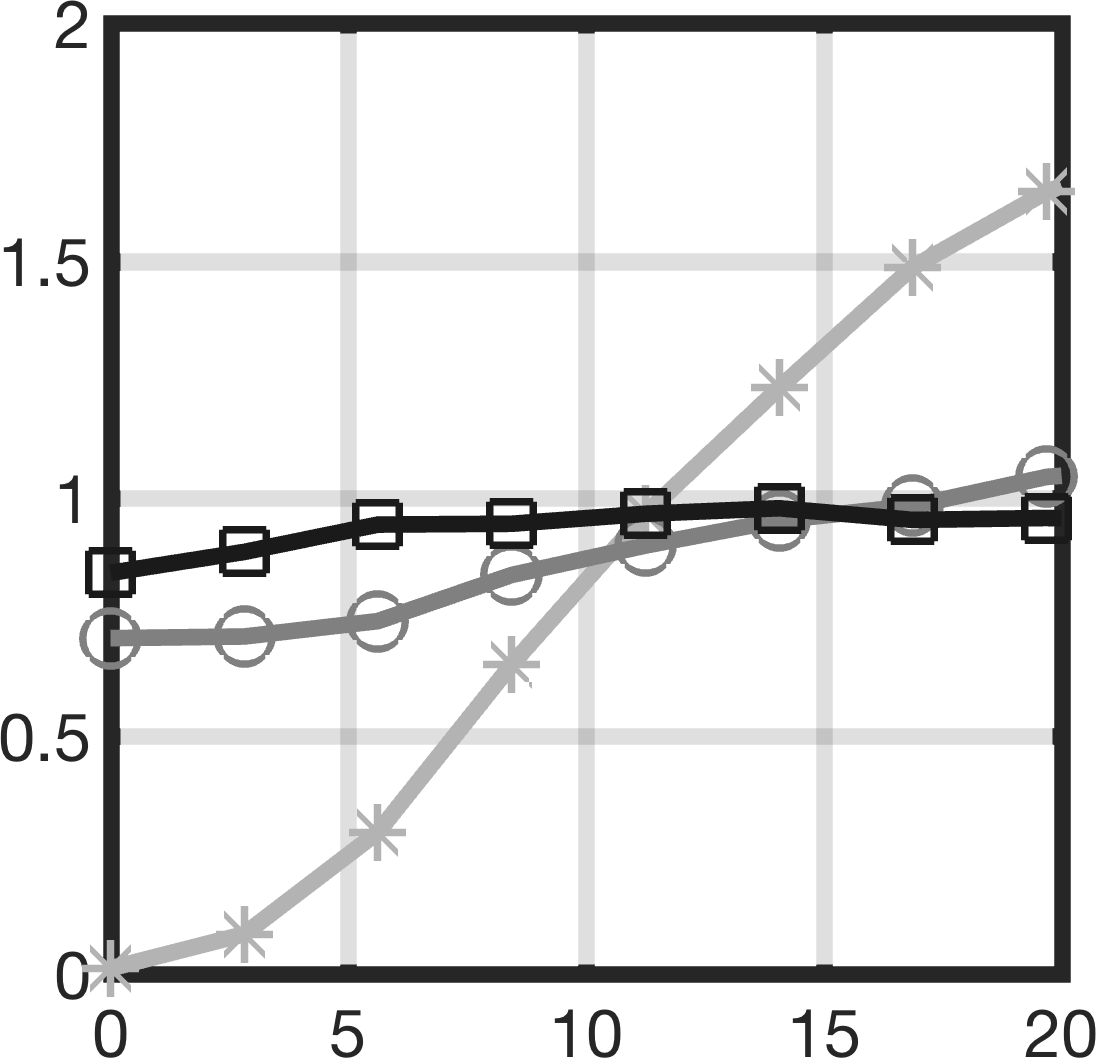} \\ 
x: RVE, y: $\theta$  (deg)
\end{center}
\end{minipage}
\end{framed}
\end{center}
\end{minipage}
\hskip0.1cm
\begin{minipage}{6.4cm}
\begin{center}
\begin{framed}
Full wave, Dual resolution \\
\begin{minipage}{2.8cm}
\begin{center}
\includegraphics[width=2.8cm]{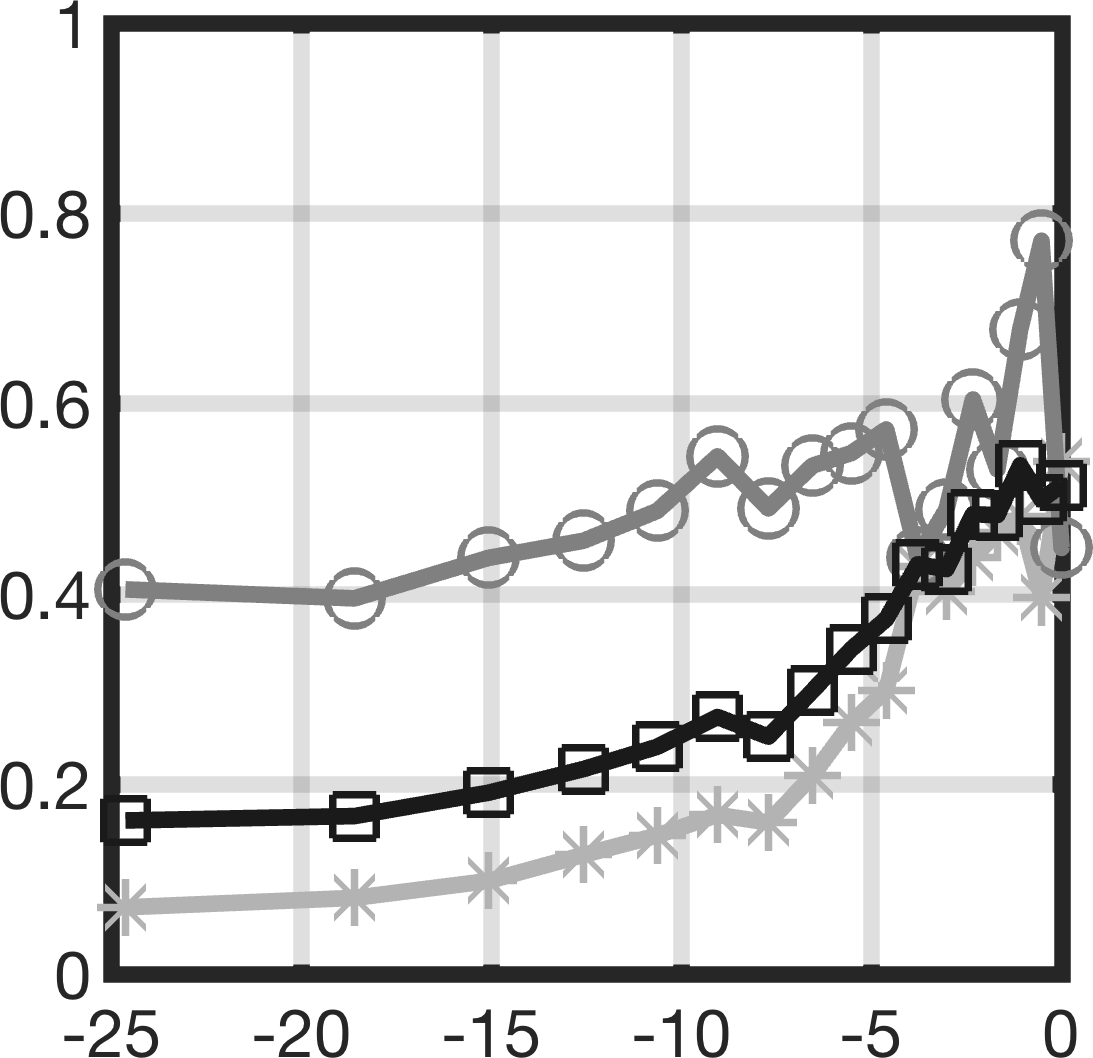}  \\
x: $\eta$ (dB), y: ROE
\end{center}
\end{minipage}
\begin{minipage}{2.8cm}
\begin{center}
\includegraphics[width=2.8cm]{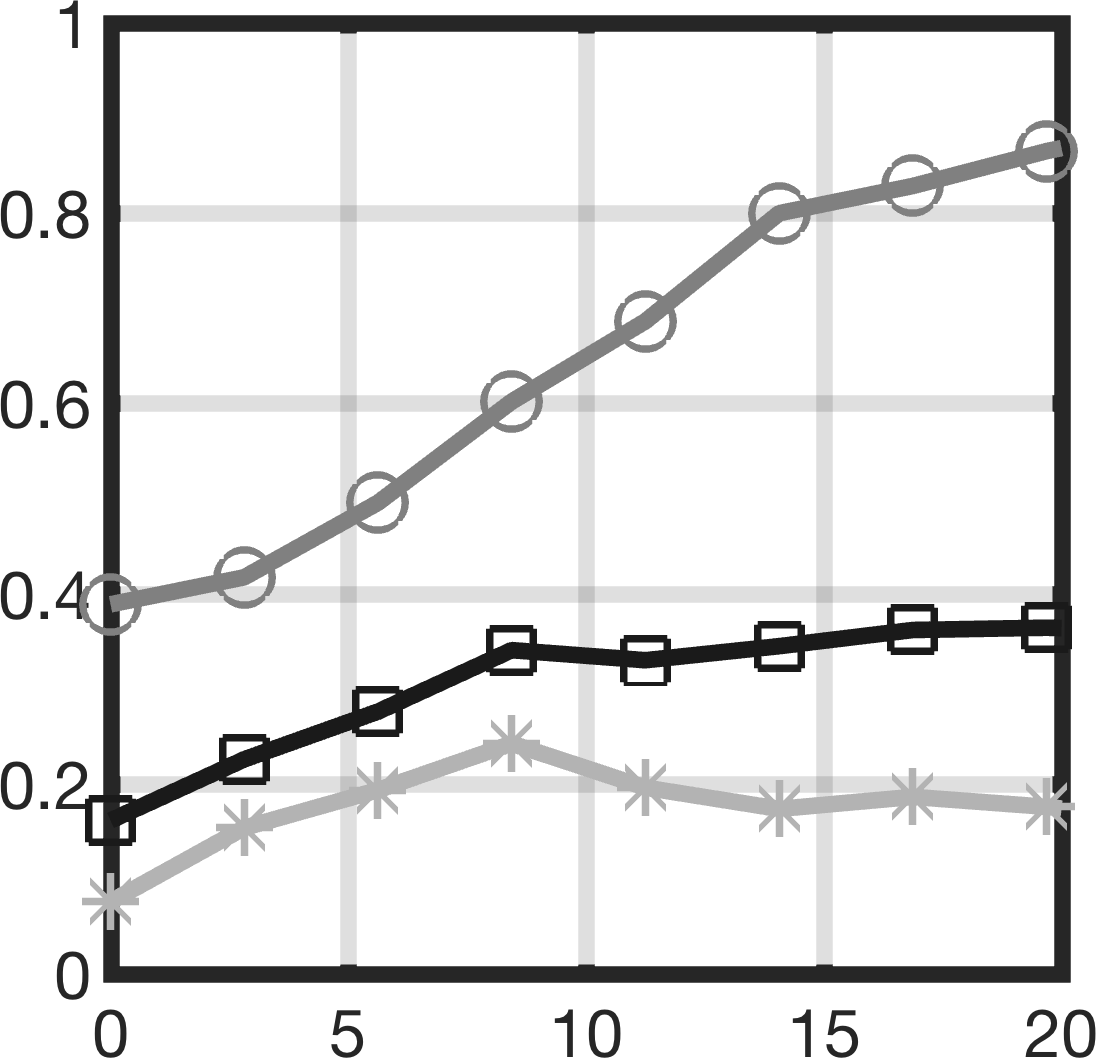} \\
x: ROE, y: $\theta$  (deg)
\end{center}
\end{minipage}
\\ 
\begin{minipage}{2.8cm}
\begin{center}
 \includegraphics[width=2.8cm]{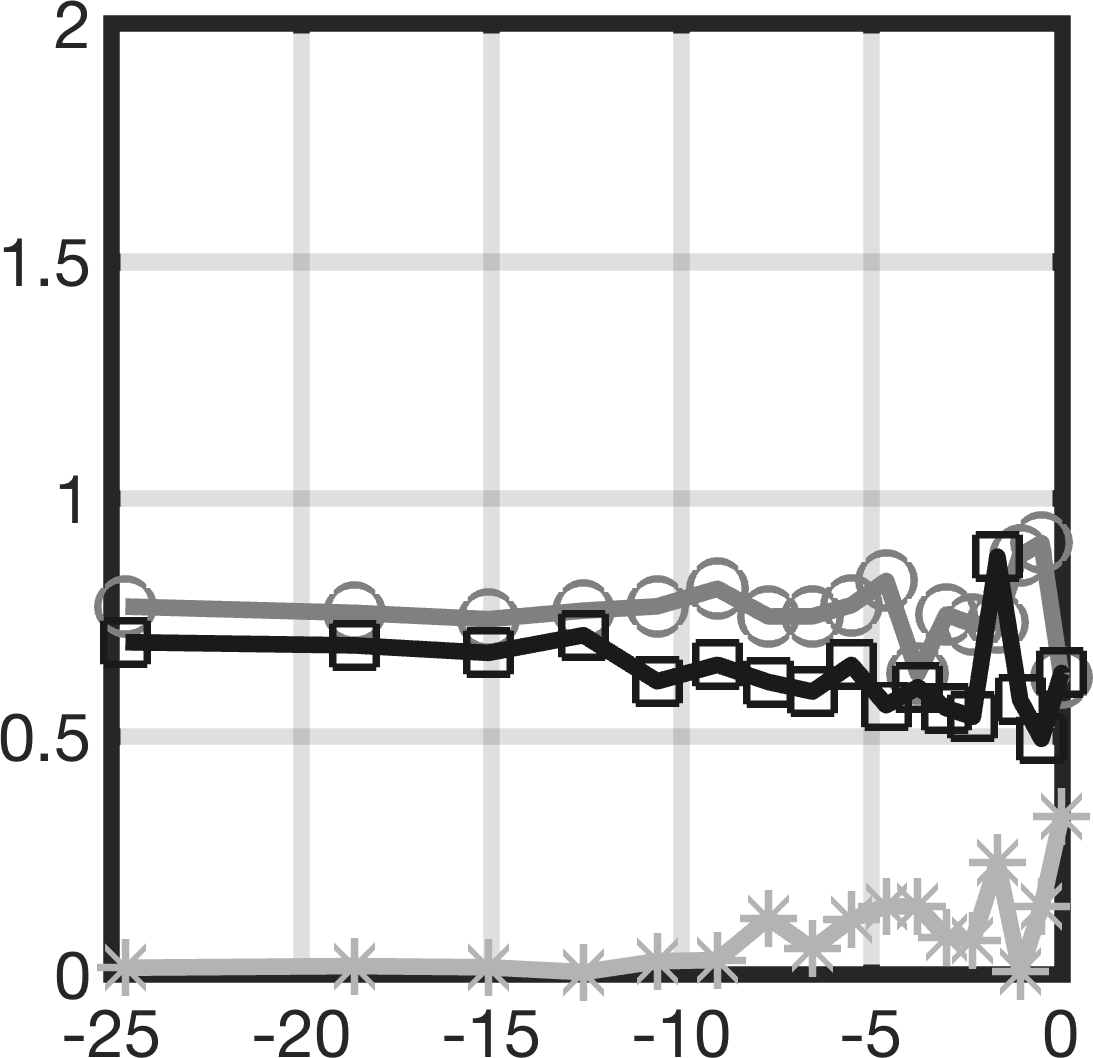}  \\
x: $\eta$ (dB), y: RVE
\end{center}
\end{minipage}
\begin{minipage}{2.8cm}
\begin{center}
\includegraphics[width=2.8cm]{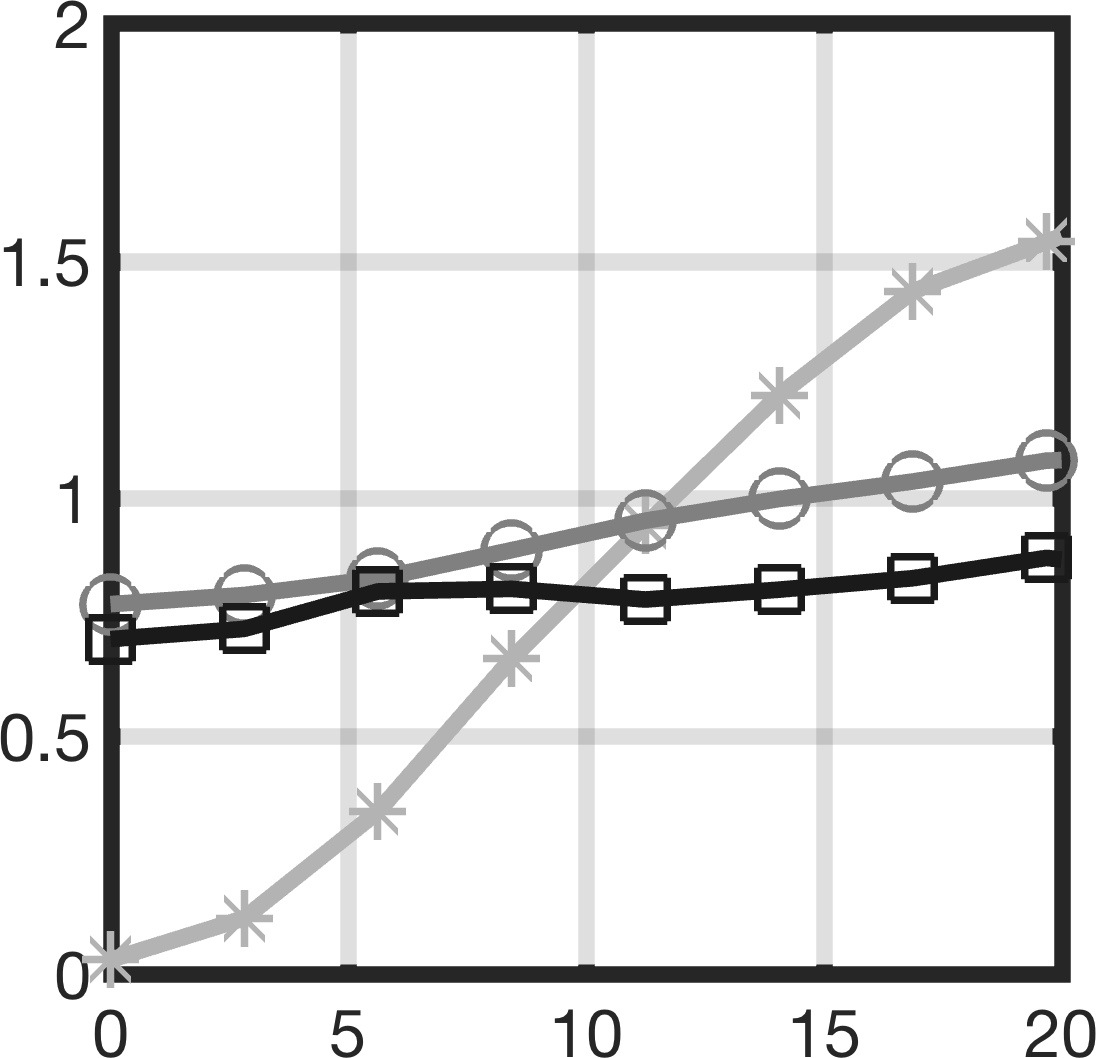} \\ 
x: RVE, y: $\theta$  (deg)
\end{center}
\end{minipage}
\end{framed}
\end{center}
\end{minipage} \\
\vskip0.1cm
\begin{minipage}{6.4cm}
\begin{center}
\begin{framed}
Projected data, Single resolution \\
\begin{minipage}{2.8cm}
\begin{center}
\includegraphics[width=2.8cm]{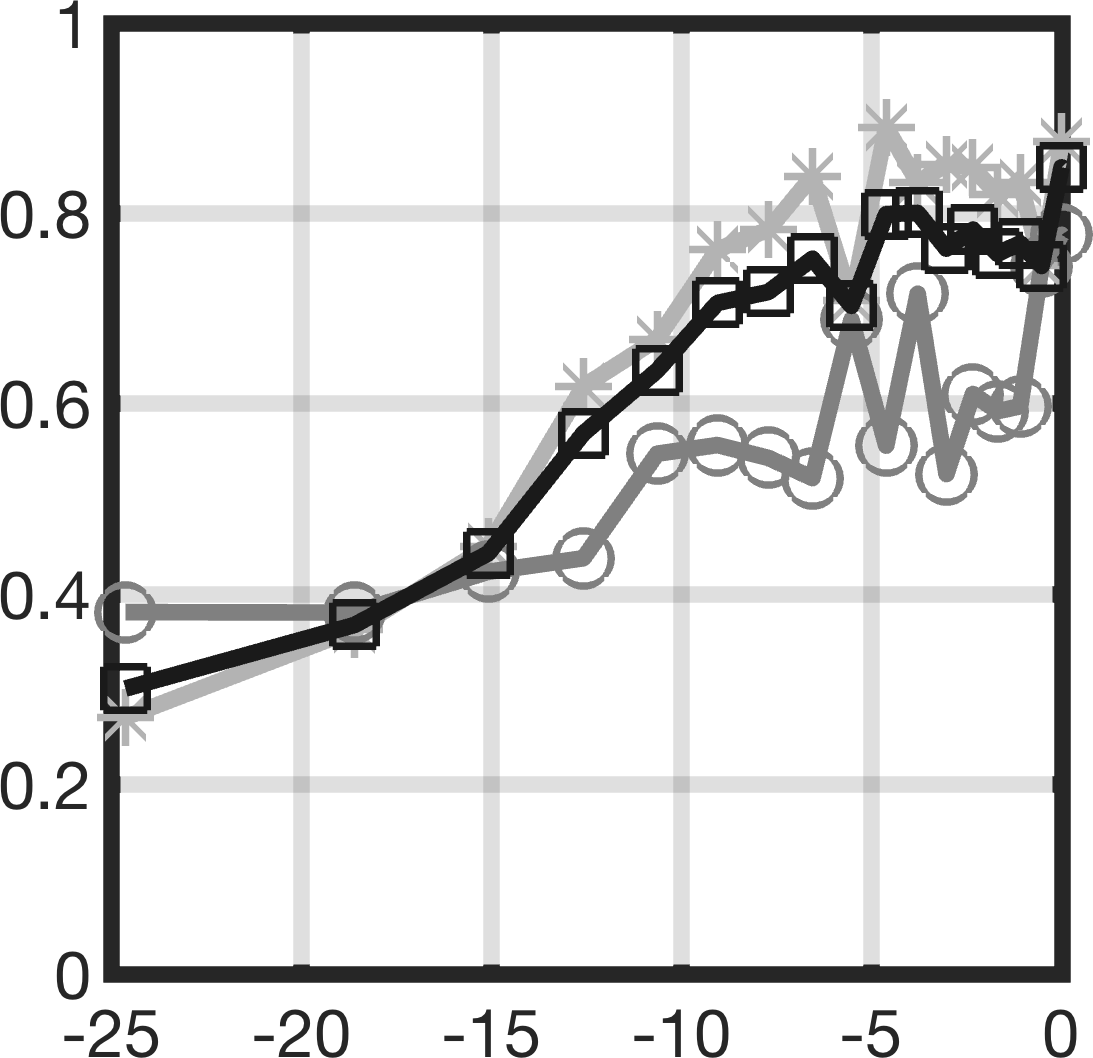}  \\
x: $\eta$ (dB), y: ROE
\end{center}
\end{minipage}
\begin{minipage}{2.8cm}
\begin{center}
\includegraphics[width=2.8cm]{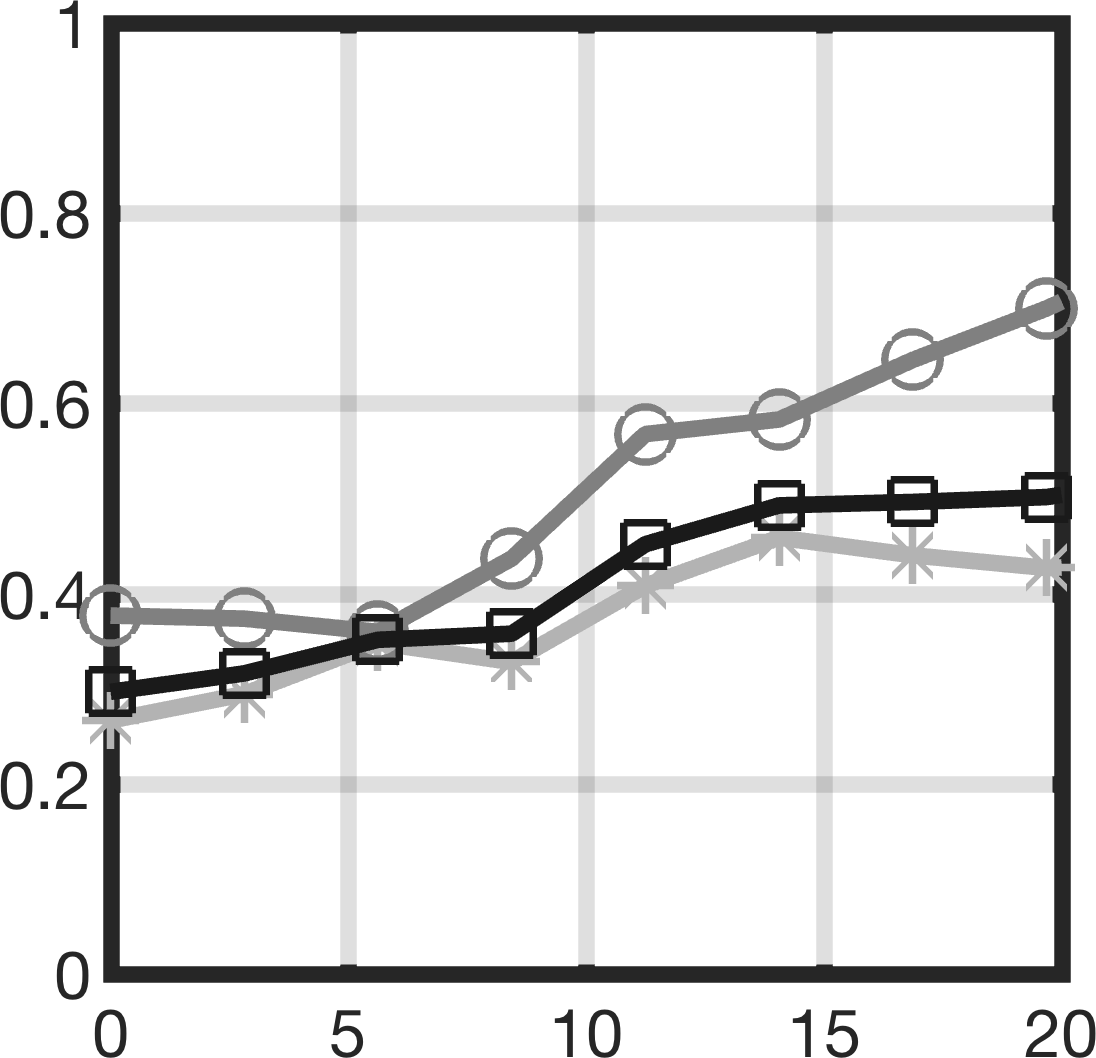} \\
x: ROE, y: $\theta$  (deg)
\end{center}
\end{minipage}
\\
\begin{minipage}{2.8cm}
\begin{center}
 \includegraphics[width=2.8cm]{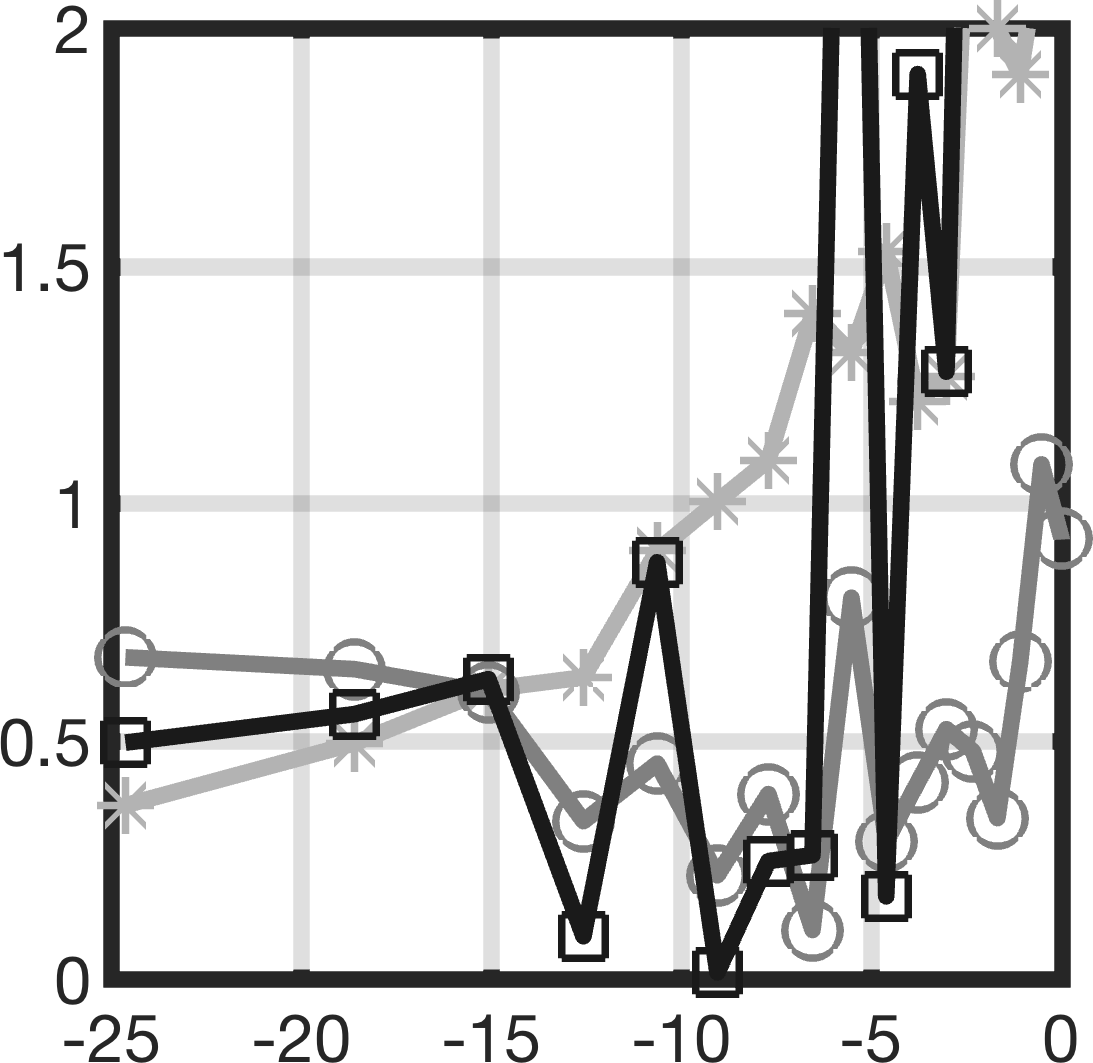}  \\
x: $\eta$ (dB), y: RVE
\end{center}
\end{minipage}
\begin{minipage}{2.8cm}
\begin{center}
\includegraphics[width=2.8cm]{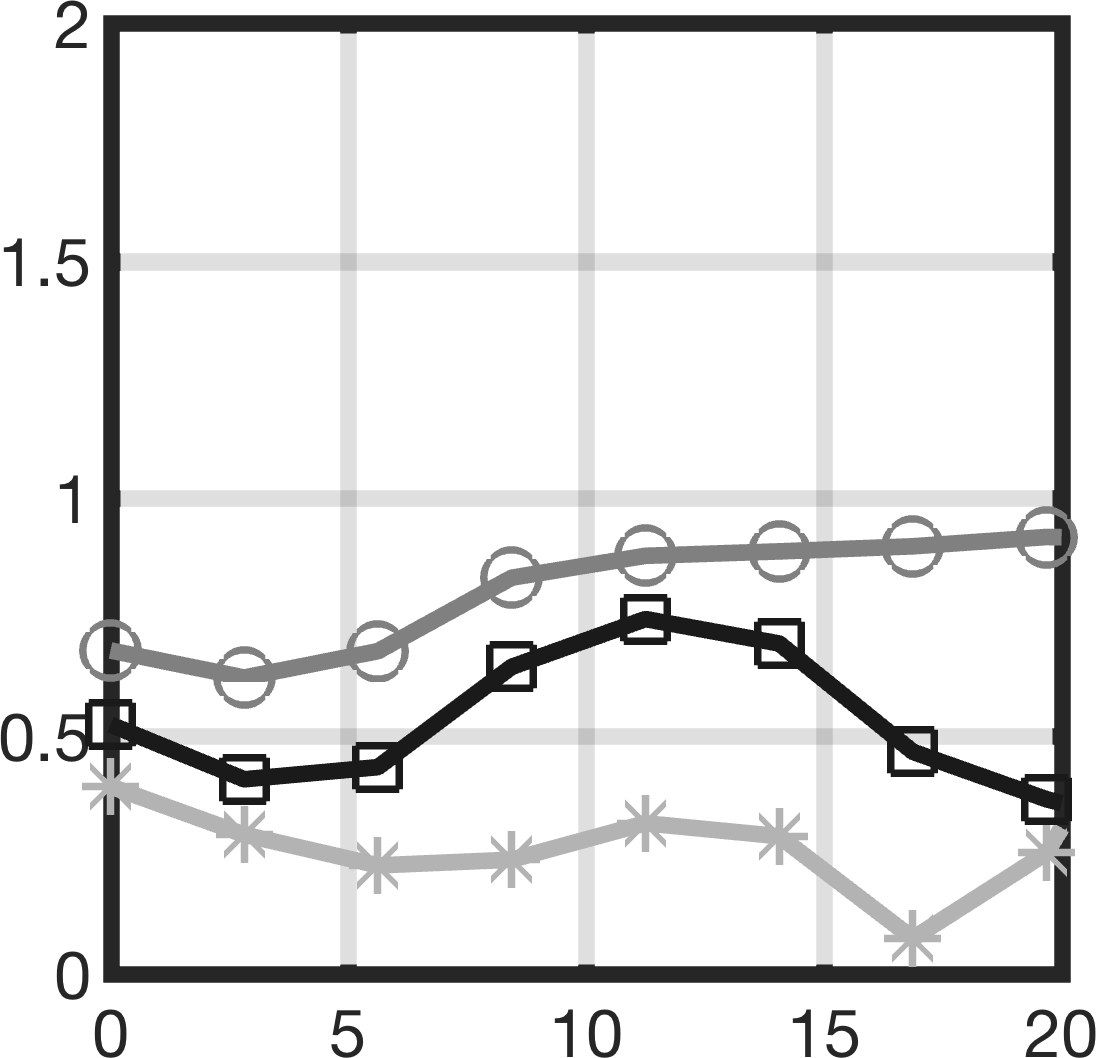} \\ 
x: RVE, y: $\theta$  (deg)
\end{center}
\end{minipage}
\end{framed}
\end{center}
\end{minipage}
\hskip0.1cm
\begin{minipage}{6.4cm}
\begin{center}
\begin{framed}
Projected data, Dual resolution \\
\begin{minipage}{2.8cm}
\begin{center}
\includegraphics[width=2.8cm]{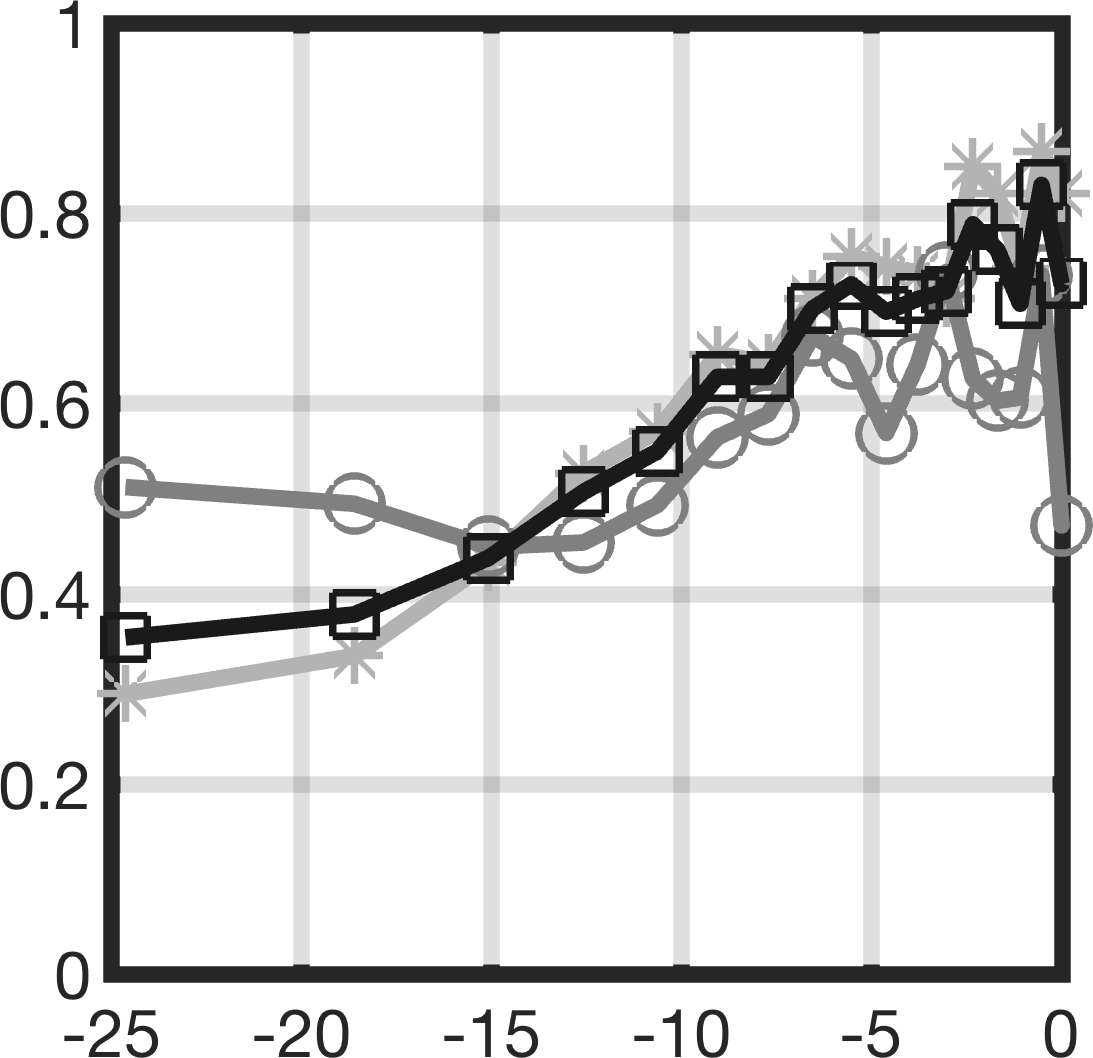}  \\
x: $\eta$ (dB), y: ROE
\end{center}
\end{minipage}
\begin{minipage}{2.8cm}
\begin{center}
\includegraphics[width=2.8cm]{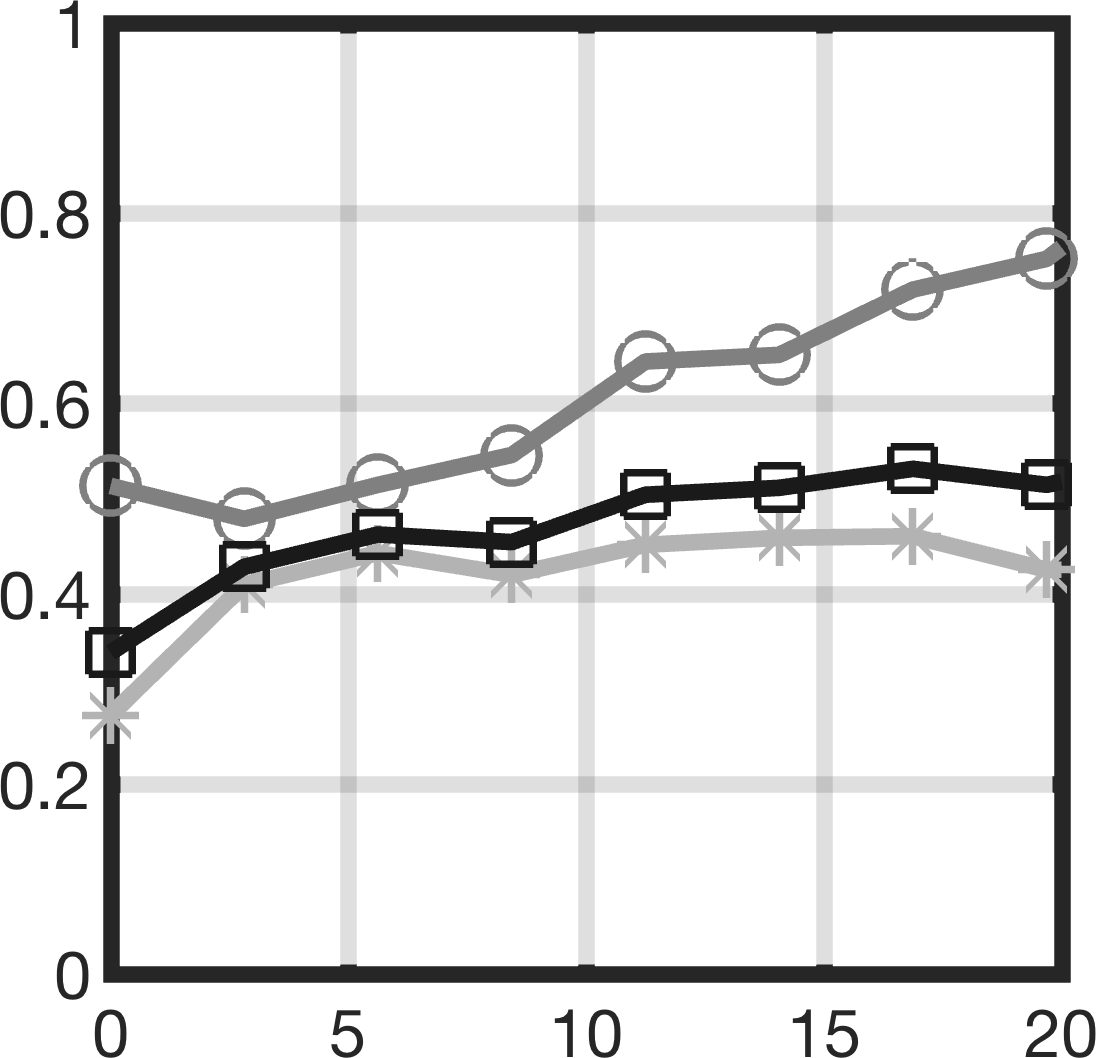} \\
x: ROE, y: $\theta$  (deg)
\end{center}
\end{minipage}
\\
\begin{minipage}{2.8cm}
\begin{center}
 \includegraphics[width=2.8cm]{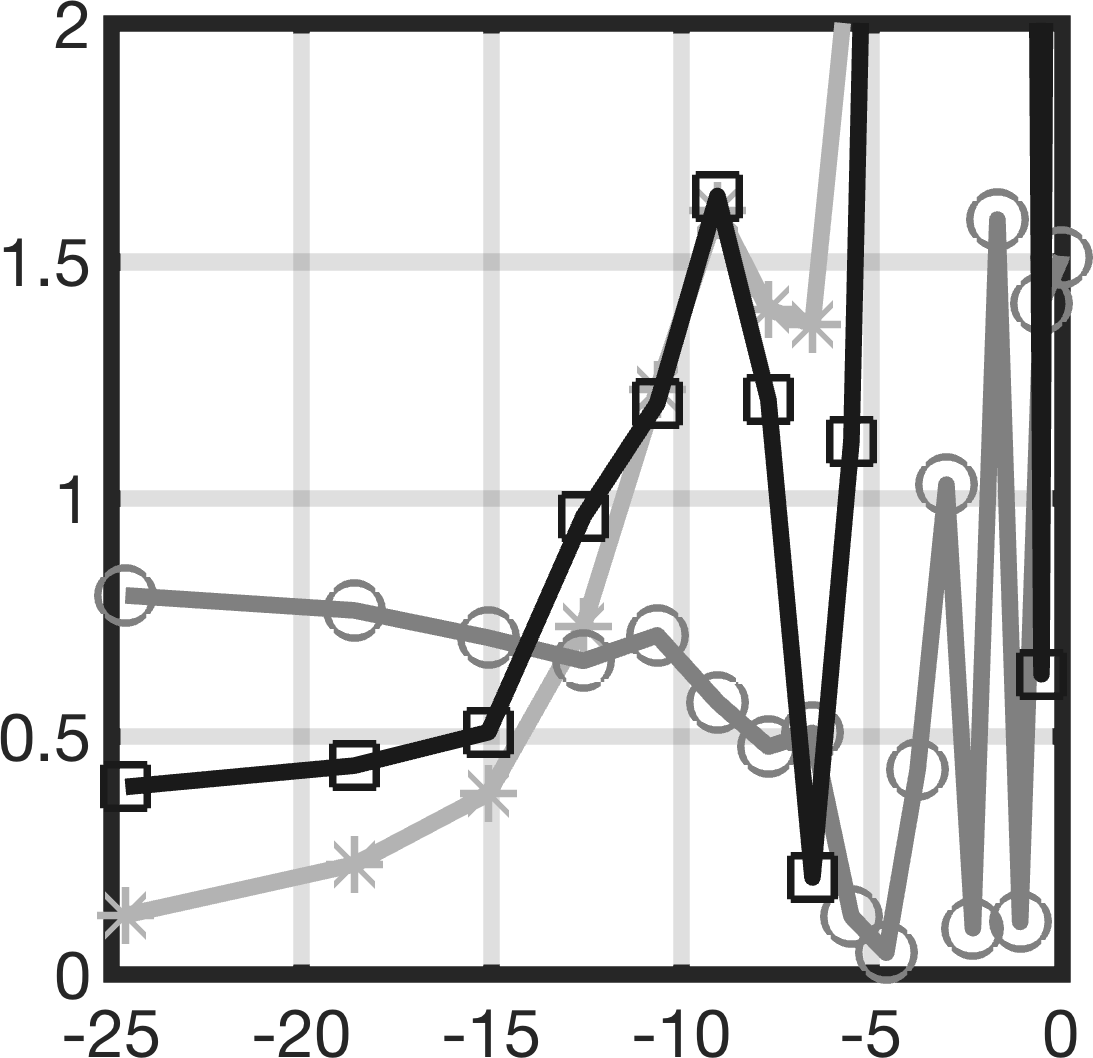}  \\
x: $\eta$ (dB), y: RVE
\end{center}
\end{minipage}
\begin{minipage}{2.8cm}
\begin{center}
\includegraphics[width=2.8cm]{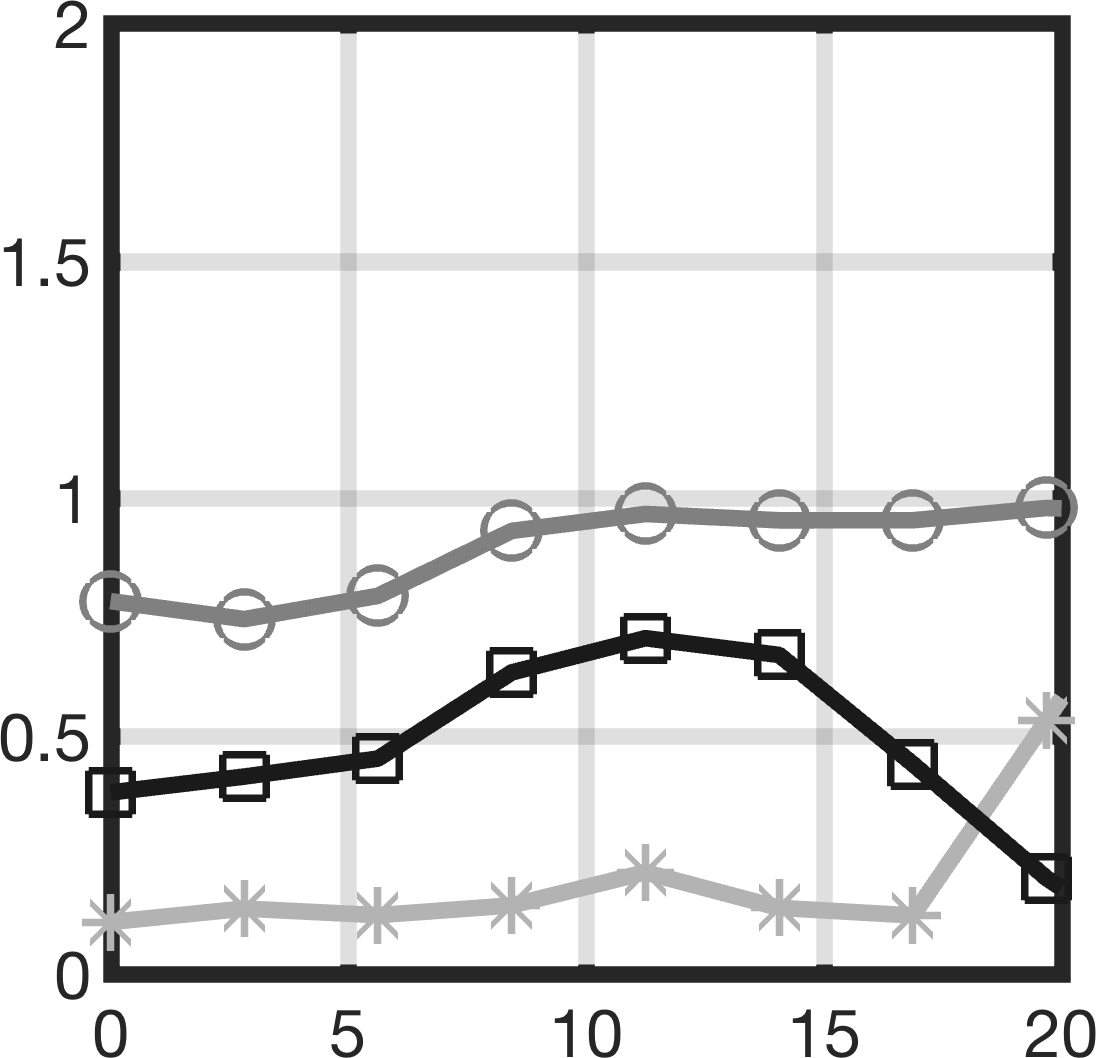} \\ 
x: RVE, y: $\theta$  (deg)
\end{center}
\end{minipage}
\end{framed}
\end{center}
\end{minipage}
\end{center}
\caption{  ROE and RVE  results for the signal configuration ({\bf B}). {\bf Surface layer}: light grey line, star marker,  {\bf Voids}: darker grey line, circle marker, {\bf Total value}: black line, square marker.  \label{b_results}}
\end{scriptsize}
\end{figure*}

\begin{figure*}\begin{scriptsize}
\begin{center}
\begin{minipage}{6.4cm}
\begin{center}
\begin{framed}
Full wave, Single resolution \\
\begin{minipage}{2.8cm}
\begin{center}
\includegraphics[width=2.8cm]{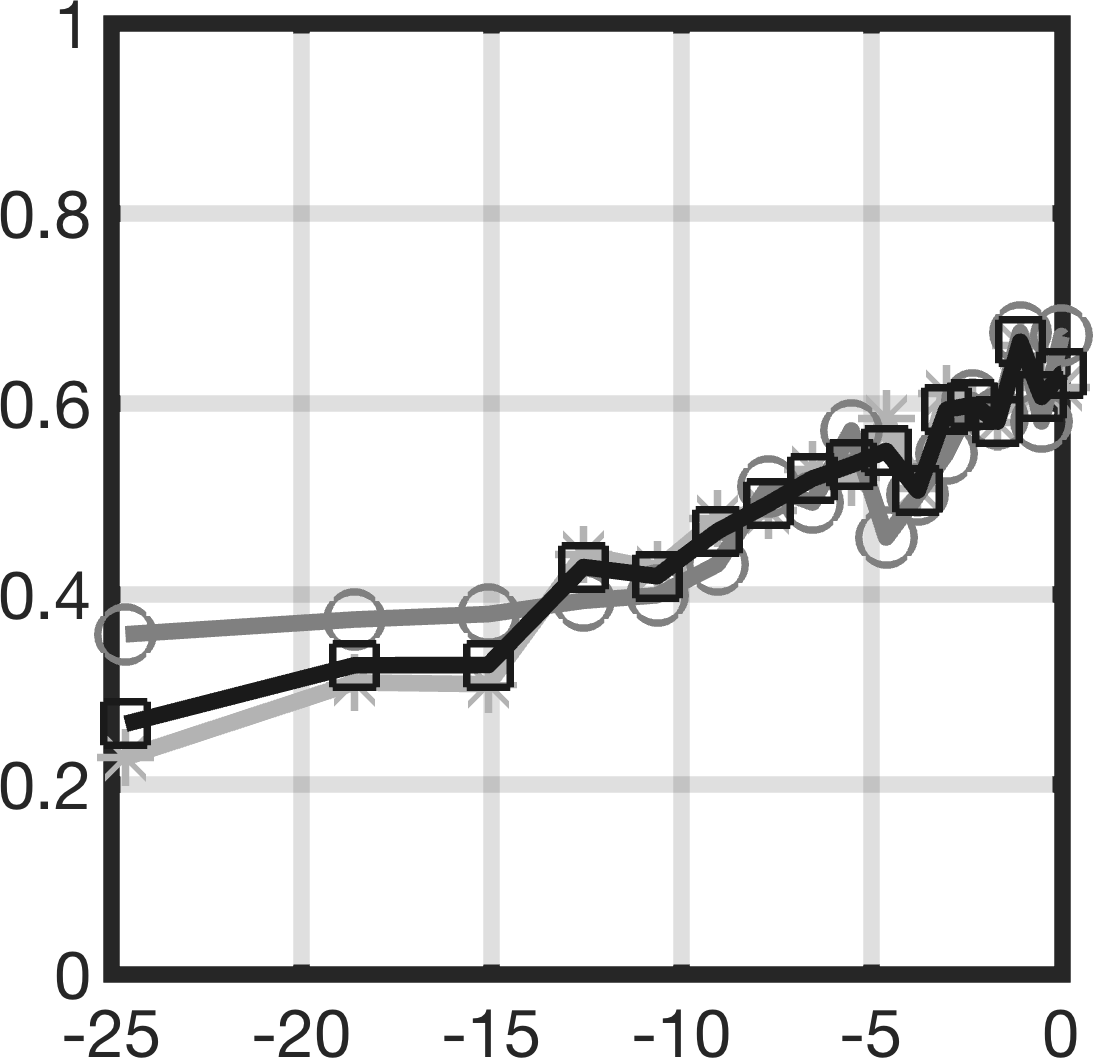}  \\
x: $\eta$ (dB), y: ROE
\end{center}
\end{minipage}
\begin{minipage}{2.8cm}
\begin{center}
\includegraphics[width=2.8cm]{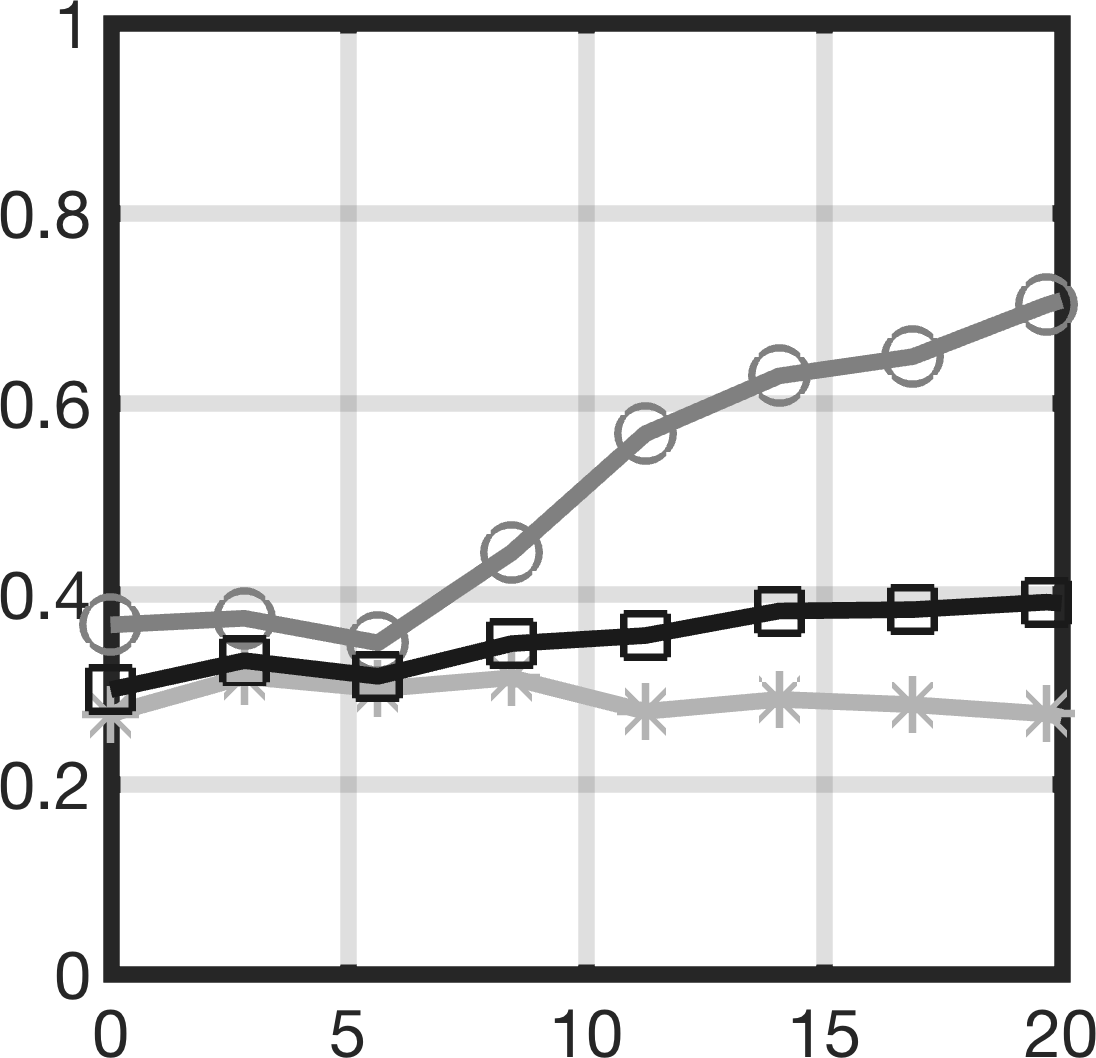} \\
x: ROE, y: $\theta$  (deg)
\end{center}
\end{minipage}
\\
\begin{minipage}{2.8cm}
\begin{center}
 \includegraphics[width=2.8cm]{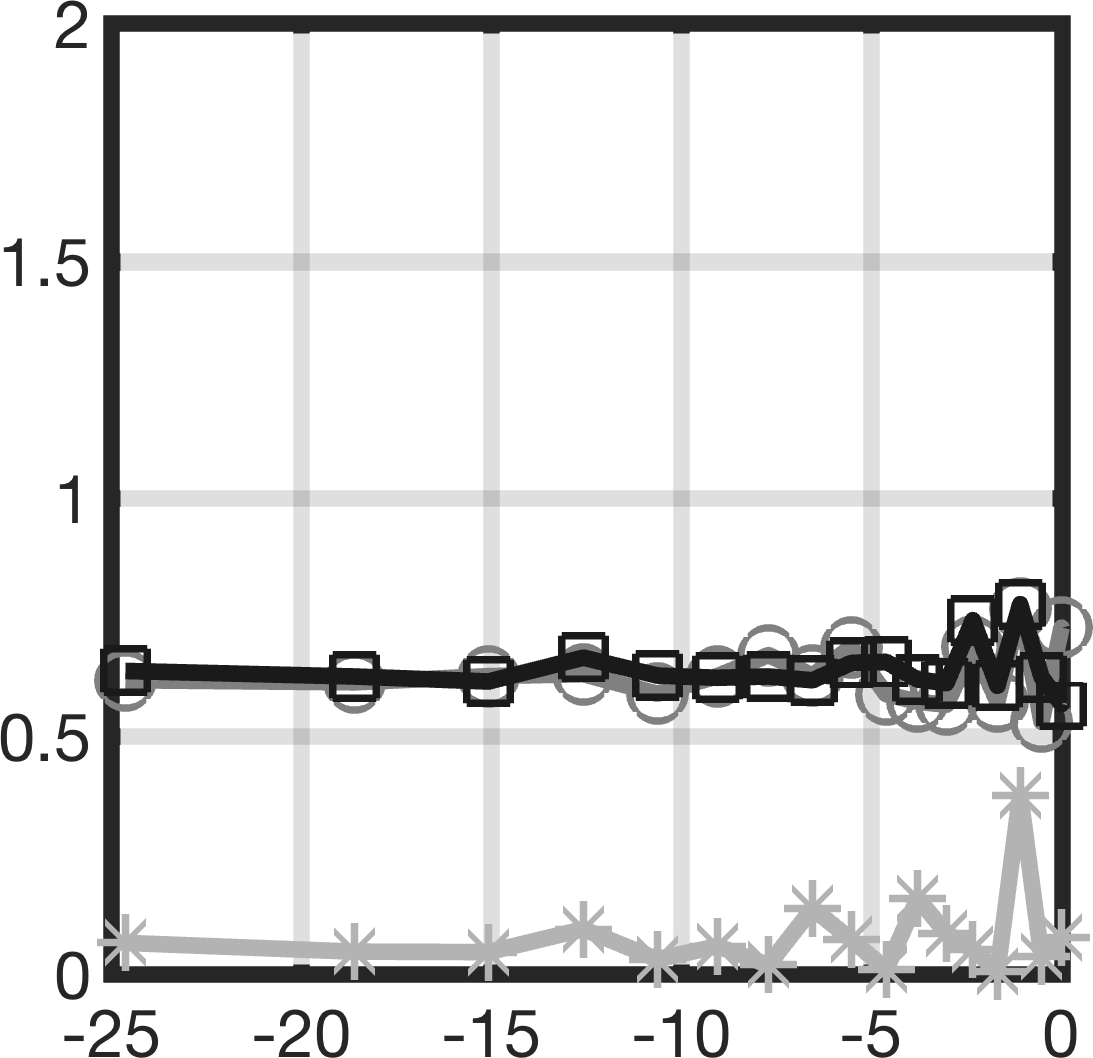}  \\
x: $\eta$ (dB), y: RVE
\end{center}
\end{minipage}
\begin{minipage}{2.8cm}
\begin{center}
\includegraphics[width=2.8cm]{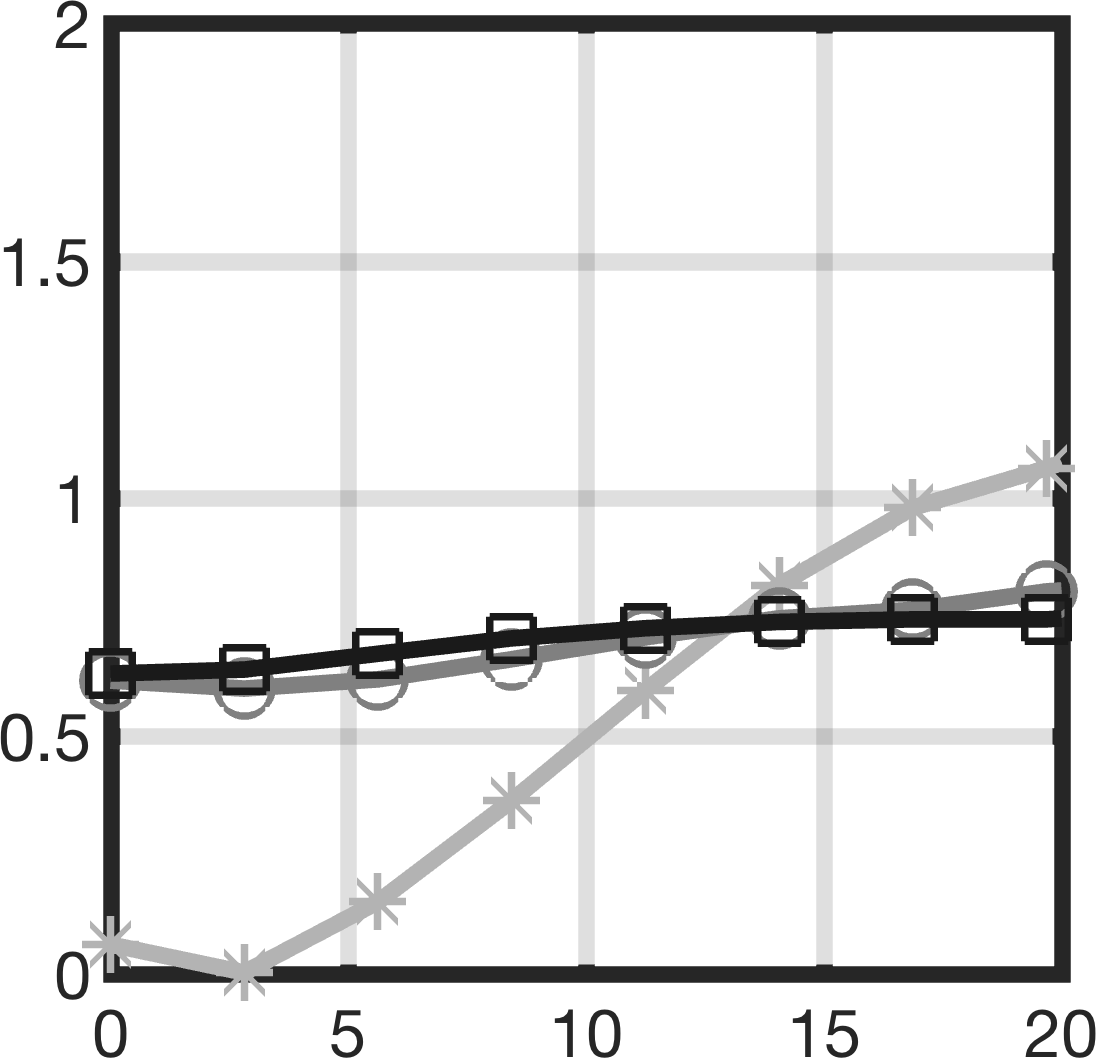} \\ 
x: RVE, y: $\theta$  (deg)
\end{center}
\end{minipage}
\end{framed}
\end{center}
\end{minipage}
\hskip0.1cm
\begin{minipage}{6.4cm}
\begin{center}
\begin{framed}
Full wave, Dual resolution \\
\begin{minipage}{2.8cm}
\begin{center}
\includegraphics[width=2.8cm]{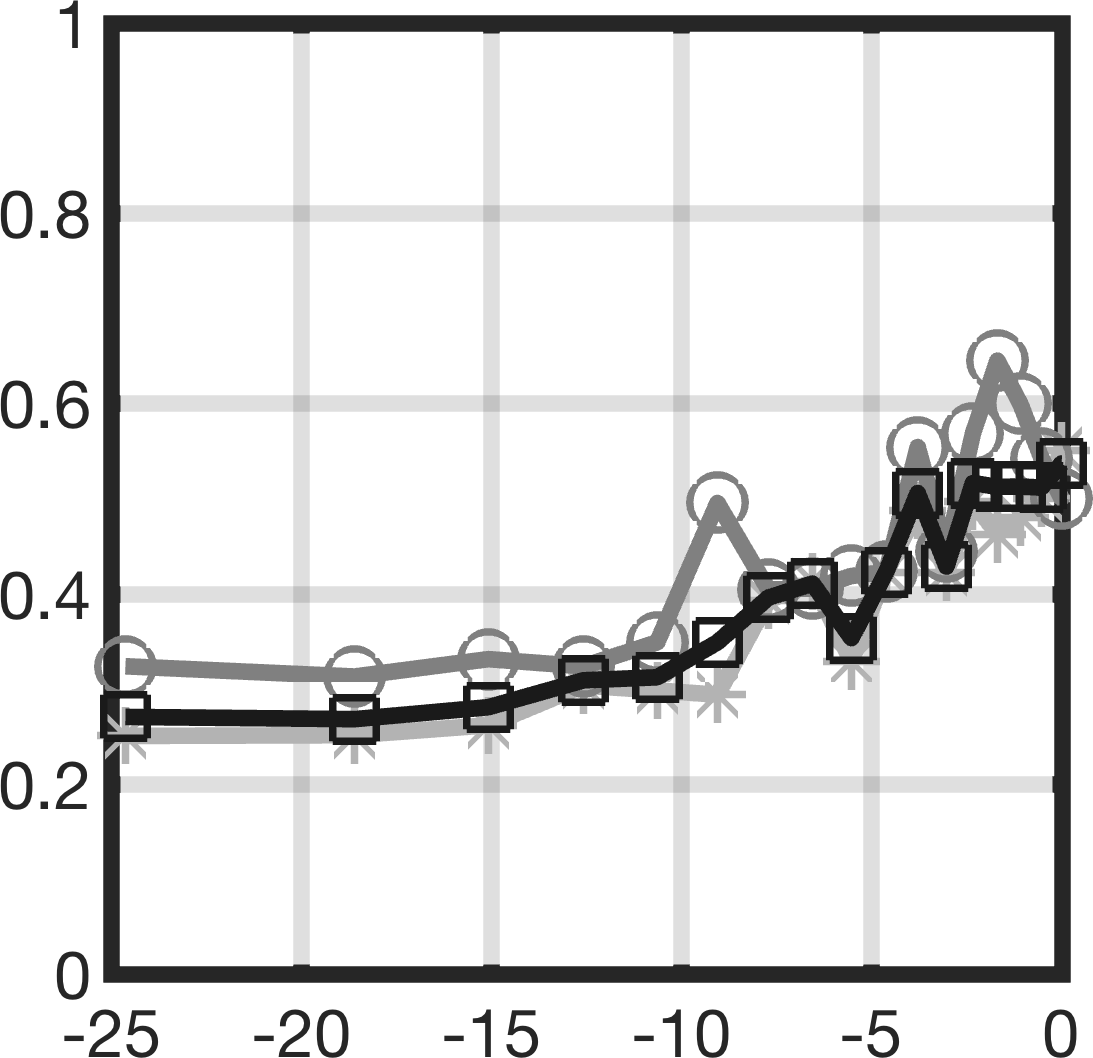}  \\
x: $\eta$ (dB), y: ROE
\end{center}
\end{minipage}
\begin{minipage}{2.8cm}
\begin{center}
\includegraphics[width=2.8cm]{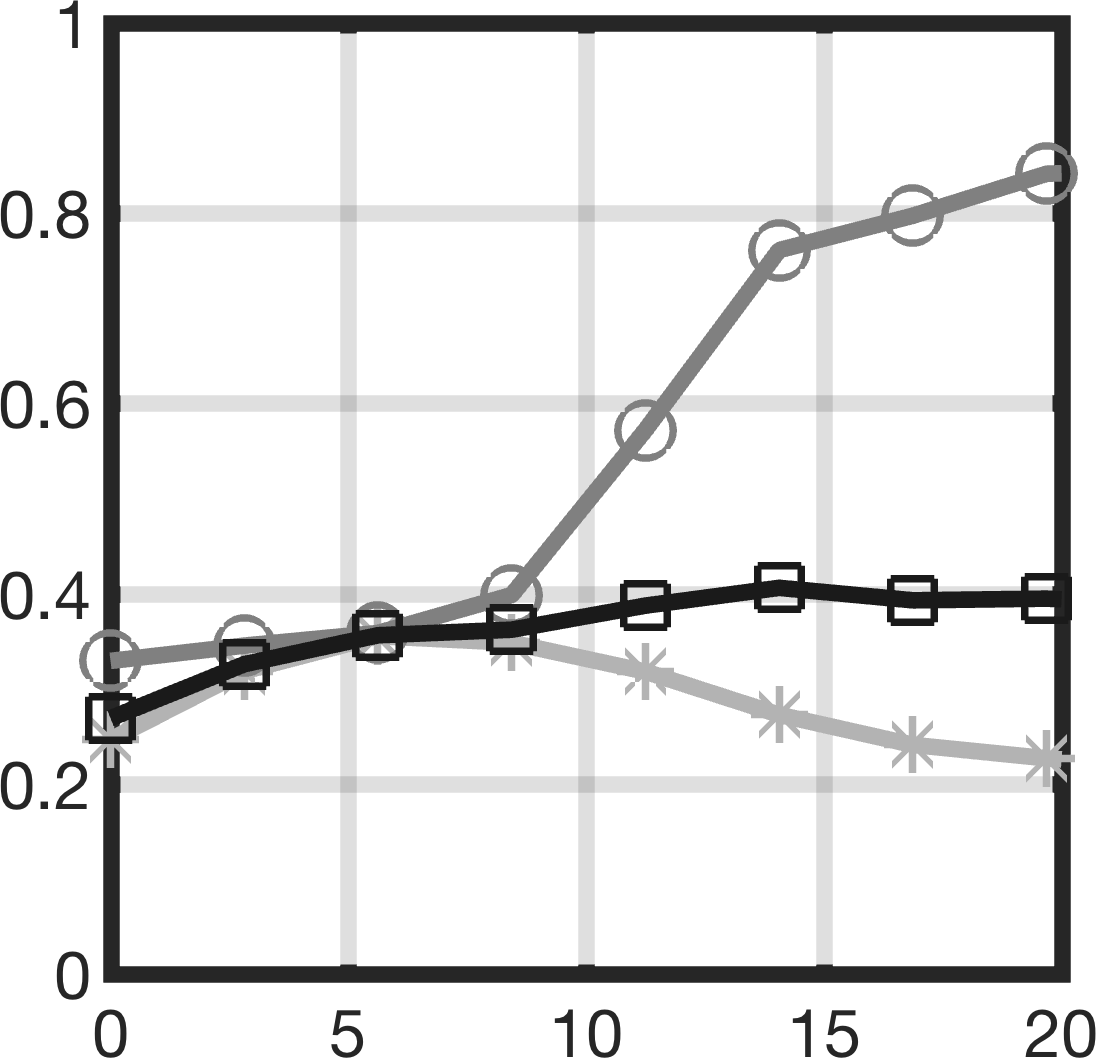} \\
x: ROE, y: $\theta$  (deg)
\end{center}
\end{minipage}
\\ 
\begin{minipage}{2.8cm}
\begin{center}
 \includegraphics[width=2.8cm]{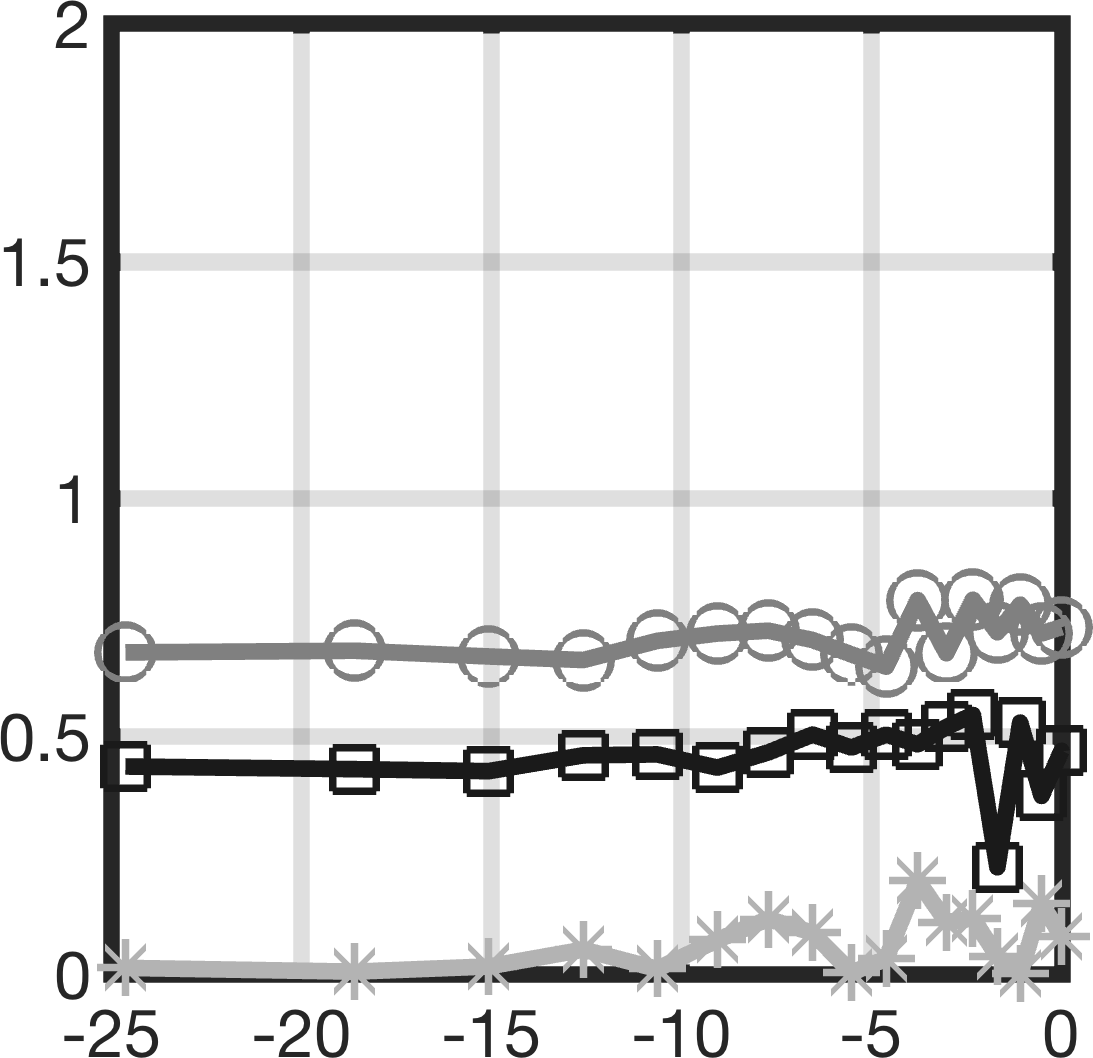}  \\
x: $\eta$ (dB), y: RVE
\end{center}
\end{minipage}
\begin{minipage}{2.8cm}
\begin{center}
\includegraphics[width=2.8cm]{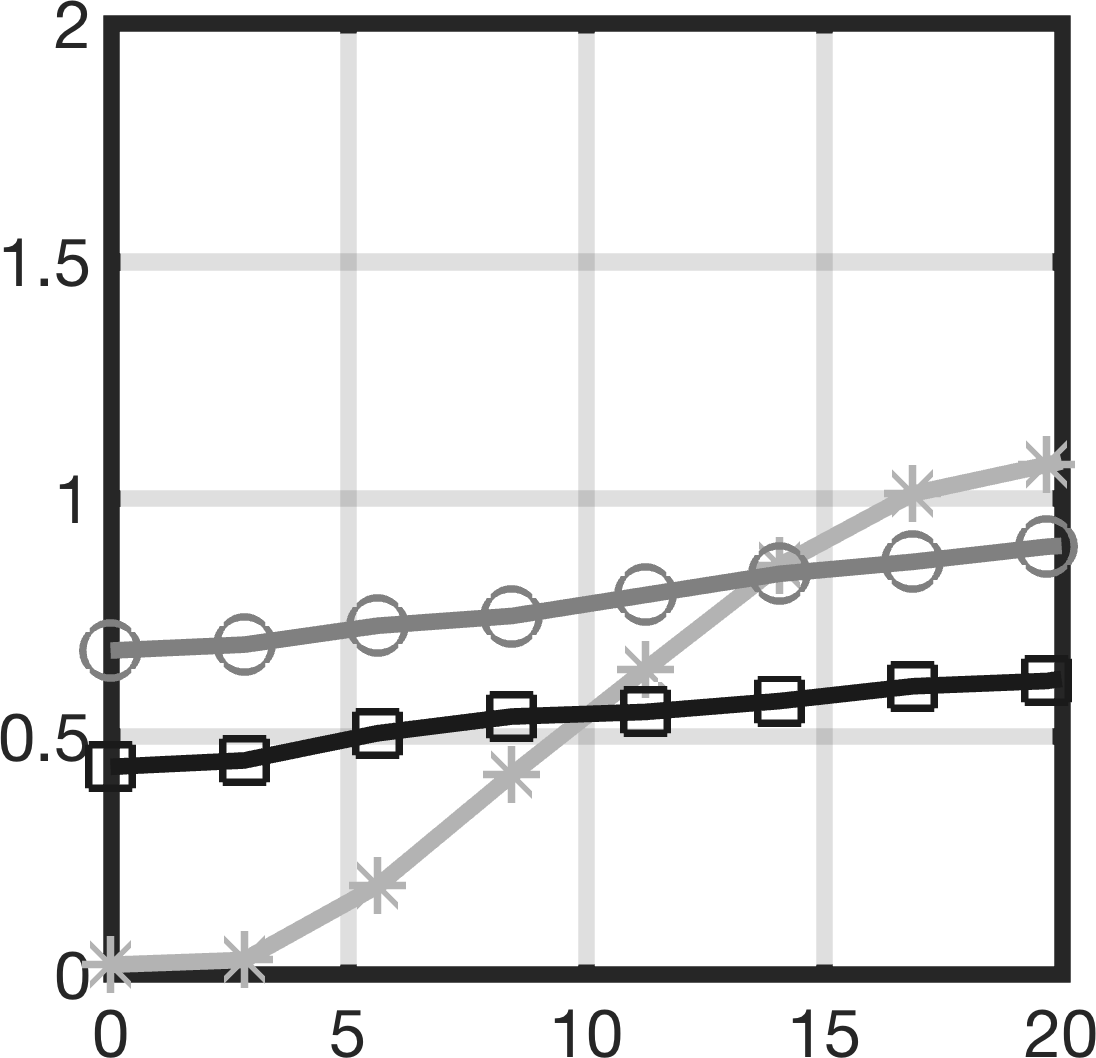} \\ 
x: RVE, y: $\theta$  (deg)
\end{center}
\end{minipage}
\end{framed}
\end{center}
\end{minipage} \\
\vskip0.1cm
\begin{minipage}{6.4cm}
\begin{center}
\begin{framed}
Projected data, Single resolution \\
\begin{minipage}{2.8cm}
\begin{center}
\includegraphics[width=2.8cm]{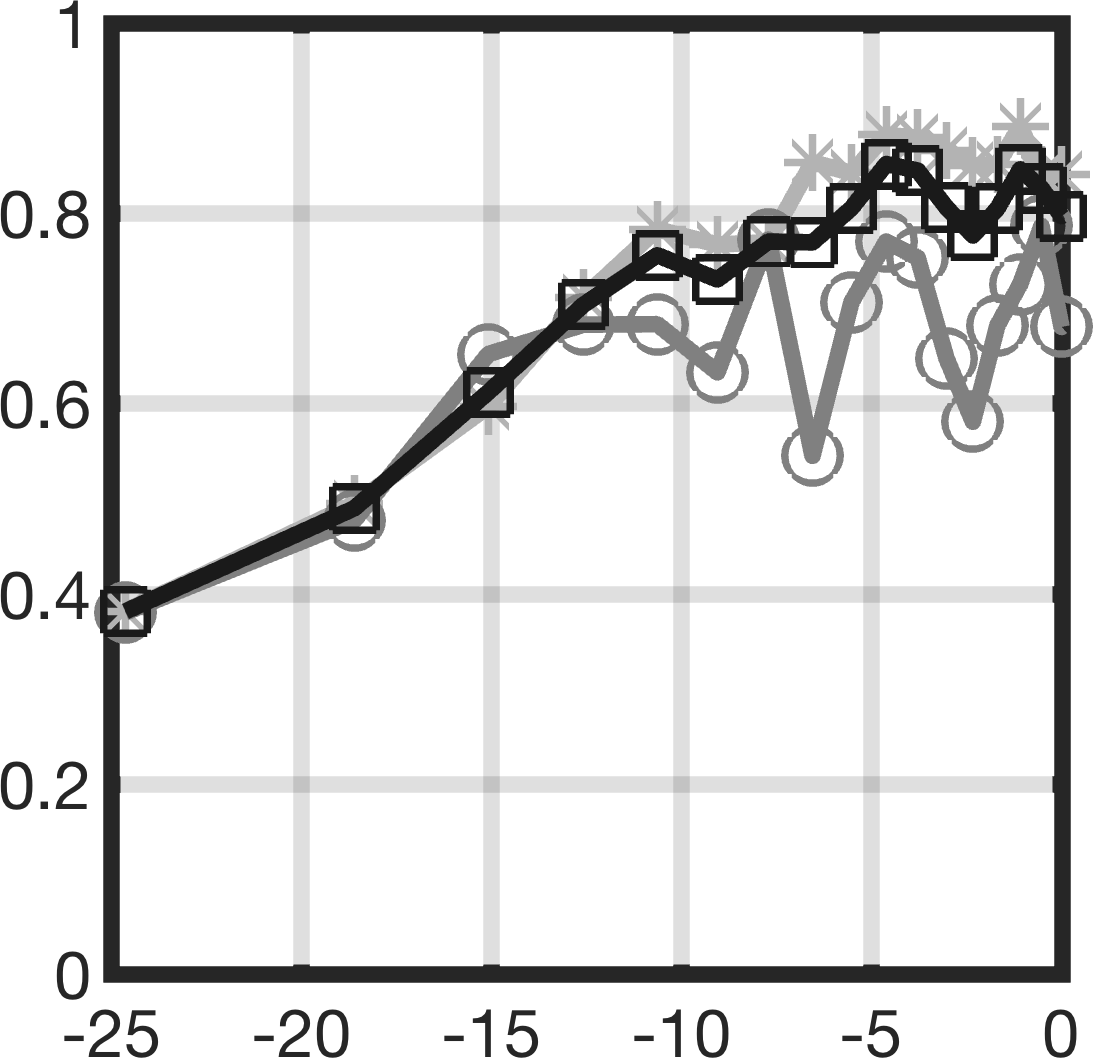}  \\
x: $\eta$ (dB), y: ROE
\end{center}
\end{minipage}
\begin{minipage}{2.8cm}
\begin{center}
\includegraphics[width=2.8cm]{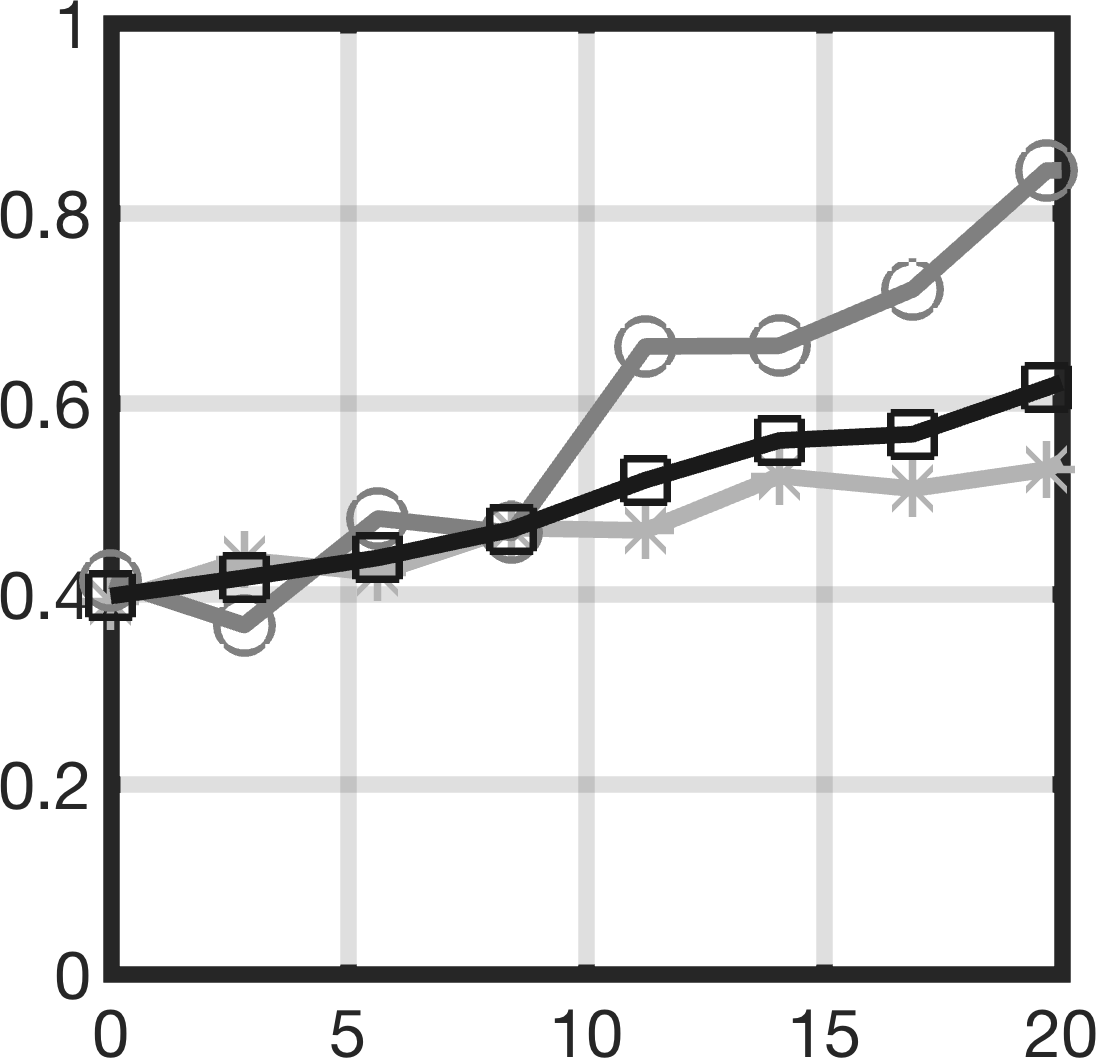} \\
x: ROE, y: $\theta$  (deg)
\end{center}
\end{minipage}
\\
\begin{minipage}{2.8cm}
\begin{center}
 \includegraphics[width=2.8cm]{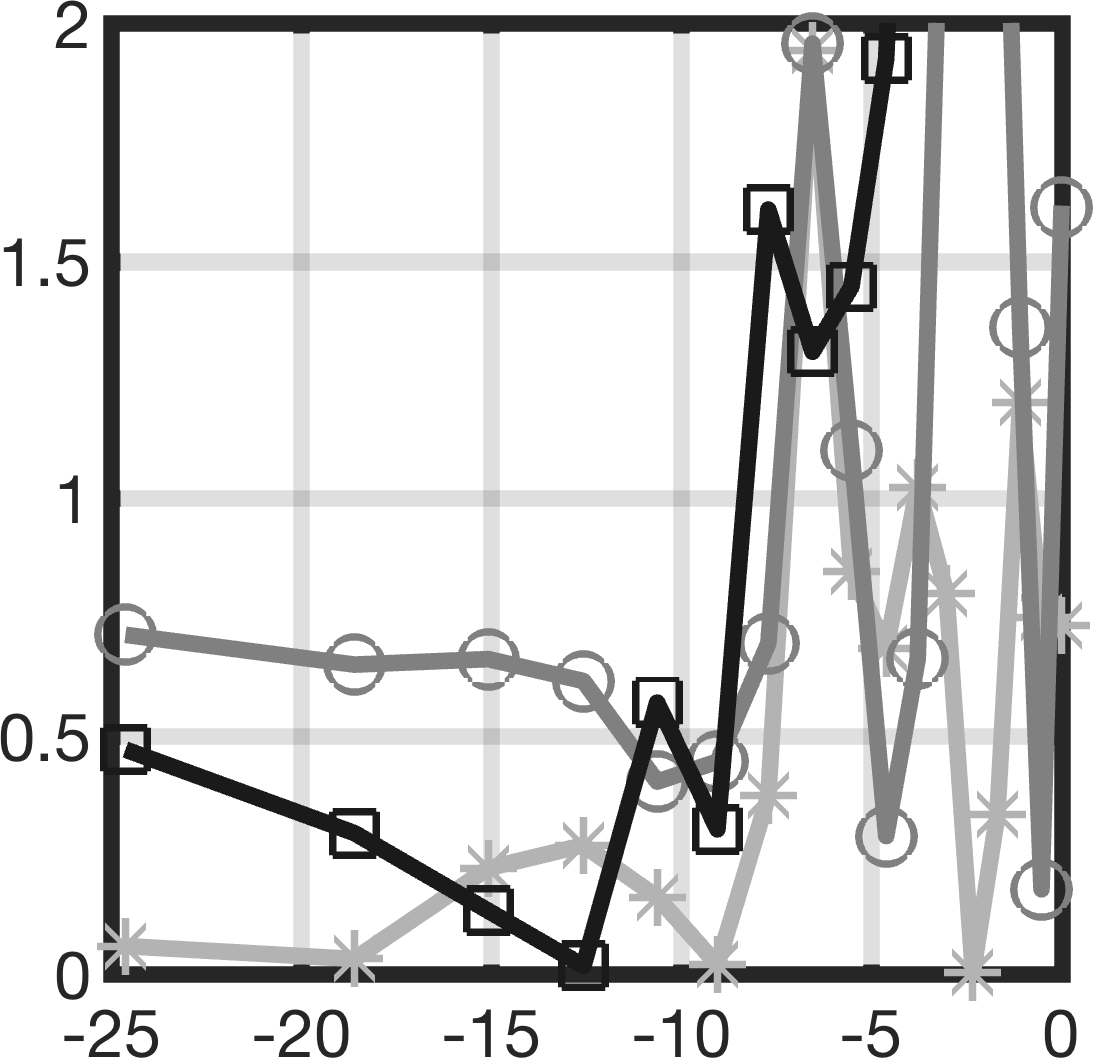}  \\
x: $\eta$ (dB), y: RVE
\end{center}
\end{minipage}
\begin{minipage}{2.8cm}
\begin{center}
\includegraphics[width=2.8cm]{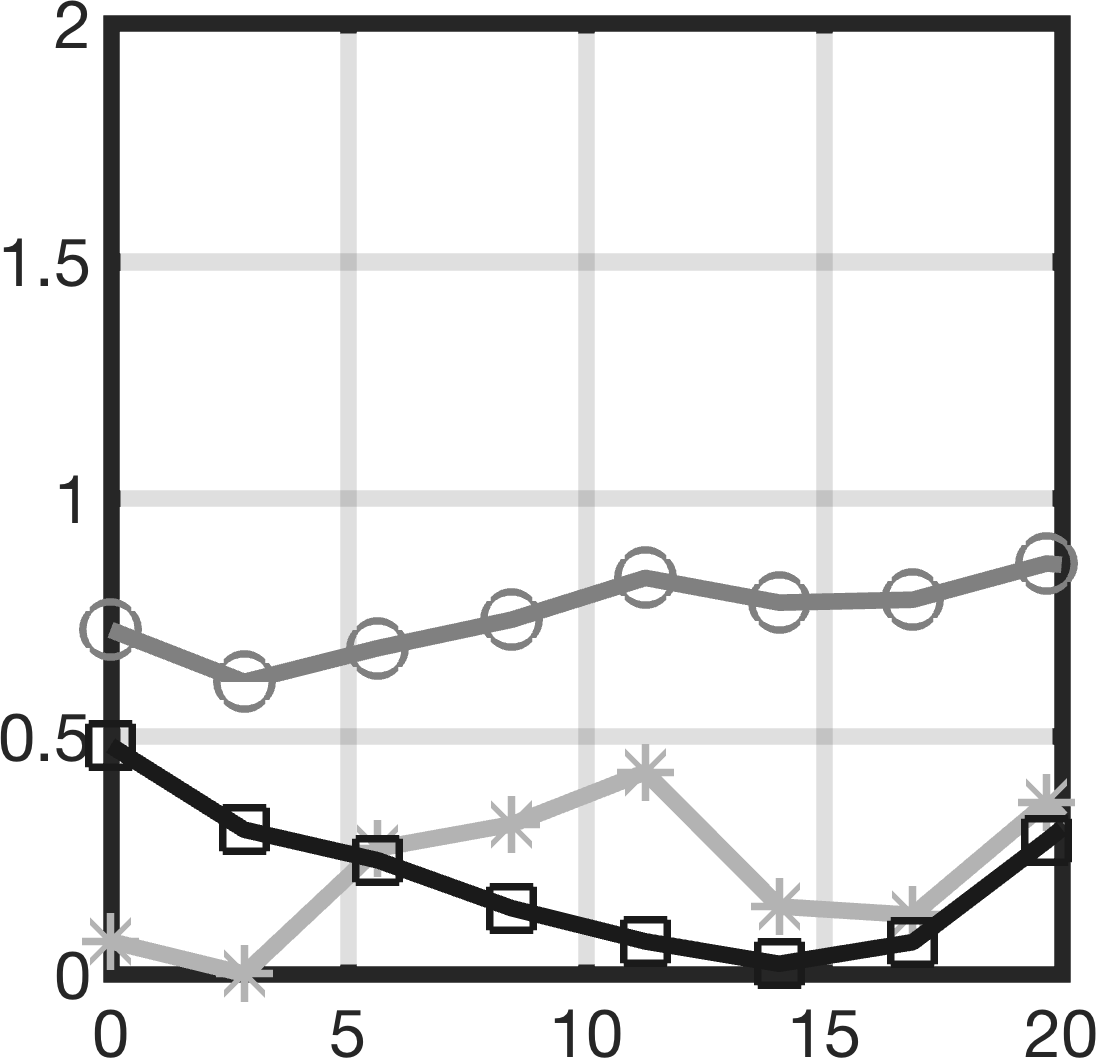} \\ 
x: RVE, y: $\theta$  (deg)
\end{center}
\end{minipage}
\end{framed}
\end{center}
\end{minipage}
\hskip0.1cm
\begin{minipage}{6.4cm}
\begin{center}
\begin{framed}
Projected data, Dual resolution \\
\begin{minipage}{2.8cm}
\begin{center}
\includegraphics[width=2.8cm]{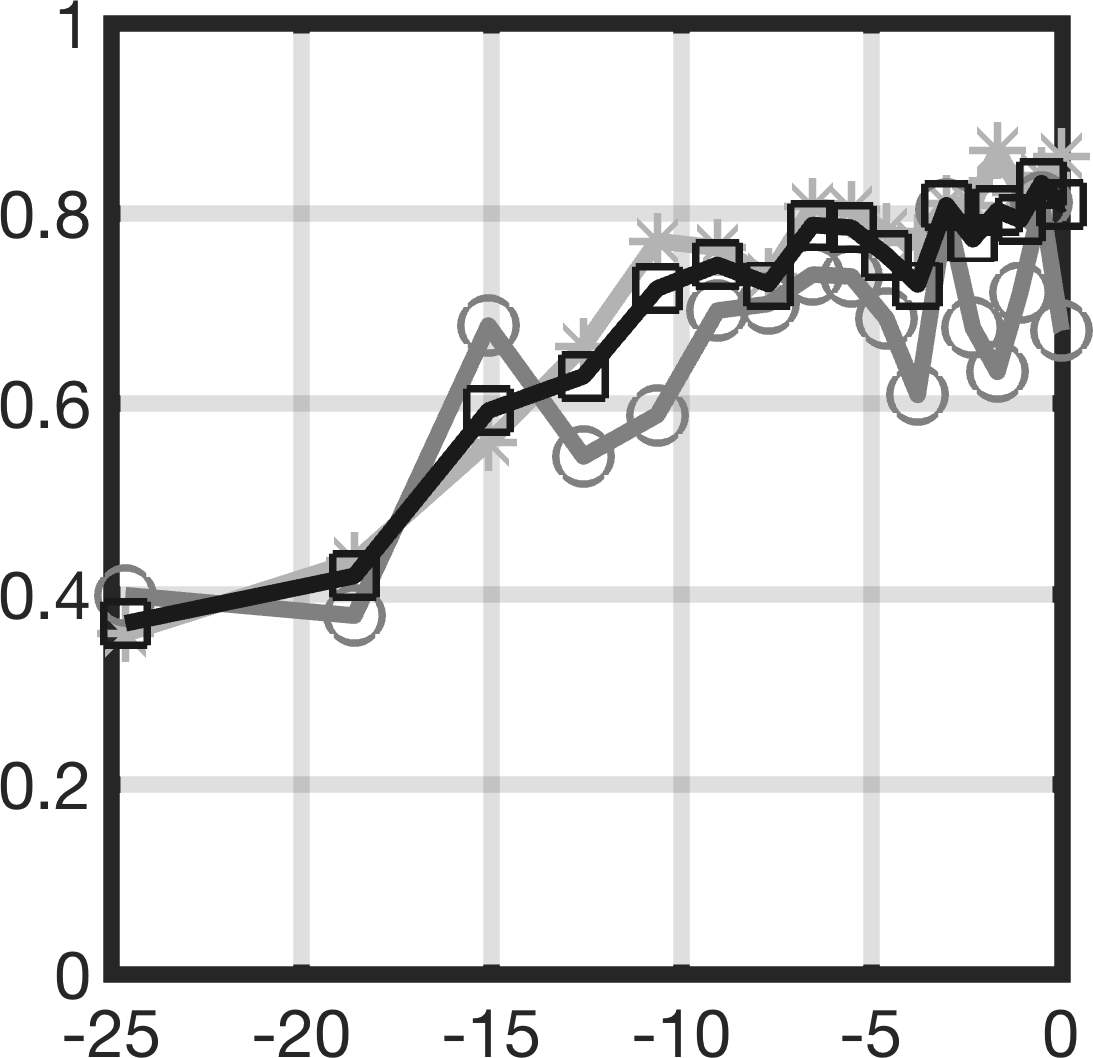}  \\
x: $\eta$ (dB), y: ROE
\end{center}
\end{minipage}
\begin{minipage}{2.8cm}
\begin{center}
\includegraphics[width=2.8cm]{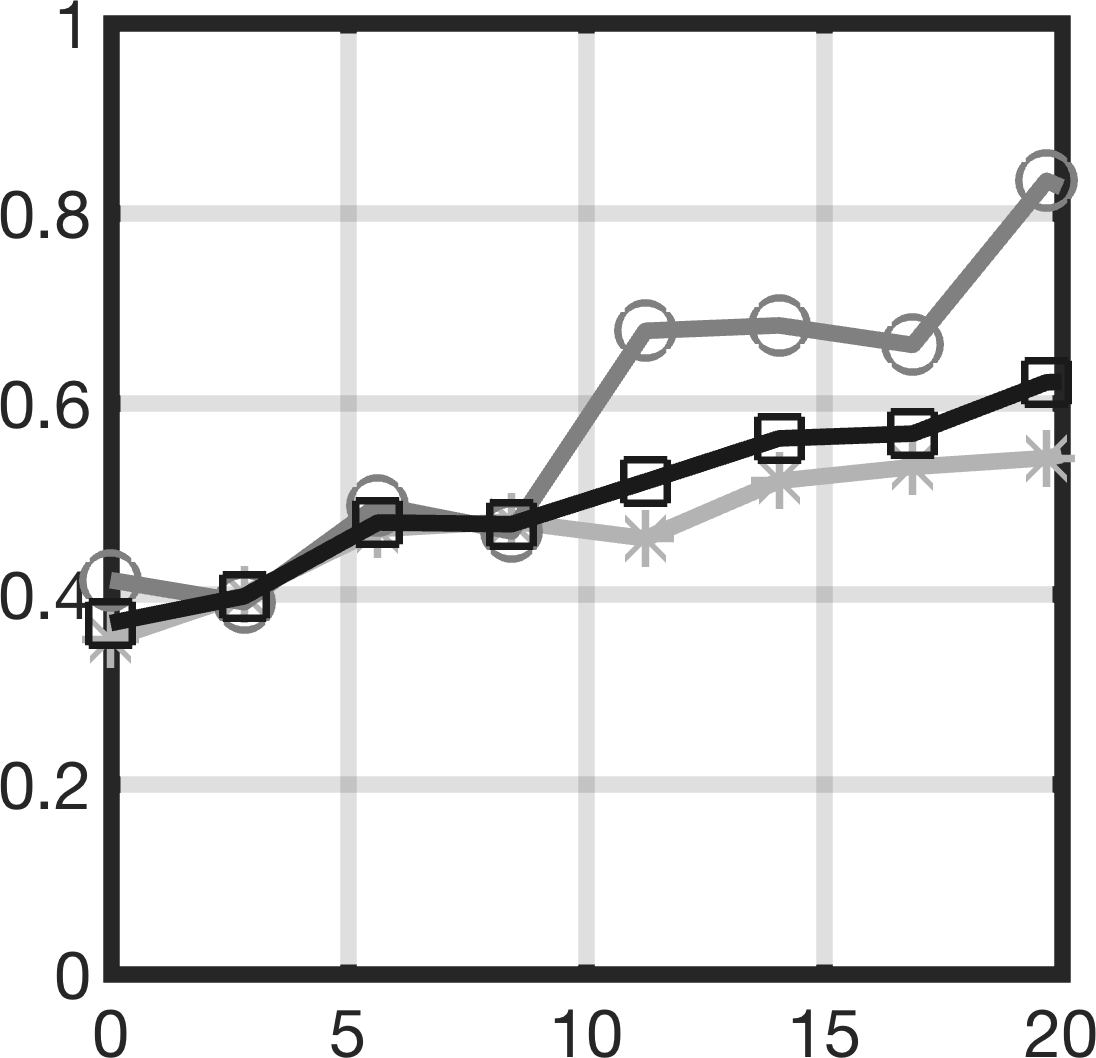} \\
x: ROE, y: $\theta$  (deg)
\end{center}
\end{minipage}
\\
\begin{minipage}{2.8cm}
\begin{center}
 \includegraphics[width=2.8cm]{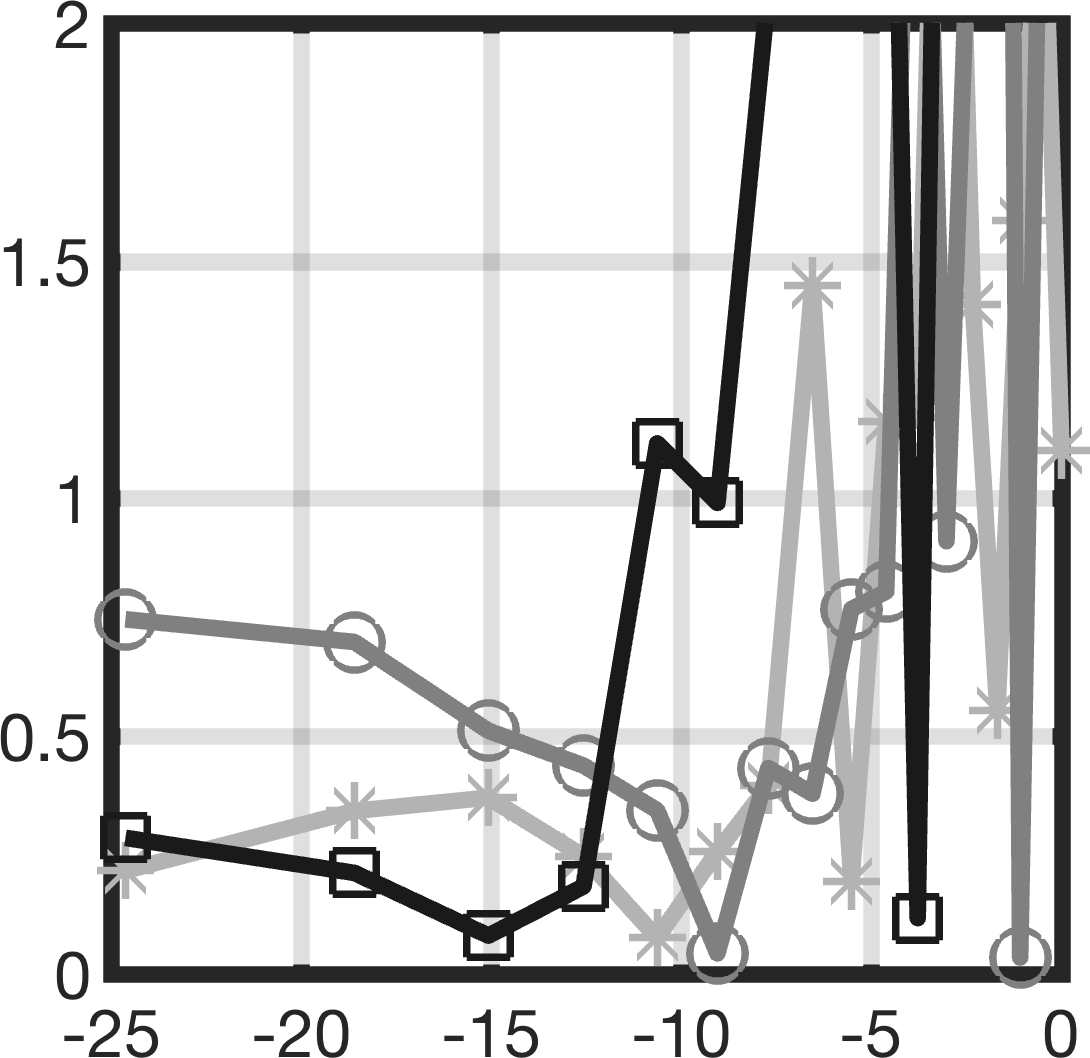}  \\
x: $\eta$ (dB), y: RVE
\end{center}
\end{minipage}
\begin{minipage}{2.8cm}
\begin{center}
\includegraphics[width=2.8cm]{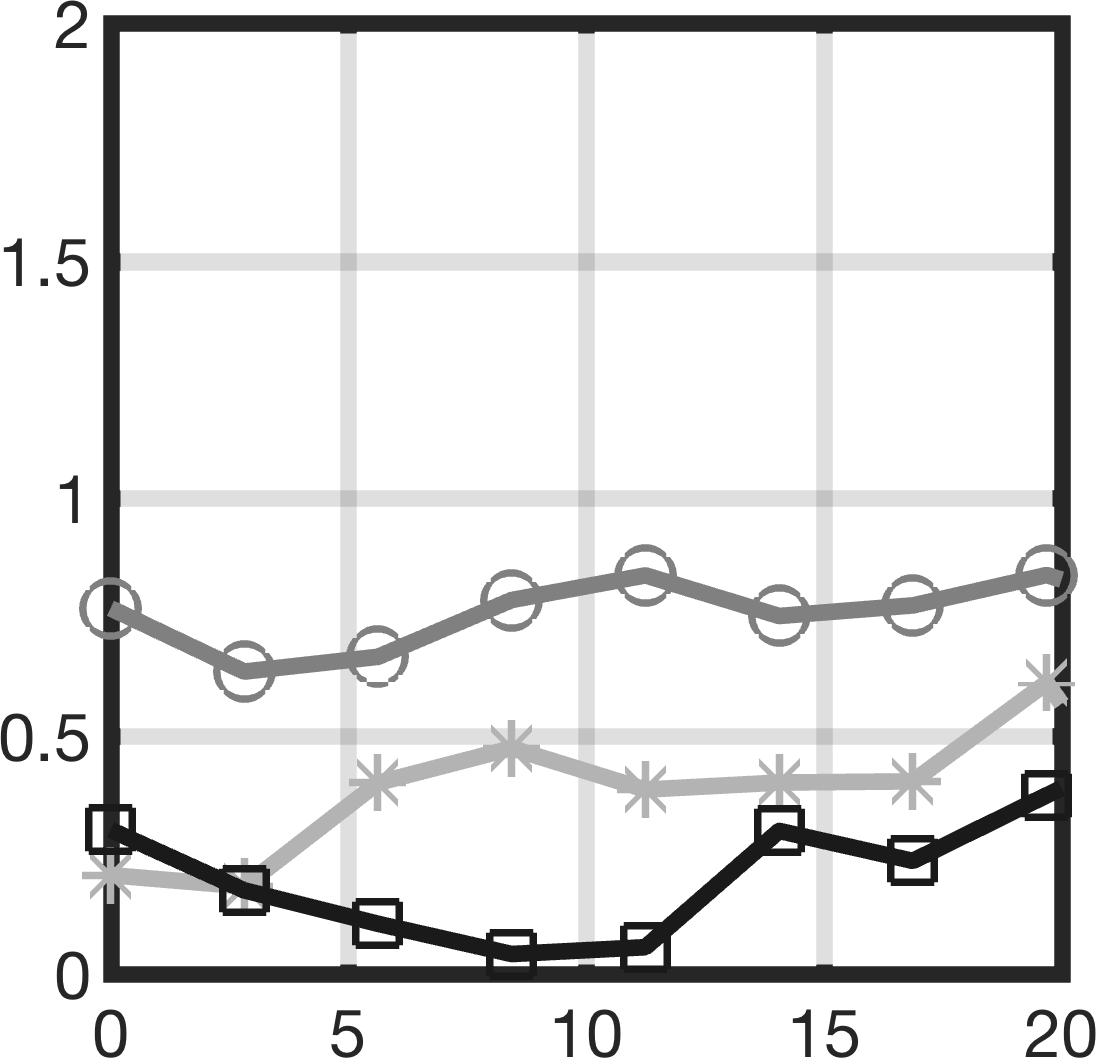} \\ 
x: RVE, y: $\theta$  (deg)
\end{center}
\end{minipage}
\end{framed}
\end{center}
\end{minipage}
\end{center}
\caption{ ROE and RVE  results for the signal configuration ({\bf C}). {\bf Surface layer}: light grey line, star marker,  {\bf Voids}: darker grey line, circle marker, {\bf Total value}: black line, square marker. \label{c_results}}
\end{scriptsize}
\end{figure*}

The square $\Omega = [-1,1] \times [-1,1]$ was utilized as the test domain of the numerical experiments. The parameters  $t,{\vec x}$, $\varepsilon_r$, $\sigma$ and ${\mathsf c}= \varepsilon_r^{-1/2}$ (velocity)  of the governing equation (\ref{pde1}) can be scaled to SI-units, respectively, via the expressions $(\mu_0 \varepsilon_0)^{1/2} s t  $, $s {\vec x}$, $\varepsilon_0 \varepsilon_r$, $(\varepsilon_0 / \mu_0)^{1/2} s^{-1} \sigma$, and  $(\varepsilon_0  \mu_0)^{-1/2} {\mathsf c}$ (${\mathsf c} = 1$ for $\varepsilon_r = 1$), where $\varepsilon_0 = 8.85 \cdot 10^{-12}$ F/m, $\mu_0 = 4 \pi \cdot 10^{-7}$ b/m and $s$ is a spatial scaling factor (meters).   

The center part of the domain included a tomography target $\mathcal{D}$  (Figure \ref{test_domain}) with an irregular shape and approximate diameter of $d \approx 0.28$.  The interior of $\mathcal{D}$ included a surface layer with the thickness around 0.02 and three inclusions (voids) with maximum diameter of  0.01--0.09. The relative permittivity values were chosen to be 4, 3 and 1 for the interior part, cover layer and voids, respectively.   Outside $\mathcal{D}$, the relative permittivity was set to be one, i.e., that of air or vacuum. In $\mathcal{D}$, the conductivity causing a signal energy loss was assumed to be a nuisance parameter of the form  $\sigma = 5 \varepsilon_r$.  The remaining part  of $\Omega$ was assumed to be lossless, i.e.\ $\sigma = 0$.  The  parameter values  for three alternative choices of $s$ ({\bf I})--({\bf III})   have been given in Table \ref{scaling_values_domain}. For ({\bf I}),  the  minimal radar bandwidth needed for lossless signal pulse transmission is 10 MHz, coninciding the bandwidth of the CONSERT instrument \cite{kofman2015}; for ({\bf II}), it is 4 MHz, which we consider more relevant regarding small spacecraft applications; and for  ({\bf III}),  it is 10  GHz, i.e., within range of  microwave tomography applications \cite{bourqui2012}.

The signal was assumed to have been transmitted and received  at 0.32 diameter circular orbit $\mathcal{C}$ centered at the origin using the Blackman-Harris window \cite{irving2006,harris1978,nuttall1981} as the source function 
$\tilde{f}(t)  =  0.359 -   0.488 \cos \left ( {20 \pi t} \right) \nonumber   +   0.141 \cos \left ( {40 \pi t} \right)  - 0.012 \cos \left ( {60 \pi t}\right)$ for $t \leq 0.1$ (0.67 ns), and $\tilde{f}(t) = 0$, otherwise. The length of the temporal interval was chosen to be  $T=1.3$. The time interval between each data sampling point was set to be $0.01$ corresponding to a 1.25 oversampling rate relative to the Nyquist criterion, i.e.,  the density of sampling points was $2.5 = 2 \times 1.25$ times the highest frequency in the signal pulse. The  signal specifications scaled according to ({\bf I})--({\bf III}) have been included  in Table \ref{scaling_values_signal}.  

Three different spatial signal configurations ({\bf A})--({\bf C}) (Table \ref{signal_configurations})  were tested.  Configurations ({\bf A}) and ({\bf B}) were monostatic. They modeled a situation in which a single orbiter moving along $\mathcal{C}$  records  backscattering data. In ({\bf A}) a dense distribution of 128 measurement points was used corresponding to oversampling by a factor of 1.6  with respect to the Nyquist criterion at $\mathcal{C}$. Configuration ({\bf B}) included 32 points, i.e.,  undersampling by the factor 0.4.  Configuration ({\bf C}) was otherwise similar to ({\bf B}) but bistatic: two orbiters transmitted the signal pulse. The velocity of the second orbiter was assumed to be ten times that of the first one, leading to mixing of the line segments between the orbiters. The orbiter-to-orbiter signal paths for ({\bf C}) have been illustrated in Figure \ref{test_domain}.   

A simulated data vector ${\bf y}$ was produced with a finite element mesh of 97517 nodes and 194232 triangular elements. The Jacobian matrix and  the data estimate ${\bf y}_0$ were obtained using a different mesh ${\mathcal{T}}$ (99121 nodes, 197440 triangles) in order to avoid overly good inverse estimates \cite{kaipio2004}. In both cases, the temporal increment of the leap-frog iteration was set to be $\Delta t = 2.5 \cdot 10^{-4}$. In the mesh generation process, an initial mesh ${\mathcal{T}''}$ was generated first. After that, ${\mathcal{T}'}$ and finally ${\mathcal{T}}$ were obtained by using the regular mesh refinement principle recursively, i.e., each element edge was split into two equivalent halves.  

Gaussian noise was added to the data. At each measurement point $\vec{p}_i$, the standard deviation $\nu$ of the noise was set to be of the form  $ 
\nu = 1.645 \cdot 10^{{\eta}/{10}} \max_t u( t, \vec{p}_i)$, where $\eta$ is the desired total noise level in decibels. The absolute value of $\eta$ was regarded as an estimate for the total peak-to-peak signal to noise ratio. Due to the scaling, approximately 5\% of the noise vector entries exceeded the decibel value $\eta$ with respect to the signal peak $\max_t u( t, \vec{p}_i)$. This noise model was  motivated by the noise estimates of \cite{kofman2015} in which the peak level is used as the reference. Besides the additive noise, the sensitivity of the inversion to positioning inaccuracies was investigated by rotating the measurement points by (polar) angle $\theta$ in between the data simulation and reconstruction phases (Figure \ref{test_domain}).  

The reconstructions were computed via (1) single and (2) dual scale processing. The number of nested resolution levels was limited to two based on the size of the permittivity fluctuations which was close to the element size in $\mathcal{T}''$. In (1), the reconstruction was found using the mesh ${\mathcal{T}'}$  and, in (2), both ${\mathcal{T}'}$  and  ${\mathcal{T}''}$ were used. The {\em a priori} guess ${\bf x}_0$ for the inversion procedure corresponded to a constant $\varepsilon_r = 4$ within $\mathcal{D}$.  In the coarse-to-fine iteration, three steps were performed with $\alpha = \beta = 0.01$ for the full data and $\alpha = 10^{-7}$ and $\beta = 1$ for the projected data. These parameter values were approximately  in the center of the range of working values based on preliminary numerical tests. 

The accuracy of the inversion was examined through the relative overlap and value error (ROE and RVE), i.e., the percentages \begin{equation} \hbox{ROE} = 100 \Big( 1 - \frac{\hbox{Area}( \mathrm{A}) }{ \hbox{Area}(\mathrm{S})} \Big) \quad \hbox{RVE} = 100 \Big( 1 -   \frac{\int_{\mathrm{S}}  \varepsilon^\ast_r \, d \mathrm{S}}{ \int_{\mathrm{S}}  \varepsilon_r  \, d \mathrm{S}  }\Big).  \end{equation} Here   $\varepsilon_r$ and   $\varepsilon^\ast_r$ denote the actual and estimated permittivity, respectively, and  $\mathrm{A} = \mathrm{S} \cap \mathrm{R}$ is the overlap between the set $\mathrm{S}$ to be recovered and the set $\mathrm{R}$ in which a given reconstruction is smaller than a limit such that $\hbox{Area}(\mathrm{R}) = \hbox{Area}(\mathrm{S})$. ROE and RVE were measured with respect to the noise level $\eta$ and the angular positioning error $\theta$.

The computations were performed using the Matlab R2017a software and a laptop equipped with a 2,2 GHz Intel Core i7 CPU and 8 GB of 1600 MHz DDR3 RAM. The single thread mode of the CPU was used, i.e., all the processes were run in a single CPU core.

\section{Results}

The results of the numerical experiments have been included in Figures \ref{residual_norm_fig}--\ref{c_results} as well as Tables \ref{mse_table} and \ref{matrix_sizes}. Figure \ref{residual_norm_fig} shows the convergence of inverse iteration as a function of iteration steps. Figure  \ref{reconstructions_comparison}  visualizes  the reconstructions and the overlap sets for configuration ({\bf A}) at noise levels $\eta = -25$ dB and $\eta = -11$ dB.  The minimum squared error \cite{lehmann2006} values for those reconstructions can be found in Table \ref{mse_table}. Also the MSE of the coarse level estimate obtained with the dual resolution scheme has been evaluated.  The coarse and fine level fluctuations of the dual scale reconstruction have been visualized in Figure \ref{reconstructions_levels}.  Those together with the MSE values show that, in the low noise case ($\eta = -25$ dB), the final reconstruction was more accurate than the coarse level estimate. The coarse one was slightly superior, when the higher noise ($\eta = -11$ dB) and full data were used. Based on the MSEs, it is also obvious that the single scale  approach is more sensitive to noise than the multigrid technique. Table \ref{matrix_sizes} includes the vector and matrix sizes as well as the computation times required by the inversion processes, showing that the proposed inversion strategy was, in general, advantageous regarding both the speed and the memory-efficiency of the inversion. The convergence curves for the relative residual norm have been included in Figure \ref{matrix_sizes}.

Figures \ref{a_results}--\ref{c_results} illustrate  the  sensitivity of the reconstructions to the total noise level $\eta$ and to the angular measurement error $\theta$ for Configurations ({\bf A})--({\bf C}).  The dual resolution approach was found to be slightly superior regarding the total ROE and RVE (the black line in Figures \ref{a_results} -- \ref{c_results}), when full wave data were used. With projected data such a difference was not observed. The reconstructions were, in general, more accurate for the surface layer than for the deep part of the interior (voids) with respect to both ROE and RVE. 

The results (Figure \ref{b_results}) obtained with the sparse signal configuration ({\bf B})  were close in accuracy to those (Figure \ref{a_results}) obtained with the dense one ({\bf A}).  The foremost accuracy for the deep interior part  of $\mathcal{D}$ (voids) was achieved  with the sparse two-orbiter configuration ({\bf C}). It seemed to drive the balance of the reconstruction towards the deep interior, as  ({\bf B}) led to superior results with respect to the surface layer.  

The noise level $\eta$ was observed to affect the reconstruction quality above -15 dB level. When full wave data was used, the dual resolution reconstructions were more tolerant to the noise than those obtained with the single resolution approach: the total ROE and RVE were, in general, less volatile and stayed on a lower level between -15 dB and -10 dB noise in the former case. The clearest difference with respect to the volatility was observed for the sparse configuration ({\bf B}).  For projected data, the volatility of RVE  was extreme  above the -15 dB level with both single and dual resolution approaches. An angular  measurement error $\theta$ between 0 and 5 degrees was observed to be small regarding the inversion results. An error larger than 5 degrees was led to clearly increased inversion errors in all the analyzed cases. Those seemed to be roughly comparable to above -10 dB noise with respect to the inverse estimates.  Of the signal configurations ({\bf A})--({\bf C}), the most robust one with respect to noise and angular errors was ({\bf C}) in which the number of signal sources is two.

\section{Discussion}

This paper concentrated on tomographic multigrid-based \cite{braess2007} inversion of waveform electromagnetic signals within a nested finite element mesh structure. Waveform imaging  has a wide range of applications varying from biomedical tomography  \cite{grzegorczyk2012,meaney2010,fear2002, ruiter2012, opielinski2013, ranger2012} and non-destructive material testing \cite{yoo2003,chai2010,chai2011,acciani2008} to astro/geophysical  applications. Here, as the foremost application, we considered reconstructing the interior structure of an asteroid \cite{pursiainen2016, herique2016, su2016, kofman2015, kofman2007} using small orbiting spacecraft. Due to the strict  limitations of space missions, e.g., the payload weight, it is likely that a sparse set of waveform measurements performed by one or two small satellites will have to be used as data in a real planetary space mission. 

A multigrid inverse approach utilizing the total variation regularization  \cite{clarkson1933,scherzer2008,stefan2008,kaipio2004} was proposed and tested. The primary goal in using more than one scale was to enhance the noise-robustness of the inversion process for sparse data compared to the single level case. The secondary goal was to reduce the system matrix size and the time needed for the inversion process via compression of the fine resolution level coefficients. The proposed technique can be interpreted as an extension of the classical total variation method to multiple resolution levels. It was chosen as it allows reconstructing and regularizing an arbitrary distribution in a similar manner at each level under scarce {\em a priori} information.  In the numerical  experiments, the single and dual resolution approaches were compared, three different signal configurations ({\bf A}), ({\bf B}) and ({\bf C}) including dense and sparse measurements were investigated, and both full wave and projected data were used.  

\subsection{Results}

Reconstructing the interior structure of the target $\mathcal{D}$ was found to be feasible with all the tested inversion approaches,  when the noise level $\eta$ and angular error $\theta$ were below $-15$ dB and 5 degrees, respectively. In the case of projected data, these were found to be the approximate upper limits for workable inversion. With full wave data, appropriate inverse estimates were obtained  also above these levels. The results suggest that the proposed dual resolution reconstruction technique is advantageous regarding the total ROE and RVE. In particular, it improves the noise tolerance (robustness) between noise levels -15 dB and -10 dB, especially, for sparse measurements. Of the tested signal configurations, the foremost tolerance for measurement errors was achieved with {({\bf C})}, i.e., with two signal sources (orbiters). 

\subsection{Application areas}

\subsubsection{Radio tomography}

The results obtained are promising from the viewpoint of future planetary missions. Namely, it was documented for CONSERT experiment that the noise peak level measured for the 90 MHz radio signal transmitted through the nucleus of the comet 67P Churyumov-Gerasimenko did not significantly exceed the  -20 dB level \cite{kofman2015}. Since full wave inversion was found to be feasible without severe artifacts at least up to -15 dB noise peak level, it seems a potential approach for a real mission, strengthening our recent findings \cite{pursiainen2016}.  Moreover, the reconstructions obtained can also be considered robust with respect to a couple of degrees angular positioning errors, which will be likely in a real mission. 

An additive Gaussian white noise model was utilized. The standard deviation was chosen so that  5\%  of the generated noise entries exceeded  the decibel value $\eta$ which was used as an estimate for the noise peak level. The motivation to use a generous Gaussian white noise model was the lack of {\em a priori} knowledge related to the true modeling errors and the resulting noise peaks. Thus, the noise level approximates roughly the total error magnitude. Further  analysis of the CONSERT data \cite{kofman2015} might help to develop a more sophisticated noise model for future studies. Whether the total variation regularization method is optimal for a real noise distribution will also need to be studied in the future. 

The comparison between the sparse signal configurations ({\bf B}) and ({\bf C}) suggests that the robustness of the inversion will improve, if in addition to the backscattering data, through going (orbiter-to-orbiter)  signals \cite{pursiainen2016} can be measured.  Hence, the most reliable results may be expected, if more than one satellite can be used. In principle, this should be possible as the preliminary plans of the AIM mission \cite{michel2016} already included two small satellites and bistatic radio tomography measurements \cite{herique2016}. 

The classical low-frequency georadar studies \cite{daniels2004} apply a single dipole antenna rod coinciding the model  presented in this paper. For this reason and for the simplicity of the configuration, we suggest that a satellite pair utilizing such  antennas might be sufficient for reconstructing the interior. A  double or triple dipole might increase the reliability of the measurements, since in that case, the signal amplitude could be recorded regardless of the polarization. Equipping a spacecraft, e.g., a small satellite, with a multidirectional antenna might, however, be more difficult and expensive.  An antenna attached to a small spacecraft needs to be a relatively simple and light one.  We suggest here a  half-wavelength dipole design with center frequency between 20--100 MHz (length 1.5--7.5 m), that is, a similar or somewhat lower value compared to the 90 MHz of CONSERT instrument \cite{kofman2007,kofman2015} in order to achieve an appropriate signal penetration. We see that emphasizing the signal penetration capability is important, since the relatively high permittivity  and the unknown macroporosity structures of asteroid interiors can cause significant signal attenuation via scattering effects.  Due to the high permittivity, the 4 MHz bandwidth corresponding to the scaling option ({\bf II}) would yield a range resolution, i.e., the signal velocity divided by two times the bandwidth, close to that of  CONSERT (10 MHz bandwidth). 

\subsubsection{Microwave tomography}

In addition to radio tomography, the present multigrid methodology is a potential approach also in other fields of waveform  tomographic reconstruction, for example, ultrasonic and microwave tomography (MWT). At the 10 GHz microwave frequency,  the permittivity and conductivity of the current target domain has a correspondence to fatty tissues \cite{pethig1987} and wood (balsam fir)   \cite{bucur2013} both of which can be inspected via microwave imaging. 
Similar parameters close to the target's permittivity and
conductivity at 10 GHz microwave frequency are fat ($\sigma=0.58$
S/m, $\varepsilon_r=4.6$), breast fat ($\sigma=0.74$ S/m,
$\varepsilon_r=3.88$) and bone marrow ($\sigma=0.57$ S/m,
$\varepsilon_r=4.6$) as no other part in the body has conductivity
lower than 1 S/m and relative permittivity lower than 10
\cite{CNR2017}. 

MWT is useful in biomedical applications when there is high dielectric contrast between the
structures, such as bone, soft tissue and fatty tissue. In
biological structures, proportional values to the surface/target
permittivity ratio given in Table \ref{scaling_values_domain} can
be found in certain areas (not necessarily matching conductivity
values). For example, in the treatment of fractured bone in
extremities, where it is important to be able to differentiate body
and soft-tissue elements such as muscle, vessels, nerves and skin,
MWT would provide an assessment on the composition of these
elements in case of injury \cite{semenov2009}.
As far as the surface/target permittivity ratio is concerned, MWT
could also be used in the detection of blood flow perfusion, which
would be helpful in also cardiac and brain imaging. In MWT of a
human body torso, the relative permittivity and conductivity
changes between different volume compartments at 10 GHz frequency match roughly
with those of the current target domain, including skin ($\sigma=8$
S/m, $\varepsilon_r=31$), blood ($\sigma=13$ S/m,
$\varepsilon_r=45$), heart ($\sigma=11.8$ S/m, $\varepsilon_r=42$),
muscle ($\sigma=10.6$ S/m, $\varepsilon_r=42.7$), deflated lung
($\sigma=10.1$ S/m, $\varepsilon_r=38$), and air ($\sigma=0$ S/m,
$\varepsilon_r=1$) \cite{CNR2017}.  

\subsection{Outlook}

In the future work, the proposed multigrid approach might be developed  and compared to other recent noise-robust approaches  \cite{wen2015,gu2014}. The present forward approach might, for example, be utilized in combination with an alternative inversion  strategy.  An important future direction regarding radio tomography would be to use a realistic 3D rubble pile asteroid model \cite{deller2015, deller2016}, e.g., to optimize the number of resolution levels and the amount of compression. 

In general, further numerical analysis will be necessary regarding, e.g., polarization effects which were not modeled in this study. Also the antenna design is a topic which needs to be addessed in the future research \cite{nielsen2001}.  Finally,  the present methodology might be validated using microwave data. This seems a promising topic, as for example, the recent research in microwave breast imaging supports the noise limits discovered in this paper; In \cite{zeng2011}, the -15 dB has been suggested as a tolerable amplitude error and -11 dB as a turning point above which the inversion artifacts begin to increase. 

\section*{Acknowledgements}
  
This work was supported by the Academy of Finland Key Project 305055 and the AoF Centre of Excellence in Inverse Problems Research.

\appendix

\subsection{Numerical solution of the wave equation}
\label{section:forward_simulation}

Defining test functions $v  : [0,T] \to \mathcal{V} \subset H^1(\Omega)$ and ${\vec w} : [0,T] \to \mathcal{W} \subset L_2(\Omega)$ with $\mathcal{V} = \hbox{span}\{ \varphi_1,\varphi_2, \ldots, \varphi_n \}$ and $\mathcal{W}=\hbox{span} \{\chi_1, \chi_2, \ldots, \chi_m \}$ the weak form can be written in the Ritz-Galerkin discretized  form \cite{braess2007}, that is, 
{\setlength\arraycolsep{2 pt} \begin{eqnarray}
\label{system1}
\frac{\partial}{\partial t}   {\bf A} {\bf q}^{(k)} - {\bf B}^{(k)}  {\bf  p} + {\bf T}^{(k)} {\bf q}^{(k)}  & = & 0,   \\
\frac{\partial}{\partial t}  {\bf C} {\bf p} + {\bf R} {\bf p}  + {\bf S} {\bf p} + \sum_{k = 1}^\mathrm{d} {{\bf B}^{(k)}}^T {\bf q}^{(k)} & = & {\bf f}, 
\label{system2}
\end{eqnarray}}
with ${\bf p} = (p_1,p_2,\ldots, p_n)$, ${\bf q}^{(k)} = (q^{(k)}_1, q^{(k)}_2, \ldots, q^{(k)}_m)$, ${\bf f} \in \mathbb{R}^{n}$, ${\bf A} \in \mathbb{R}^{m \times m}$,  ${\bf B} \in \mathbb{R}^{m \times n}$, ${\bf C} \in \mathbb{R}^{n  \times n}$, ${\bf S} \in \mathbb{R}^{n  \times n}$, ${\bf T} \in \mathbb{R}^{m  \times m}$. Here,  ${\bf A}$ and ${\bf T}^{(k)} = \zeta^{(k)} {\bf A}$ are diagonal matrices with non-zero entries determined by $A_{i,i}  =  \int_{\mathrm{T}_i}  \,  \hbox{d} \Omega$. The right-hand  side of the second equation is given by $f_{i}  =  \int_\Omega {f} \, \varphi_i  \, \hbox{d} \Omega$.  The matrix ${\bf B}^{(k)}$ is a projection matrix of the form  $B_{i,j}^{(k)}  =  \int_{\mathrm{T}_i} \, \vec{e}_k \cdot \nabla \varphi_j \, \hbox{d} \Omega$, and ${\bf C}$, ${\bf R}$ and ${\bf S}$ are mass matrices weighted by $\varepsilon_r$, $\sigma$ and $\xi$, respectively, as given by  $C_{i,j}  =  \int_\Omega \varepsilon_r \, \varphi_i \varphi_j  \, \hbox{d} \Omega$, 
$R_{i,j} =  \int_\Omega \sigma \, \varphi_i \varphi_j  \, \hbox{d} \Omega$ and 
$S_{i,j}  =  \int_\Omega \xi \, \varphi_i \varphi_j  \, \hbox{d} \Omega$. 
The matrices  ${\bf S}$ and ${\bf T}^{(k)}$ correspond to a split-field perfectly matched layer (PML), i.e.,  the set $\{ {\vec x} \in \Omega \, | \, \varrho_1 \leq \max_k | x_k | \leq \varrho_2 \}$ which  eliminates reflections from the boundary $\partial \Omega$ back to the inner part of $\Omega$ \cite{schneider2016}. For the PML parameters, $\xi({\vec x}) = \varsigma$, if $\varrho_1 \leq \max_k | x_k | \leq \varrho_2$, and $\zeta^{(k)} ({\vec x})= \varsigma$, if $\varrho_1 \leq | x_k |  \leq \varrho_2$, and  $\xi({\vec x}) = \zeta^{(k)}({\vec x}) = 0$, otherwise. 

To discretize the time interval $[0,T]$, we utilize $\Delta t$ spaced regular grid of $N$ time points and the standard difference approximations for the time derivative which  substituted into equations (\ref{system1}) and (\ref{system2}) lead to the the following system 
\begin{eqnarray}
\label{leap-frog1}
{\bf q}^{(k)}_{\ell + \frac{1}{2}} & = & {\bf q}^{(k)}_{\ell - \frac{1}{2}} \! +  \! \Delta t  {\bf  A}^{-1} \Big(     {\bf B}^{(k)}  {\bf  p}_{\ell} \! - \! {\bf T}^{(k)} {\bf q}^{(k)}_{\ell - \frac{1}{2}} \Big),  \\ 
 {\bf p}_{\ell + 1}  & = & {\bf p}_{\ell}  \! + \!  \Delta t {\bf  C}^{-1} \Big( {\bf f}_\ell \! - \! {\bf R} {\bf p}_\ell \! - \!  {\bf S} {\bf p}_\ell \! - \! \sum_{k = 1}^\mathrm{d} {{\bf B}^{(k)}}^T {\bf q}^{(k)}_{\ell + \frac{1}{2}}  \Big), 
\label{leap-frog2}  
\end{eqnarray}
for $\ell = 1, 2, \ldots, N$. Simulating the signal propagation via (\ref{leap-frog1})--(\ref{leap-frog2})  is known as the leap-frog time integration method \cite{schneider2016,bossavit1999,yee1966} .

\subsubsection{Auxiliary system}
\label{section:linearized_forward_simulation}

 In order to obtain the expressions (\ref{approx_1}) and (\ref{approx_2}), we define the following auxiliary source vector  \begin{equation}
\label{aku_ankka}
 {\bf h}_\ell^{(i,j)}    =   \frac{\partial {\bf C}}{\partial c_j} {\bf Q}^{(i)} \, {\bf b}_\ell,    
\end{equation} where ${\bf b}_\ell  =   {\bf C}^{-1} ( {\bf R} {\bf p}_\ell +  {\bf S} {\bf p}_\ell +  \sum_{k = 1}^\mathrm{d} {{\bf B}^{(k)}}^T {\bf q}^{(k)}_{\ell + \frac{1}{2}})$, the matrix ${\bf Q}^{(i)} \in \mathbb{R}^{n \times n}$ has one nonzero entry ${Q}^{(i)}_{i,i} = 1$, and  $( {\partial {\bf C}}/{\partial c_j} )_{i_1,i_2} = \int_{\mathrm{T}'_j} \varphi_{i_1} \varphi_{i_2} \, \hbox{d} \Omega$ is nonzero if the $j$-th element includes nodes $i_1$ and $i_2$. Furthermore, we define the auxiliary system 
{\setlength\arraycolsep{2 pt} \begin{eqnarray}
\label{linearized1}
{\bf r}^{(i, j, k)}_{\ell + \frac{1}{2}} & = & 
{\bf r}^{(i, j, k)}_{ \ell - \frac{1}{2}} +  \Delta t  {\bf  A}^{-1} \Big(  {\bf B}^{(k)}   {\bf d}^{(i, j)}_{\ell}  - {\bf T}^{(k)} {\bf r}^{(i, j, k)}_{\ell - \frac{1}{2}} \Big),  \\ 
{\bf d}^{(i, j)}_{\ell+1} & = & {\bf d}^{(i, j)}_{\ell}+  \Delta t {\bf  C}^{-1} \Big( {\bf h}_\ell^{(i,j)} - {\bf R}  {\bf d}^{(i, j)}_{\ell} - {\bf S} {{\bf d}^{(i,j)}_\ell} - \sum_{k = 1}^\mathrm{d} {{\bf B}^{(k)}}^T {{\bf r}^{(i,j,k)}_{\ell + \frac{1}{2}}} \Big).
\label{linearized2} 
\end{eqnarray}}
This follows from  (\ref{leap-frog1})--(\ref{leap-frog2})  simply by substituting ${\bf h}_\ell^{(i,j)}$  as the source. Due to the sparse structure of $({\partial {\bf C}}/{\partial c_j})$,  the vector ${\bf h}^{(i,j)}$ differs from zero only if the $i$-th node belongs to the element $\mathrm{T}'_j \in \mathcal{T}'$.

\subsubsection{Regularized deconvolution in forward modeling}
\label{deconvolution}

For the extensive number of the source vectors ${\bf h}_\ell^{(i,j)}$, solving all the system of the form (\ref{linearized1})--(\ref{linearized2}) via the leap-frog iteration would be extremely slow compared to the computation of the systems of the form (\ref{leap-frog1})--(\ref{leap-frog2}). A faster way to approach  forward modeling is to utilize  regularized deconvolution \cite{pursiainen2016} through the following steps ({\bf 1})--({\bf 4}):

\begin{description} \item[(1)] Place the source ${\bf \tilde{f}}$ at the point $\vec{p}_{1}$ and solve the system (\ref{leap-frog1})--(\ref{leap-frog2}) using the leap-frog iteration. Based on the solution, calculate and store ${\bf \tilde{h}}$ of the form (\ref{aku_ankka}) for a given node $\vec{r}$ and element $\mathrm{T}$ of the mesh $\mathcal{T}'$. 
\item[(2)] Place ${\bf \tilde{f}}$ at $\vec{p}_2$. Solve (\ref{leap-frog1})--(\ref{leap-frog2}) and store ${\bf \tilde{p}}$ of the form (\ref{leap-frog1}) at $\vec{r}$.  
\item[(3)] Estimate the Green's function ${\bf \tilde{g}}$ satisfying ${\bf \tilde{p}}={\bf \tilde{g}} \ast {\bf \tilde{f}} $ using Tikhonov regularized deconvolution with a suitably chosen regularization parameter $\delta$ \cite{kaipio2004}. 
\item[(4)] Based on the estimated ${\bf \tilde{g}}$ and the reciprocity of the signal wave  \cite{altman1991}, approximate ${\bf \tilde{d}}$, the solution of (\ref{linearized2})  at $\vec{p}_2$, through the convolution ${\bf \tilde{d}} ={\bf \tilde{g}} \ast {\bf \tilde{h}}$.
\end{description}
Here, ${\bf \tilde{f}}, {\bf \tilde{h}}, {\bf \tilde{p}}, {\bf \tilde{g}}$, and ${\bf \tilde{d}}$ denote vectors whose entries contain the pointwise time evolution of the corresponding variable. Backscattering data will be obtained, if $\vec{p}_1 = \vec{p}_2$.

\begin{figure}
\begin{center}
\begin{minipage}{6.5cm} \begin{center} \includegraphics[width=4.1cm]{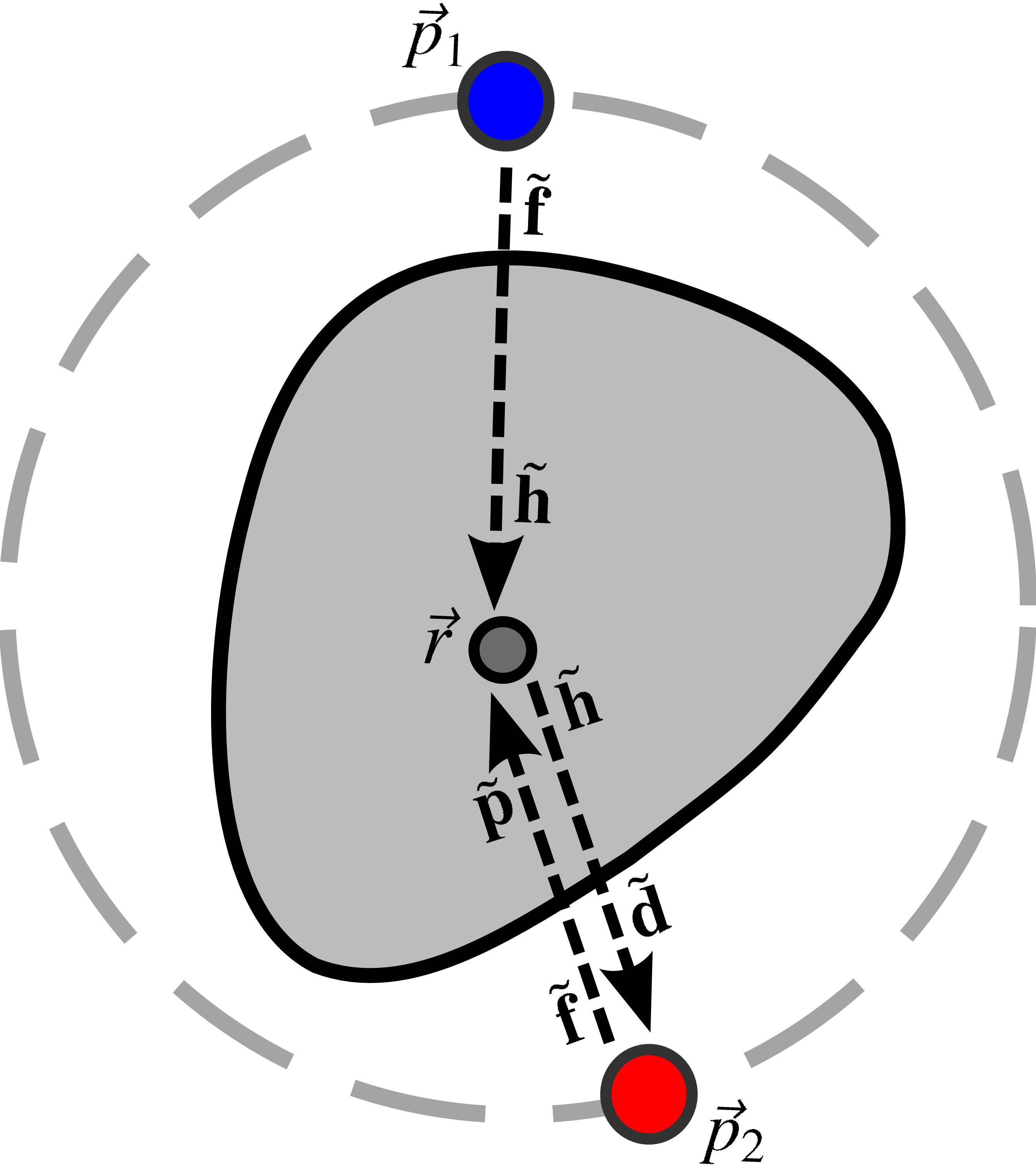} \end{center} \end{minipage}   
\end{center}
\caption{A schematic picture visualizing the regularized deconvolution-based solution of  (\ref{linearized1})--(\ref{linearized2}).  {\bf (1)} Place the source ${\bf \tilde{f}}$ at the point $\vec{p}_{1}$ and solve the system (\ref{leap-frog1})--(\ref{leap-frog2}) using the leap-frog iteration. Based on the solution, calculate and store ${\bf \tilde{h}}$ of the form (\ref{aku_ankka}) for a given node $\vec{r}$ and element $\mathrm{T}$ of the mesh $\mathcal{T}'$. 
{\bf (2)} Place ${\bf \tilde{f}}$ at $\vec{p}_2$. Solve (\ref{leap-frog1})--(\ref{leap-frog2}) and store ${\bf \tilde{p}}$ of the form (\ref{leap-frog1}) at $\vec{r}$.  
{\bf (3)} Estimate the Green's function ${\bf \tilde{g}}$ satisfying $ {\bf \tilde{p}} = {\bf \tilde{g}} \ast {\bf \tilde{f}}$ using Tikhonov regularized deconvolution with a suitably chosen regularization parameter $\delta$ \cite{kaipio2004}. 
{\bf (4)} Based on the estimated ${\bf \tilde{g}}$ and the reciprocity of the signal wave  \cite{altman1991}, approximate ${\bf \tilde{d}}$, the solution of (\ref{linearized2})  at $\vec{p}_2$, through the convolution ${\bf \tilde{d}} = {\bf \tilde{g}} \ast {\bf \tilde{h}} $.  \label{constellation_image}}
\end{figure}

\subsubsection{Multigrid approach to forward modeling}
\label{multiresolution}

The number of the terms in the sum ${\partial {\bf p}_\ell}/{\partial c_j}  = \sum_{\vec{r}_i \in \mathrm{T}'_j, \,  i \leq n}  {\bf d}_\ell^{(i,j)}$ depends on the density of the finite element mesh $\mathcal{T}$, that is, the number of nodes $\vec{r}_i \in \mathcal{T}$ belonging to $ \mathrm{T}'_j$. In order to lower this number, and thereby also reduce the computational work in forming the model, we redefine the source (\ref{aku_ankka}) with respect to the coarse mesh $\mathcal{T}'$ as \begin{equation} {\bf h'}_\ell^{(i,j)}    =  \frac{\partial {\bf C'}}{\partial c_j} {\bf Q'}^{(i)} \, {\bf b}_\ell^{(i)}. \end{equation}  Here, the entries of ${\bf C'}$ are of the form $C'_{i,j}  =  \int_\Omega \varepsilon_r \, \varphi'_i \varphi'_j  \, \hbox{d} \Omega$   with $\varphi'_i $  and $\varphi'_j$ denoting piecewise linear nodal basis functions  of $\mathcal{T}'$ and ${\bf Q'}^{(i)}\in \mathbb{R}^{N \times n}$ has a single nonzero entry ${\bf Q'}^{(i)}_{i,i} = 1$.  Denoting by  ${\bf d'}_\ell^{(i,j)}$ the regularized deconvolution-based solution (Section \ref{deconvolution}) of (\ref{linearized1})--(\ref{linearized2})  with respect to   ${\bf h'}_\ell^{(i,j)}    =   ({\partial {\bf C'}}/{\partial c_j}) {\bf b'}_\ell^{(i)}$, one obtains the estimate (\ref{approx_2}), that is, the basis of the present multigrid approach.

\bibliographystyle{IEEEtran}
\bibliography{references,references1,references2,mt_mscthesis_refs,takala_pursiainen_artikkeli_references}

\begin{IEEEbiography}[{\includegraphics[height=1in,height=1.25in,clip,keepaspectratio]{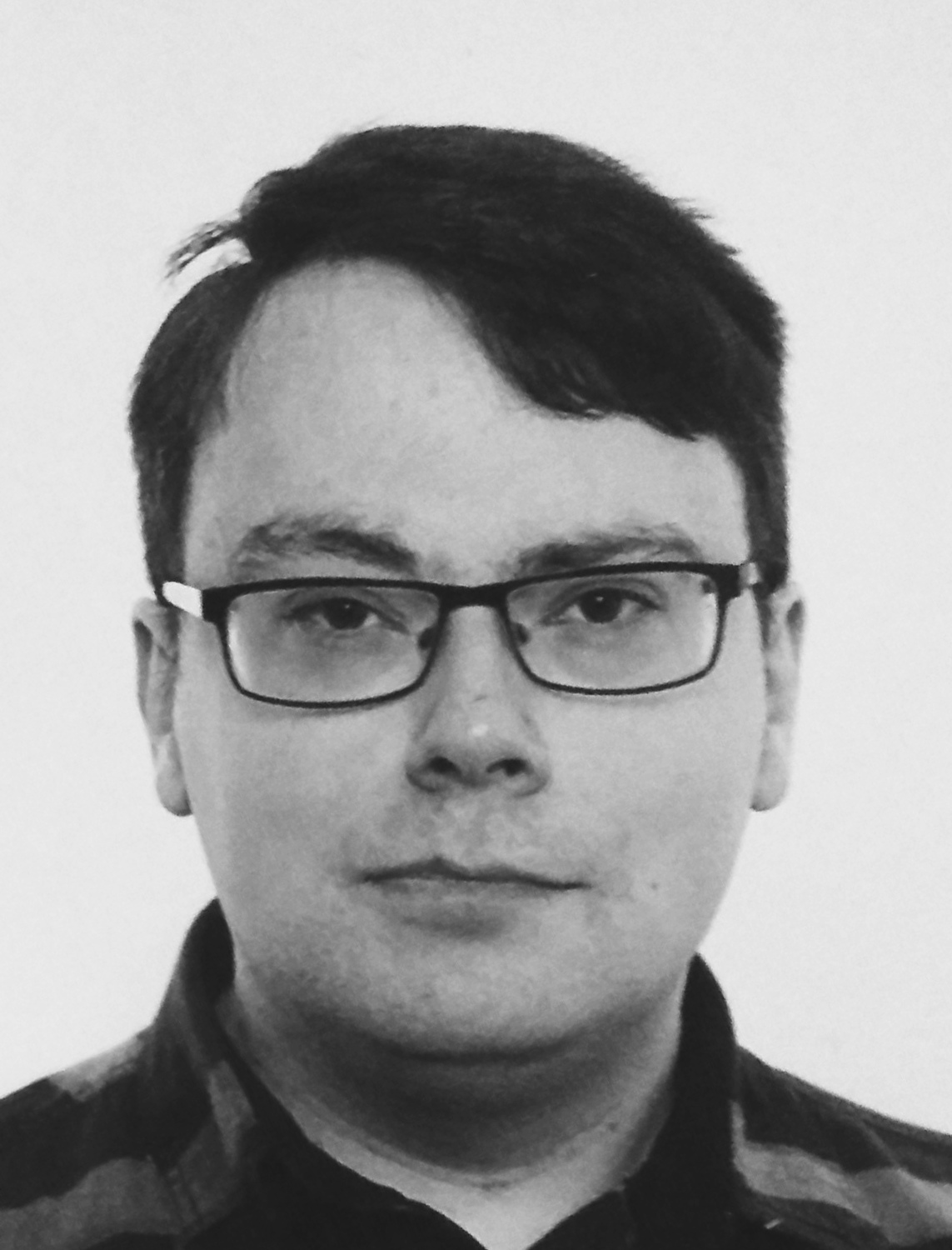}}]{Mika Takala} received the B.Sc. degree in electrical engineering from the Tampere University of Technology (TUT), Tampere, Finland in 2015, and the M.Sc.(tech.) degree from TUT in 2016. His master's thesis in the field of embedded systems concentrated on implementation of signal preprocessing modules with High-Level Synthesis for waveform inversion applications. 
In 2016 he started working at the Laboratory of Mathematics, TUT, as a PhD student. He currently works on his PhD research, which is closely related to geophysical inversion strategies witha  view towards computing aspects and embedded systems. He also works as a software architect at Granite Devices, Inc., Tampere, Finland.
\end{IEEEbiography}
\begin{IEEEbiography}[{\includegraphics[height=1in,height=1.25in,clip,keepaspectratio]{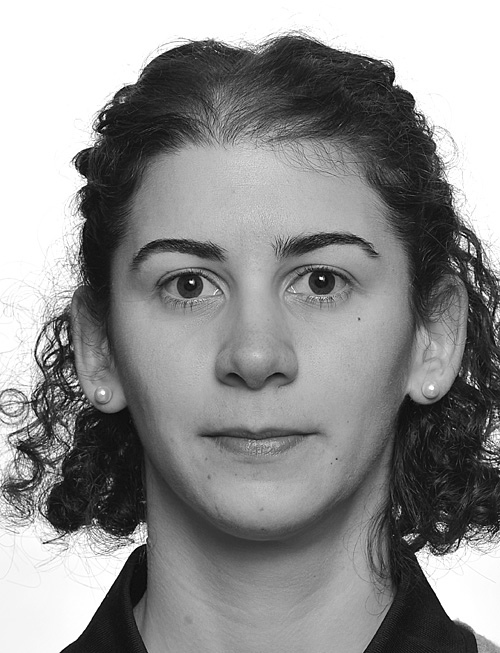}}]{Defne Us} received the B.Sc. degree in electrical and electronics engineering in 2010 from Bilkent University, Ankara, Turkey and M.Sc. degree in biomedical engineering in 2013 from the Tampere University of Technology, Tampere, Finland. She has worked towards Ph.D. degree within the Laboratory of Signal Processing between 2014-2017. She is now continuing her degree with laboratory of Mathematics. Her research interests include metal artifact reduction methods in tomography and medical image reconstruction.
\end{IEEEbiography}
\begin{IEEEbiography}[{\includegraphics[height=1in,height=1.25in,clip,keepaspectratio]{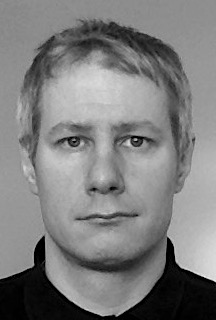}}]{Sampsa Pursiainen} received his MSc(Eng) 
and PhD(Eng) degrees (Mathematics)  in the Helsinki University of Technology (Aalto University since 2010), Espoo, Finland, in 2003 and 2009. He focuses on various forward and inversion techniques of applied mathematics. In 2010--11, he stayed at the Department of Mathematics, University of Genova, Italy collaborating also with the Institute for Biomagnetism and Biosignalanalysis (IBB), University of M\"{u}nster, Germany. In 2012--15, he worked at the  Department of Mathetmatics and System Analysis, Aalto University, Finland and also at the Department of Mathematics, Tampere University of Technology, Finland, where he currently continues as an Assistant Professor. 
\end{IEEEbiography}

\end{document}